%% file: CS_BOOK_STRUCTSPARSE.tex
\documentclass[11pt,a4paper]{extarticle}
\input{preamble}
\usepackage[left=1cm,right=1cm,top=1.5cm,bottom=1.5cm]{geometry}

\begin{document}

\title{Structured Sparsity: Discrete and Convex approaches}
\author{Anastasios Kyrillidis, Luca Baldassarre, Marwa El-Halabi, Quoc Tran-Dinh and Volkan Cevher \\
EPFL, Lausanne, Switzerland. }

\maketitle
\abstract{
Compressive sensing (CS) exploits sparsity to recover sparse or compressible signals from dimensionality reducing, non-adaptive sensing mechanisms. 
Sparsity is also used to enhance interpretability in machine learning and statistics applications: While the ambient dimension is vast in modern data analysis problems, the relevant information therein typically resides in a much lower dimensional space.
However, many solutions proposed nowadays do not leverage the true underlying structure.
Recent results in CS extend the simple sparsity idea to more sophisticated {\em structured} sparsity models, which describe the interdependency between the nonzero components of a signal, allowing to increase the interpretability of the results and lead to better recovery performance.
In order to better understand the impact of structured sparsity, in this chapter we analyze the connections between the discrete models and their convex relaxations, highlighting their relative advantages. 
We start with the general group sparse model and then elaborate on two important special cases: the dispersive and the hierarchical models. 
For each, we present the models in their discrete nature, discuss how to solve the ensuing discrete problems and then describe convex relaxations.
We also consider more general structures as defined by set functions and present their convex proxies. 
Further, we discuss efficient optimization solutions for structured sparsity problems and illustrate structured sparsity in action via three applications.}
\tableofcontents
\input{intro}
\input{prelim}
\input{group}
\input{dispersive}

\input{hierarchical}
\input{submodular}
\input{algorithms}
\input{experiments}



\bibliographystyle{plain}
\bibliography{biblio}

\end{document}

%% file: preamble.tex
\usepackage{palatino}
\usepackage{makeidx}         
\usepackage{multicol}        
\usepackage{url}
\usepackage{float}
\usepackage{epsfig}
\usepackage{color}
\usepackage{dsfont}
\usepackage{amsmath,epsfig,citesort,cite,multirow}
\usepackage{graphicx}
\usepackage{amssymb}
\usepackage{algorithm}
\usepackage{algorithmic}
 \usepackage{tikz} 
 \usetikzlibrary{arrows,automata}
\usepackage{enumerate}
\usepackage{amsxtra}
\usepackage{epsf}
\usepackage{caption}
\usepackage{subcaption}
\usepackage{psfrag}
\usepackage{hyperref}
\usepackage{bbm}
\usepackage[mathcal]{euscript}
\usepackage{booktabs}
\usepackage{wrapfig}

\newcommand{\signal}{\mathbf{x}}

\newcommand{\sensing}{\boldsymbol{\Phi}}
\newcommand{\noise}{\boldsymbol{\varepsilon}}

\newcommand{\obs}{\mathbf{u}}
\newcommand{\xtrue }{\mathbf{x}^{\star}}

\newcommand{\abs}[1]{\left\vert#1\right\vert}

\newcommand{\Real}{\mathbb R}

\newcommand{\A}{{\bf A}}

\newcommand{\R}{ \mathbb{R} }     
\newcommand{\reals}{\mathbb R}
\newcommand{\sparse}{\Sigma}
\newcommand{\numsam}{m}
\newcommand{\N}{\mathcal{N}}

\newcommand{\atomic}{\mathcal{A}}
\newcommand{\atom}{\mathbf{a}}

\newcommand{\vectornorm}[1]{\|#1\|}

\def\S{\mathcal{S}}

\newcommand{\Id}{\mathbf{I}}
\newcommand{\1}{\mathbbm{1}}
\newcommand{\BB}{\mathbb B} 
\newcommand{\GG}{\mathfrak G} 
\newcommand{\G}{\mathcal G} 
\newcommand{\ngroups}{M}

\newcommand{\sss}{\mathcal{S}} 
\newcommand{\T}{\mathcal{T}}
\newcommand{\constraint}{\mathcal{M}_\sparsity}
\newcommand{\constraintG}{\mathcal{M}_{G}}

\newcommand{\dispersive}{\mathcal{D}_{\sparsity}}
\DeclareMathOperator{\supp}{supp}

\def\x{{\mathbf x}}
\def\w{\mathbf w}

\def\D{\mathbf D}

\newtheorem{prop}{Proposition}

\newcommand{\bitem}{\begin{itemize}}
\newcommand{\eitem}{\end{itemize}}

\newcommand{\argmin}{\operatornamewithlimits{argmin}}
\newcommand{\argmax}{\operatornamewithlimits{argmax}}

\newcommand{\beqn}{\begin{equation}}
\newcommand{\eeqn}{\end{equation}}
\newcommand{\balign}{\begin{align}}
\newcommand{\ealign}{\end{align}}

\def\proj { \mathcal{P} } 
\def \dim {n}      
\def \sparsity {k} 

\DeclareMathOperator{\tr}{trace}

\newtheorem{theorem}{Theorem}[section]
\newtheorem{lemma}[theorem]{Lemma}

\newtheorem{definition}[theorem]{Definition}

%% file: intro.tex
\section{Introduction}

Information in many natural and man-made signals can be exactly represented or well approximated by a sparse set of nonzero coefficients in an appropriate basis \cite{mallat1999wavelet}. Compressive sensing (CS) \cite{donoho2006compressed, candes2006compressive, stojnic2008compressed} exploits this fact to recover signals from their compressive samples through dimensionality reducing, non-adaptive sensing mechanisms. From a different perspective, sparsity is used to enhance \emph{interpretability} in machine learning \cite{girosi1998equivalence, guigue2005kernel} and statistics \cite{tibshirani1996regression} applications: While the ambient dimension is vast in modern data analysis problems, the relevant information therein typically resides in a much lower dimensional space. This conclusion has lead to several new theoretical and algorithmic developments in different communities, including theoretical computer science \cite{gilbert2010sparse}, neuronal imaging \cite{gramfort2009improving,jenatton2011multiscale}, gene expression inference \cite{subramanian2005gene,obozinski2011group}, bioinformatics \cite{rapaport2008classification, zhou2010association} and X-ray crystalography \cite{millane1990phase, candes2013phase, jaganathan2013sparse}. In all these disciplines, given a data set to analyze, one is usually interested in the \emph{simplest} model that well explains the observations. 

Apart from the rather mature theory describing the sparse signal approximation problem, its success lies also in the computational tractability of such task. In the CS regime, Donoho \cite{donoho2006compressed} and Candes et\ al.\ \cite{candes2006compressive} utilize simple convex optimization---i.e., linear programming solvers or quadratic programming methods---and matrix isometry assumptions over sparsely restricted sets to compute a sparse approximation from a limited set of linear measurements in polynomial time. Alternatively and in contrast to the conventional convex relaxation approaches, iterative greedy algorithms \cite{needell2009cosamp,foucart2011hard,kyrillidis2011recipes} rely on the discrete structure of the sparse signal approximation problem to find a solution and are characterized by identical approximation guarantees as in \cite{donoho2006compressed, candes2006compressive}. Moreover, their overall complexity matches, if not improves, the computational cost of convex approaches. 

\subsection{Why structured sparsity?}

While many of the optimization solutions proposed nowadays result in \emph{sparse} model selections, in many cases they do not capture the true underlying structure to best explain the observations \cite{baraniuk2010model}. In fact, this {\em uninformed} {selection} process has been the center of attention in sparse approximation theory \cite{kyrillidis2012combinatorial, zeng2013novel}, because it not only prevents interpretability of results in many problems, but also fails to exploit key prior information that could radically improve estimation performance. 

\begin{figure}[tb]
\captionsetup{width=0.8\textwidth}
\begin{minipage}{0.29\columnwidth}
\centering Original
\end{minipage}
\begin{minipage}{0.29\columnwidth}
\centering Simple sparsity
\end{minipage}
\begin{minipage}{0.29\columnwidth}
\centering Structured sparsity
\end{minipage}
\centering
\includegraphics[width=0.29\columnwidth]{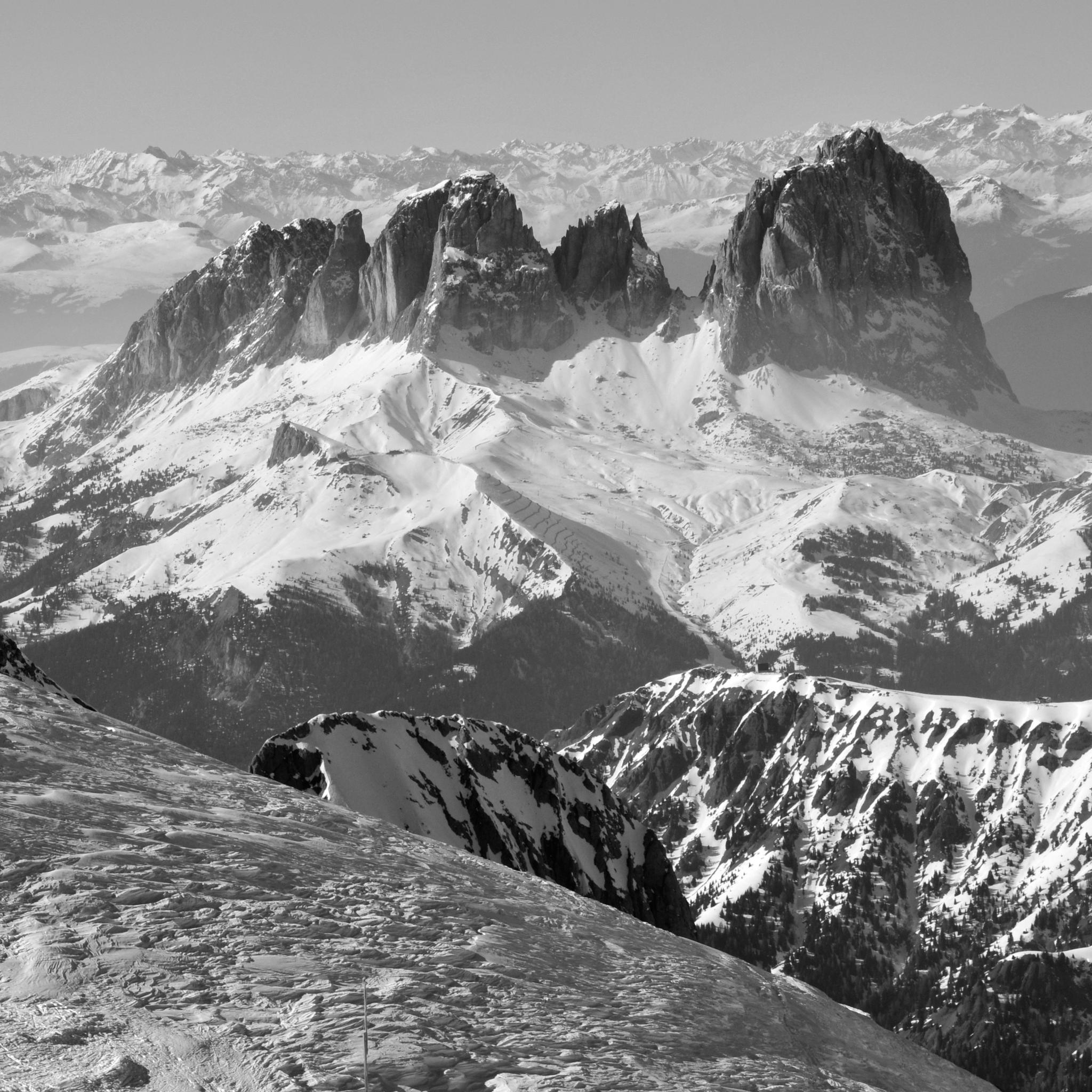} 
\includegraphics[width=0.29\columnwidth]{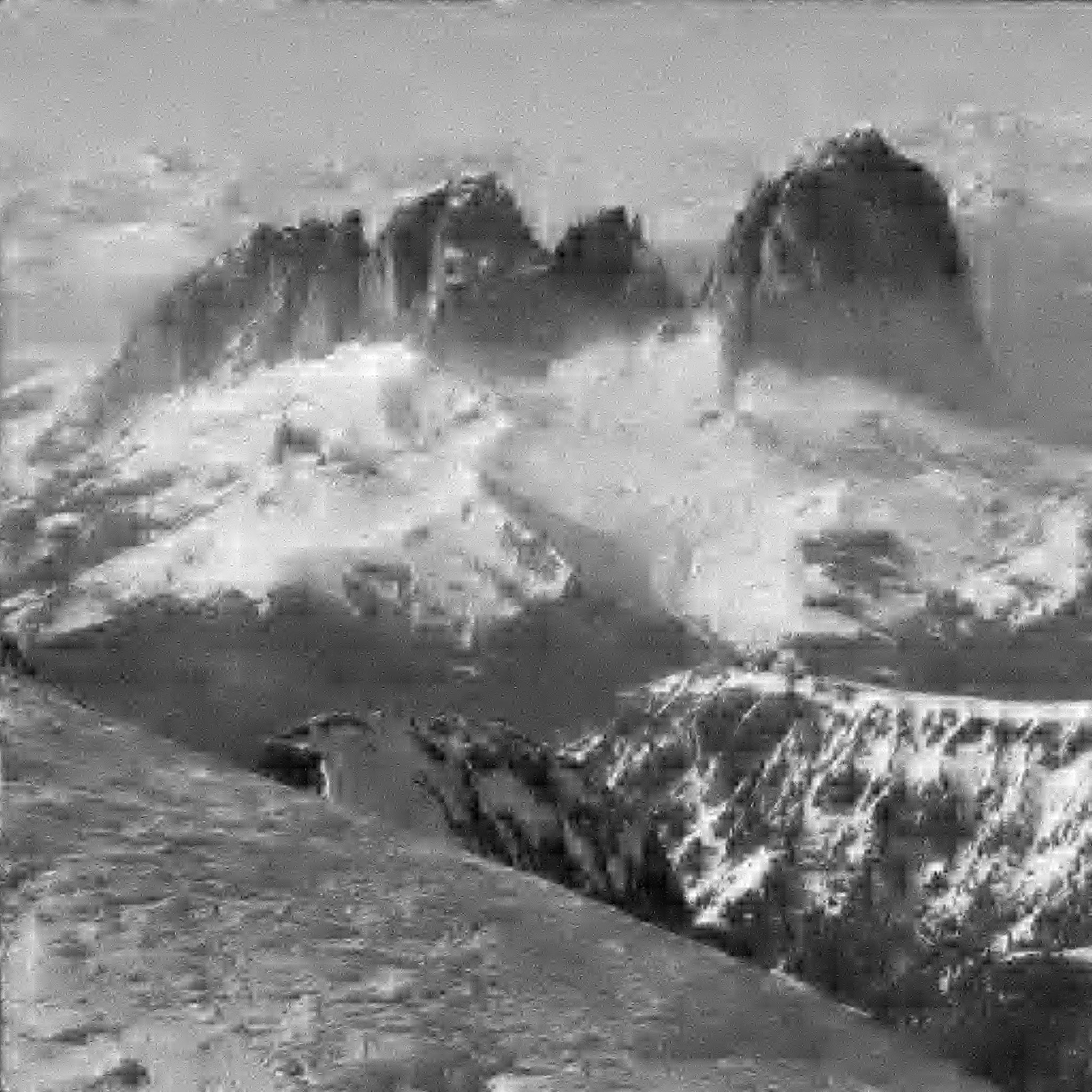} 
\includegraphics[width=0.29\columnwidth]{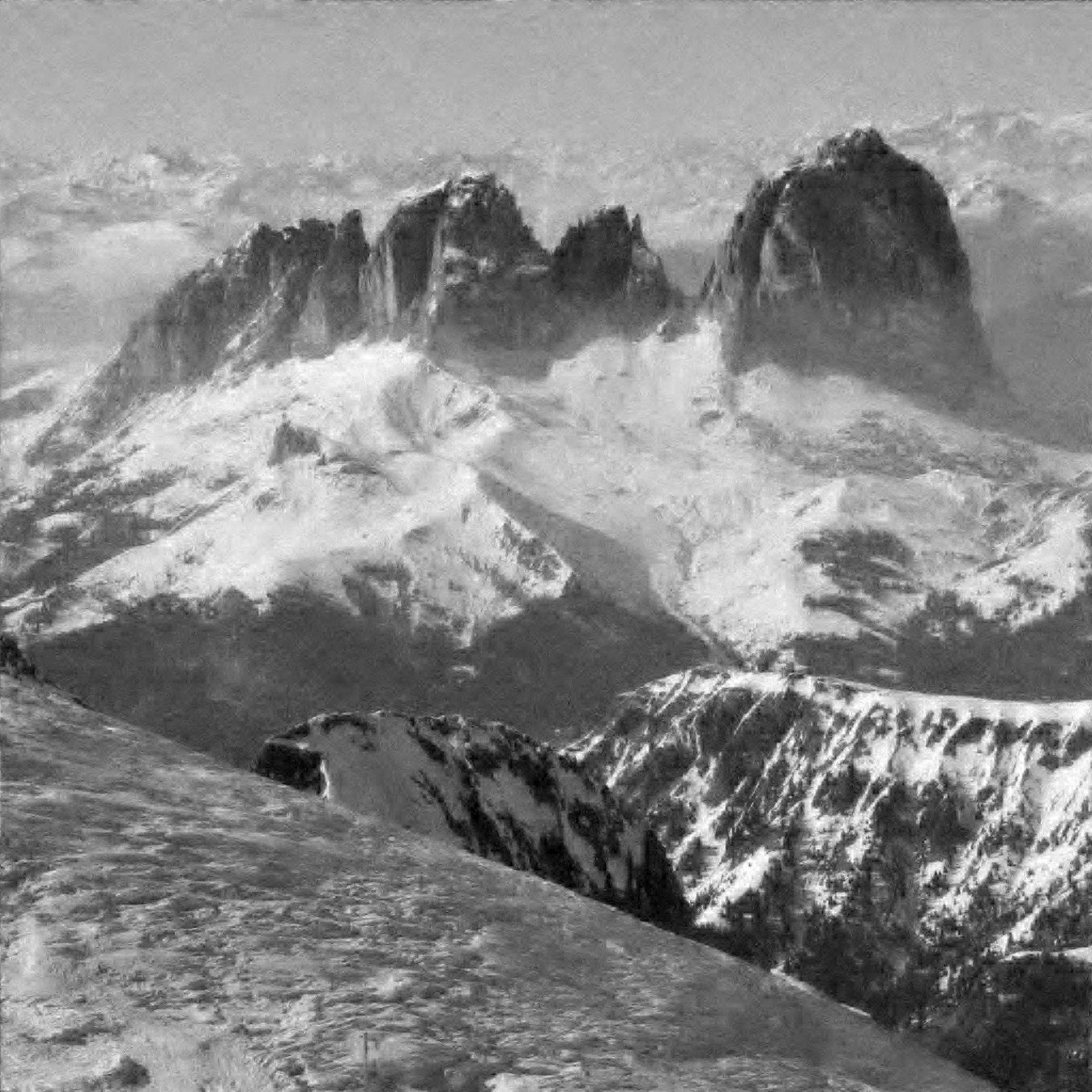} \\ \vspace{-0.15cm}
\begin{minipage}{0.29\columnwidth}
\centering \quad
\end{minipage}
\begin{minipage}{0.29\columnwidth}
\centering PSNR: $20.43$ dB
\end{minipage}
\begin{minipage}{0.29\columnwidth}
\centering PSNR: $23.56$ dB
\end{minipage}  \\ \vspace{0.1cm}
\includegraphics[width=0.29\columnwidth]{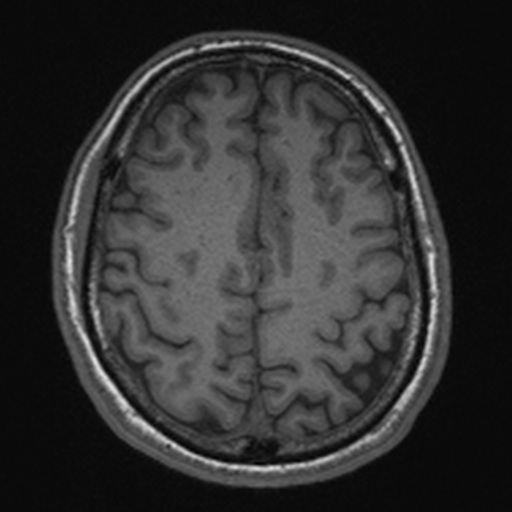} 
\includegraphics[width=0.29\columnwidth]{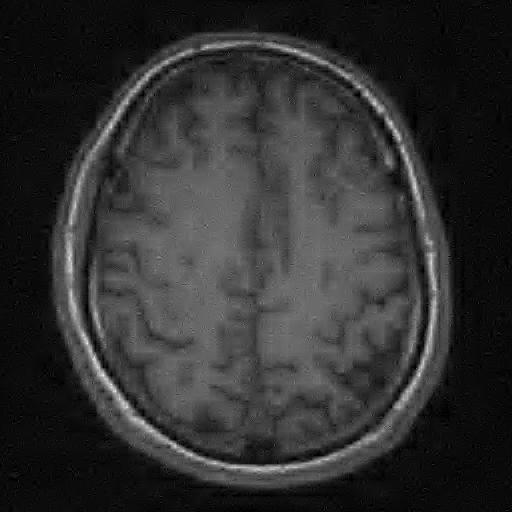} 
\includegraphics[width=0.29\columnwidth]{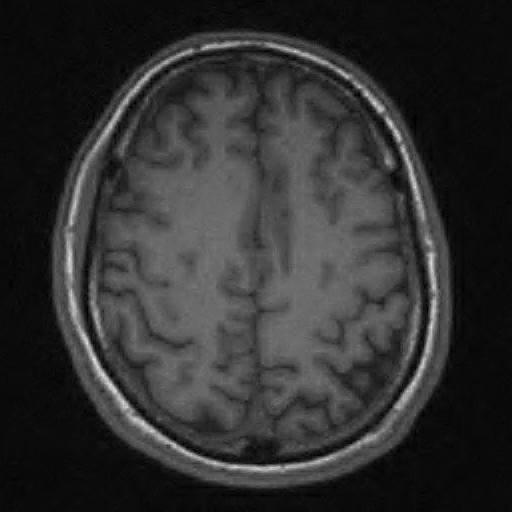} \\ \vspace{-0.15cm}
\begin{minipage}{0.29\columnwidth}
\centering \quad
\end{minipage}
\begin{minipage}{0.29\columnwidth}
\centering PSNR: $28.43$ dB
\end{minipage}
\begin{minipage}{0.29\columnwidth}
\centering PSNR: $33.24$ dB
\end{minipage}
\caption{\label{fig:image_tree} Empirical performance of simple and structured sparsity recovery on natural images. In all cases, the number of linear measurements are $5\%$ (top row) and $10\%$ (bottom row) of the actual image dimensions. \textbf{Left panel:} Original images of dimension: (Top row) $2048 \times 2048$, (Bottom row) $512 \times 512$. \textbf{Middle panel:} Conventional recovery using simple sparsity model. \textbf{Right panel:} Tree-structured sparse recovery.}
\end{figure}

Recent results in CS extend the simple sparsity idea to consider more sophisticated {\em structured} sparsity models, which describe the interdependency between the nonzero coefficients, increase the interpretability of the results and lead to better recovery performance \cite{eldar2009robust, blumensath2009sampling, baraniuk2010model, rao2012signal}. Nowadays, we have witnessed aplenty elaborate approaches that guide the selection process: (overlapping) group Lasso, fused Lasso, greedy approaches for signal approximation under tree-structure assumptions just to name a few; see \cite{yuan2006model, friedman2010note, tibshirani2005sparsity, jenatton2011structured}. To show the merits of such approaches, consider the problem of image recovery from a \emph{limited} set of measurements using the tree-structured group sparse model \cite{baraniuk1999optimal}. Figure \ref{fig:image_tree} shows the performance of structured sparsity models, compared to simple ones. 

Moreover, such a priori model-based assumptions result into more robust solutions and allow recovery with far fewer samples, e.g. $O(\sparsity)$ samples for $k$ sparse signals whose non-zero coefficients are arranged into few blocks or form a rooted connected subtree of a given tree over the coefficients  \cite{baraniuk2010model}. To highlight the importance of this property, in the case of Magnetic Resonance Imaging (MRI), reducing the total number of measurements is highly desirable for both capturing functional activities within small time periods and rendering the whole procedure less ``painful'' for the patient \cite{lustig2007sparse}. 

%% file: prelim.tex
\section{Preliminaries}
\label{sec:prelim}



Plain lowercase and uppercase letters represent scalars while boldface lowercase letters represent vectors. We reserve boldface uppercase letters for matrices. Calligraphic uppercase letters are mostly used to denote sets. The $i$-th entry of a vector $\w$ is denoted as $w_i$. 
We use superscripts such as $\w^i$ to denote the estimate at the $i$-th iteration of an algorithm. The sign of a scalar $\alpha \in \R$ is given by $\text{sign}(\alpha)$.

Given a set $\mathcal{S} \subseteq \mathcal{N} := \{1,\ldots, n\}$, the complement $\mathcal{S}^c$ is defined with respect to $\mathcal{N}$, and its cardinality as $|\mathcal{S}|$. The support set of $ \w $ is $\text{supp}(\w) = \lbrace i \in \mathcal{N}~:~w_i \neq 0 \rbrace $. Given a vector $ \w \in \R^n $, $ \w_{\mathcal{S}} $ is the projection (in $\R^n$) of $\w$ onto $\mathcal{S}$, i.e.~$\left(\w_{\mathcal{S}}\right)_{\mathcal{S}^c}= \mathbf{0} $, whereas $ \w_{|\mathcal{S}} \in \R^{|\mathcal{S}|}$ is $\w$ limited to $\mathcal{S}$ entries. We define $\sparse_\sparsity := \left \{\w ~:~ \w \in \R^n, ~|\text{supp}(\w)| \leq k \right\}$ as the set of all $\sparsity$-sparse vectors in $n$-dimensions; with a slight abuse of notation, we use $\sparse_\sparsity $ to further denote the set of $\sparsity$-sparse supports in $\mathcal{N}$. We sometimes write $\x \in \sparse_\sparsity$ to mean $\text{supp}(\x)\in \sparse_\sparsity$. 

We use $\mathbb{B}^\dim$ to represent the space of $\dim$-dimensional binary vectors and define $\iota: \Real^\dim \to \mathbb{B}^\dim$ to be the indicator function of the nonzero components of a vector in $\Real^\dim$, i.e., $\iota({\bf x})_i = 1$ if $x_i \neq 0$ and $\iota({\bf x})_i = 0$, otherwise.
We let $\1_\dim$ to be the $\dim$-dimensional vector of all ones,  $\1_{\dim,\sss}$ the $\dim$-dimensional vector of all ones projected onto $\sss$ and $\Id_\dim$ the $\dim \times \dim$ identity matrix; we often use $\Id$ when dimension is clear from the context.

\medskip
\noindent \textbf{Norms:} For completeness, we repeat some of the norm definitions presented in the Introduction chapter of this book. We define the $\ell_p^n$-norm in $n$-dimensions as: 
\begin{align*}
\|\x\|_p = \left\{
	\begin{array}{ll}
		\left( \sum_{i = 1}^n |x_i|^p \right)^{1/p}  & \mbox{if } p \in (0, \infty), \\
		\max_{i} |x_i| & \mbox{if } p = \infty.
	\end{array}
\right.
\end{align*} The $\ell_0$ pseudo-norm is defined as:
$\|\x\|_0 := |\text{supp}(\x)|.$


\medskip
\noindent \textbf{Projections operations:} Given a discrete set $\constraint$ with sparsity level $\sparsity$ and an anchor point $\x \in \R^\dim$, a key operation in our subsequent discussions is the following projection problem:
\begin{equation}
\mathcal{P}_{\constraint}(\signal) \in \argmin_{\w \in \mathbb{R}^{\dim}}\left\{\vectornorm{\w - \signal}_2^2 ~\big|~ {\rm  supp}(\w) \in \constraint \right\}.
\label{eq:proj} 
\end{equation}

\medskip
\noindent \textbf{Proximity operations:} Consider $g(\cdot): \R^n \rightarrow \R$ is a regularizer function. 
We define the proximity operator of $g(\cdot)$ as \cite[eq. (2.13)]{combettes2005signal}:
\begin{align}
\text{prox}_{\lambda}^g(\x):= \argmin_{\w \in \R^n} \left\{\frac{1}{2}\|\w - \x \|_2^2 + \lambda \cdot g(\w) \right\}, \label{eq:prox}
\end{align} where $\lambda > 0$ is the regularizer weight.

\medskip
\noindent \textbf{Optimization preliminaries:} Existing algorithmic solutions invariably rely on two structural assumptions on the objective function that particularly stand out among many others: the \emph{Lipschitz continuous gradient} assumption\and the \emph{strong regularity} condition.

\begin{definition}(Lipschitz gradient continuity)\label{def:lipschitz}
Let $f:~\R^\dim \rightarrow \R$ be a convex, smooth differentiable function. Then, $f$ is a smooth Lipschitz continuous gradient function if and only if for any $\mathbf{v},~\mathbf{w} \in \text{dom}(f)$:
\begin{align*}
\| \nabla f(\mathbf{v}) - \nabla f(\w) \|_2 \leq L \|\mathbf{v} - \w\|_2,
\end{align*} for some global constant $L > 0$.
\end{definition} 

\begin{definition}(Strong regularity condition)\label{def:regularity}
Let $f:~\R^\dim \rightarrow \R$ be a $L$-Lipschitz convex, doubly differentiable function. Then, $f$ is strongly convex if and only if:
\begin{align*}
\mu \mathbf{I} \preceq \nabla^2f(\x) \preceq L\mathbf{I},\quad \forall \x \in \text{dom}(f), 
\end{align*} for some global constant $\mu > 0$.
\end{definition}

%% file: group.tex
\section{Sparse group models}
\label{sec:groups}

\begin{figure}[ht]
\centering
\begin{minipage}[b]{0.29\linewidth}
\centering
\includegraphics[width=1\linewidth]{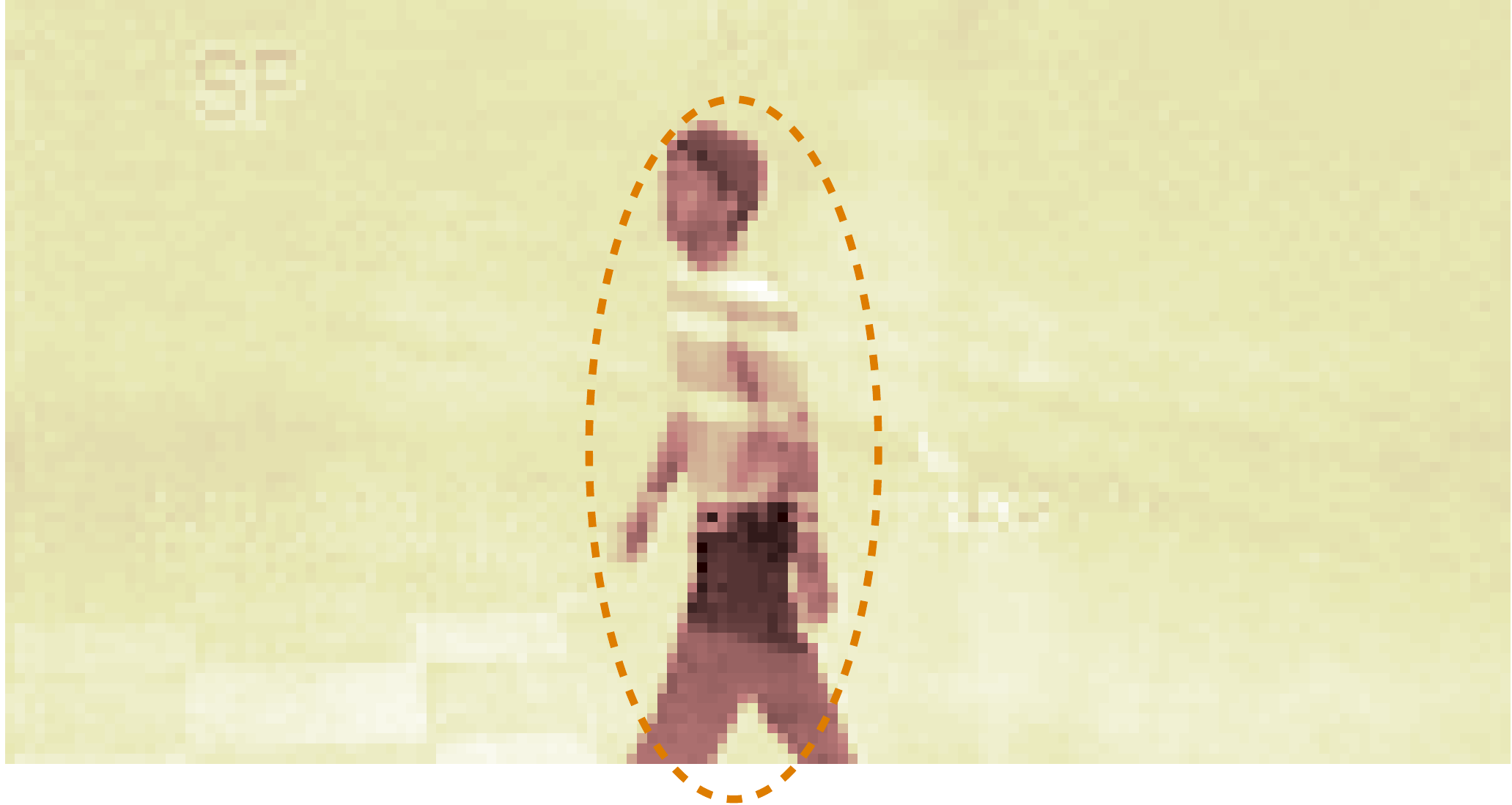} \\
\includegraphics[width=1\linewidth]{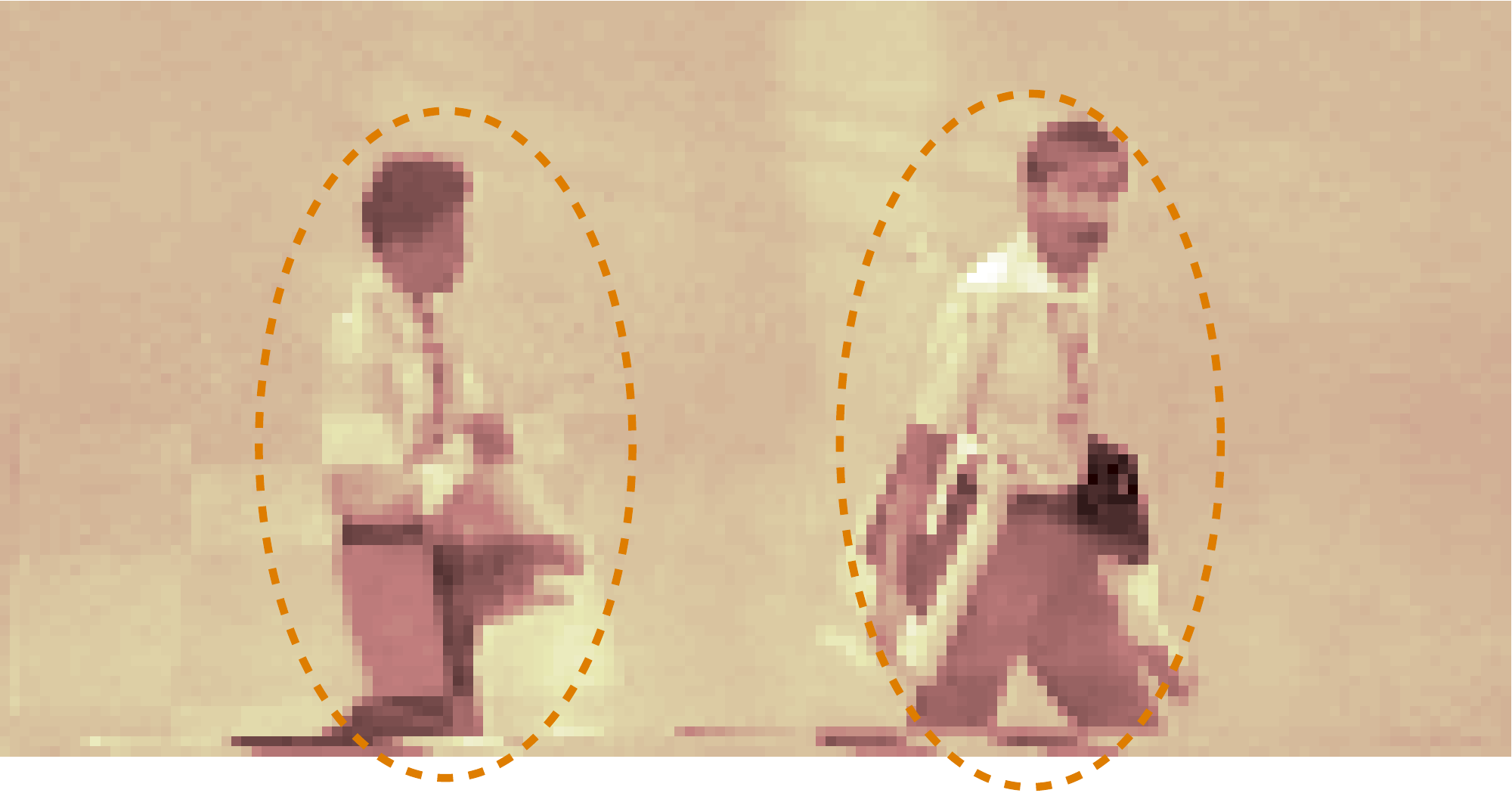} 
\caption{Image segmentation application: the signal of interest includes human activity, expressed in groups.}
\label{fig:group_example1}
\end{minipage}
\hspace{0.5cm}
\begin{minipage}[b]{0.39\linewidth}
\centering
\includegraphics[width=1.13\linewidth]{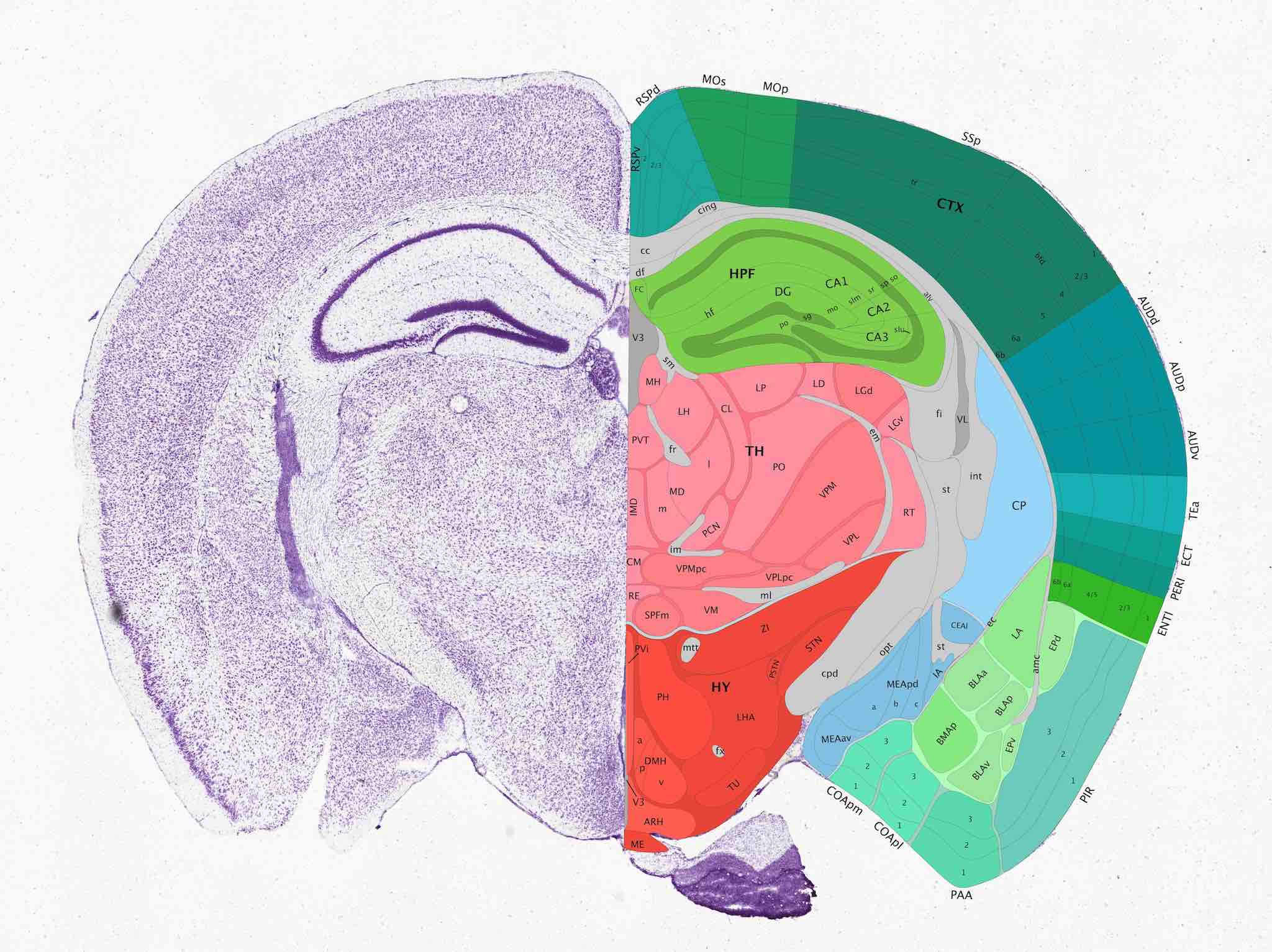} 
\caption{Example of a mouse brain where brain regions are represented as groups of voxels \cite{lein2007genome} (\copyright2014 Allen Institute for Brain Science).}
\label{fig:group_example2}
\end{minipage}
\end{figure}

We start our discussion with the \emph{group sparse} models, i.e., models where groups of variables are either selected or discarded together \cite{baraniuk2010low, jenatton2011structured, obozinski2011group, rao2012signal, rao2011convex, huang2010benefit}.
These structures naturally arise in applications such as neuroimaging \cite{gramfort2009improving,jenatton2011multiscale}, gene expression data \cite{subramanian2005gene,obozinski2011group}, bioinformatics \cite{rapaport2008classification, zhou2010association} and computer vision \cite{cevher2009sparse,baraniuk2010model}.
For example, in cancer research, the groups might represent genetic pathways that constitute cellular processes. 
Identifying which processes lead to the development of a tumor can allow biologists to directly target certain groups of genes instead of others \cite{subramanian2005gene}. 
Incorrect identification of the active/inactive groups can thus have a rather dramatic effect on the speed at which cancer therapies are developed.
Figures \ref{fig:group_example1}-\ref{fig:group_example2} illustrate some more applications of group sparse models used in practice.

Such group sparsity models---denoted as $\GG$---feature collections of groups of variables that could overlap arbitrarily; that is $\GG = \{\G_1, \ldots, \G_M\}$ where each $\G_j$ is a subset of the index set $\mathcal{N} := \{1, \ldots, \dim\}$. Arbitrary overlaps means that we do not restrict the intersection between any two sets $\G_j$ and $\G_\ell$ from $\GG$, $j \neq \ell$. 


We can represent a group structure $\GG$ as a bipartite graph, where on one side we have the $\dim$ variables nodes and on the other the $\ngroups$ group nodes. 
An edge connects a variable node $i$ to a group node $j$ if $i \in \G_j$. 
The bi-adjacency matrix $\A \in \mathbb{B}^{\dim \times \ngroups}$ of the bipartite graph encodes the group structure,
$$
\label{eq:group_structure}
A_{ij} = \bigg \{ \begin{array}{lc} 1, & \text{if}~i \in \G_j; \\ 0, & \text{otherwise.} \end{array} \; 
$$
Figure~\ref{fig:var_groups} shows an example.

\tikzstyle{vnode}=[circle,draw=black,fill=white,thick, minimum size=6pt, inner sep=0pt]
\tikzstyle{vnodeholder}=[circle,draw=black,fill=white,thick, minimum size=1pt, inner sep=0pt]
\tikzstyle{gnode}=[rectangle,draw=black,fill=white,thick, minimum size=6pt, inner sep=0pt]
\tikzstyle{gnodeholder}=[rectangle,draw=black,fill=white,thick, minimum size=1pt, inner sep=0pt]

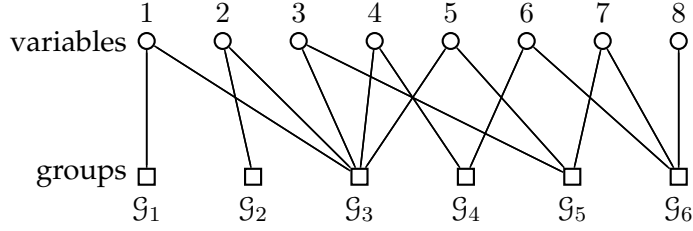
\begin{figure}
\captionsetup{width=0.8\textwidth}
\centering
\begin{tikzpicture}[-,>=stealth',shorten >=1pt,auto,node distance=1.5cm, semithick]

\node[vnode] (v1) at (0,0) [label=above:$1$] [label=left:variables] {};
\node[vnode] (v2) at (1,0) [label=above:$2$] {};
\node[vnode] (v3) at (2,0) [label=above:$3$] {};
\node[vnode] (v4) at (3,0) [label=above:$4$] {};
\node[vnode] (v5) at (4,0) [label=above:$5$] {};
\node[vnode] (v6) at (5,0) [label=above:$6$] {};
\node[vnode] (v7) at (6,0) [label=above:$7$] {};
\node[vnode] (v8) at (7,0) [label=above:$8$] {};

\node[gnode] (g1) at (0,-1.8) [label=below:$\G_1$] [label=left:groups] {};
\node[gnode] (g2) at (1.4,-1.8) [label=below:$\G_2$] {};
\node[gnode] (g3) at (2.8,-1.8) [label=below:$\G_3$] {};
\node[gnode] (g4) at (4.2,-1.8) [label=below:$\G_4$] {};
\node[gnode] (g5) at (5.6,-1.8) [label=below:$\G_5$] {};
\node[gnode] (g6) at (7,-1.8) [label=below:$\G_6$] {};
	
\draw (v1) to (g1);\draw (g1) to (v1);\draw (v2) to (g2);\draw (g2) to (v2);
\draw (v1) to (g3);\draw (v2) to (g3);\draw (v3) to (g3);\draw (v4) to (g3);\draw (v5) to (g3);\draw (g3) to (v1);\draw (g3) to (v2);\draw (g3) to (v3);\draw (g3) to (v4);\draw (g3) to (v5);
\draw (v3) to (g5); \draw (g5) to (v3);
\draw (v4) to (g4);\draw (v6) to (g4);\draw (g4) to (v4);\draw (g4) to (v6);
\draw (v5) to (g5);\draw (v7) to (g5);\draw (g5) to (v5);\draw (g5) to (v7);
\draw (v6) to (g6);\draw (v7) to (g6);\draw (v8) to (g6);\draw (g6) to (v6);\draw (g6) to (v7);\draw (g6) to (v8);

\end{tikzpicture}
\caption{\label{fig:var_groups} Consider the group structure $\GG^1$ defined by the following groups, $\G_1 = \{1\}$, $\G_2 = \{2\}$, $\G_3 = \{1, 2, 3, 4, 5\}$, $\G_4 = \{4,6\}$, $\G_5 = \{3, 5, 7\}$ and $\G_6 = \{6, 7, 8\}$. }
\end{figure}

Another useful representation of a group structure is via a {\em group graph} $(\mathcal{V}, \mathcal{E})$ where the nodes $\mathcal{V}$ are the groups $\G \in \mathfrak{G}$ and the edge set $\mathcal{E}$ contains $e_{ij}$ if $\G_i \cap \G_j \neq \emptyset$, that is an edge connects two groups that {\em overlap}. 
A sequence of connected nodes $v_1, v_2, \ldots, v_n$, is a {\em loop} if $v_1 = v_n$. For an illustrative example see Figure \ref{fig:bipartite}.


\tikzstyle{place_black}=[circle,draw=black,fill=black,thick, minimum size=6pt, inner sep=0pt]
\tikzstyle{place_red}=[circle,draw=black,fill=white,thick, minimum size=6pt, inner sep=0pt]
\begin{figure}
\captionsetup{width=0.8\textwidth}
\centering
\begin{tikzpicture}[-,>=stealth',shorten >=1pt,auto,node distance=2.8cm, semithick]
  \tikzstyle{every state}=[fill=blue!20,draw=none,text=black]
	\node[place_black] (n1) at (0,0) [label=below:$\G_1$] {};
	\node[place_black] (n2) at (0,1.5) [label=below:$\G_2$] {};
	\node[place_red] (n3) at (2.25,0.75) [label=below:$\G_3$] {};
	\node[place_black] (n4) at (4.5,0) [label=below:$\G_4$] {};
	\node[place_black] (n5) at (4.5,1.5) [label=below:$\G_5$] {};
	\node[place_red] (n6) at (6.75,0.75) [label=below:$\G_6$] {};
  
	\path (n1) edge node {} (n3);
	\path (n2) edge node {$\{2\}$} (n3);
	\path (n3) edge node {} (n4);
	\path (n3) edge node {$\{3, 5\}$} (n5);
	\path (n4) edge node {} (n6);
	\path (n5) edge node {$\{7\}$} (n6);
	
	\path (n3) edge node {$\{1\}$} (n1);
	\path (n3) edge node {} (n2);
	\path (n4) edge node {$\{4\}$} (n3);
	\path (n5) edge node {} (n3);
	\path (n6) edge node {$\{6\}$} (n4);
	\path (n6) edge node {} (n5);
\end{tikzpicture}
\caption{\label{fig:bipartite} Bipartite group graph with loops induced by the group structure $\GG^1$ described in Figure \ref{fig:var_groups}, where on each edge we report the elements of the intersection.}
\end{figure}
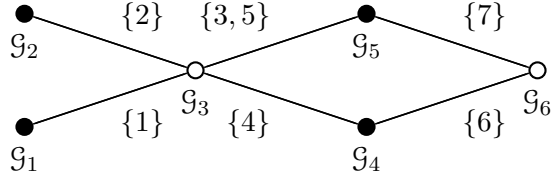

\tikzstyle{place}=[circle,draw=black,fill=white,thick, minimum size=6pt, inner sep=0pt]
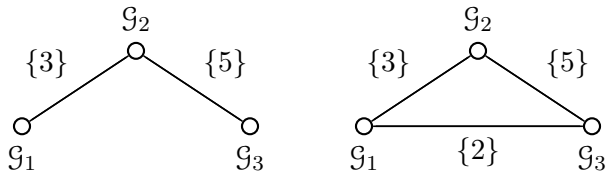
\begin{figure}[tb]
\captionsetup{width=0.8\textwidth}
\centering
\begin{tikzpicture}[-,>=stealth',shorten >=1pt,auto,node distance=1.5cm, semithick]
	\node[place] (n1) at (0,0)    [label=below:$\G_1$] {};
	\node[place] (n2) at (1.5,1) [label=above:$\G_2$] {};
	\node[place] (n3) at (3,0)    [label=below:$\G_3$] {};
	
\draw (n1) to node {$\{3\}$} (n2);
\draw (n2) to node {} (n1);
\draw (n2) to node {$\{5\}$} (n3);
\draw (n3) to node {} (n2);

	\node[place] (n4) at (4.5,0)    [label=below:$\G_1$] {};
	\node[place] (n5) at (6,1)       [label=above:$\G_2$] {};
	\node[place] (n6) at (7.5,0)    [label=below:$\G_3$] {};
\draw (n4) to node {$\{3\}$} (n5);
\draw (n5) to node {$\{5\}$} (n6);
\draw (n6) to node [label=below:$\{2\}$] {} (n4);

\draw (n5) to node {} (n4);
\draw (n6) to node {} (n5);
\draw (n4) to node {} (n6);

\end{tikzpicture}
\caption{\label{fig:loopless} Consider $\G_1 = \{1, 2, 3\}$, $\G_2 = \{3, 4, 5\}$, $\G_3 = \{5, 6, 7\}$, which can be represented by the graph in the left plot. 
If $\G_3$ were to include an element from $\G_1$, for example $\{2\}$, we would have the loopy graph of the right plot.
Note that $\GG^1$ is pairwise overlapping, but not loopless, since $\G_3, \G_4, \G_5$ and $\G_6$ form a loop.
}
\end{figure}

An important group structure is given by {\em loopless pairwise overlapping groups} because it leads to tractable projections \cite{baldassarre2013group}.
This group structure consists of groups such that each element of the ground set occurs in at most two groups and the induced graph does not contain loops. 
Therefore the group graph for these structures is actually a tree or a forest and hence bipartite; see Figure \ref{fig:loopless}.

Given the above and for a user-defined group budget $G \in \mathbb{Z}_{+}$, we define the group model $\constraintG$ as $\constraintG := \{ \bigcup_{\G_\ell \in \mathcal{I}} \G_{\ell}, \mathcal{I} \subseteq \GG, |\mathcal{I}| \leq G \}$, that is all sets of indexes that are the union of at most $G$ groups from the collection $\GG$. Then, the projection problem \eqref{eq:proj} becomes: 
\begin{equation}
\label{eq:group_approx_general}
\widehat{\bf x} =: \proj_{\constraintG}(\x) \in \argmin\limits_{{\bf z} \in \Real^\dim} \left \{ \|{\bf x} - {\bf z}\|_2^2 ~\big|~ \supp({\bf z}) \in \constraintG \right \}.
\end{equation} Moreover, one might be only interested in identifying the {\em group-support} of the approximation $\widehat{{\bf x}}$, that is the $G$ groups that constitute its support. 
We call this the group-sparse {\em model selection} problem.

\subsection{The discrete model} \label{sec:discrete_relax}

%

According to \eqref{eq:group_approx_general}, we seach for $\widehat{\bf x} \in \mathbb{R}^{\dim}$ such that $\|\widehat{\bf x} - \x\|_2^2$ is minimized, while $\widehat{\bf x}$ does not exceed a given group budget $G$. A useful notion in the group sparse model is that of the {\bf group $\ell_0$-``norm"}:
\begin{equation}
\label{eq:group_0_norm}
\|{\bf w}\|_{\GG,0} := \min\limits_{\boldsymbol{\omega} \in \mathbb{B}^\ngroups} \left \{ \sum_{j=1}^\ngroups \omega_j ~\big|~ \A \boldsymbol{\omega} \geq \iota({\bf w}) \right \};
\end{equation} here, $\A$ denotes the adjacency matrix as defined in the previous subsection, the binary vector $\boldsymbol{\omega}$ indicates which groups are active and the constraint $\A \boldsymbol\omega \geq \iota({\bf w})$ makes sure that, for every non-zero component of ${\bf w}$, there is at least one active group that covers it.
Given the above definitions, the group-based signal approximation problem \eqref{eq:group_approx_general} can be reformulated as
\begin{equation}
\label{eq:group_approx}
\hat{\bf x} \in \argmin\limits_{{\bf w} \in \Real^N} \left \{ \|{\bf w} - {\bf x}\|_2^2 ~\big|~ \|{\bf w}\|_{\GG,0} \leq G \right \}.
\end{equation}

One can easily observe that, in the case where we already know the group cover of the approximation $\hat{\bf x}$, we can obtain $\hat{\bf x}$ as $\hat{\bf x}_\mathcal{I} = {\bf x}_\mathcal{I}$ and $\hat{\bf x}_{\mathcal{I}^c} = 0$, where $\mathcal{I} = \bigcup_{\G \in \mathcal{S}^G(\hat{\bf x})} \G$, with $\mathcal{S}^G(\hat{\bf x})$ denoting the group support of $\widehat{x}$  and $\mathcal{I}^c = \N \setminus \mathcal{I}$. I.e., if we know the group support of the solution, the entries' values are naturally given by the anchor point $\x$.
The authors in \cite{baldassarre2013group} prove that the group support can be obtained by solving a discrete problem, according to the next lemma.

\begin{lemma}{\cite{baldassarre2013group}}
Given ${\bf x} \in \Real^N$ and a group structure $\GG$, the group support of the solution $\hat{\bf x}$---denoted as $\mathcal{S}^G(\hat{\bf x}) = \{ \G_j \in \GG : \omega^G_j = 1 \}$---is given by the solution $(\boldsymbol{\omega}^G, {\bf y}^G)$ of the following binary maximization problem:
\begin{equation}
\label{eq:WMC}
\max\limits_{\boldsymbol{\omega} \in \mathbb{B}^\ngroups,~{\bf y} \in \mathbb{B}^\dim} \left \{ \sum_{i=1}^N y_i x_i^2 : \A \boldsymbol{\omega} \geq {\bf y},  \sum_{j=1}^\ngroups \omega_j \leq G \right \}.
\end{equation}
\end{lemma} Here, ${\bf y}$ denotes the selected variables, while $\boldsymbol{\omega}$ the active groups. Thus, the constraint $\A \boldsymbol{\omega} \geq {\bf y}$ guarantees that, for every selected variable, there is at least one group that covers it.

The problem in \eqref{eq:WMC} can produce all the instances of the weighted maximum coverage problem (WMC) \cite{nemhauser1988integer}, where the weights for each element are given by $x_i^2$ ($1 \leq i \leq \dim$) and the index sets are given by the groups $\G_j \in \GG$ ($1 \leq j \leq \ngroups$).
Since WMC is in general NP-hard \cite{nemhauser1988integer} and given Lemma 1, finding the groups that cover the solution $\hat{\bf x}$ is in general NP-hard.

However, it is possible to approximate the solution of \eqref{eq:WMC} using the greedy WMC algorithm \cite{nemhauser1978analysis}.
At each iteration, the algorithm selects the group that covers new variables with maximum combined weight until $G$ groups have been selected. 
To the best of our knowledge, only the work in \cite{baldassarre2013group} presents a polynomial time algorithm for solving \eqref{eq:WMC} for loopless pairwise overlapping groups structures, using dynamic programming. 
The algorithm gradually explores the group graph, which in this case is a tree or a forest of trees, from the leaves towards the root, storing the best solutions found so far and dynamically updating them until the entire tree is examined. 

\subsection{Convex approaches} 
\label{sec:group_convex}
Recent works in compressive sensing and machine learning with group sparsity have mainly focused on leveraging the group structures for lowering the number of samples required for recovering signals \cite{stojnic2009reconstruction, eldar2009robust, blumensath2009sampling, baraniuk2010model, rao2012signal, huang2011learning,jacob2009group, obozinski2011group}. 

For the special case of non-overlapping groups, dubbed as the block-sparsity model, the problem of model selection does not present computational difficulties and features a well-understood theory \cite{stojnic2009reconstruction}. The first convex relaxations for group-sparse approximation \cite{yuan2006model} considered only non-overlapping groups: the authors proposed the \texttt{Group LARS} (\textbf{L}east \textbf{A}ngle \textbf{R}egre\textbf{S}sion) algorithm to solve this problem, a natural extension of simple sparsity LARS algorithm \cite{efron2004least}. Using the same algorithmic principles, its extension to overlapping groups \cite{zhao2009composite} has the drawback of selecting supports defined as the complement of a union of groups, even though it is possible to engineer the groups in order to favor certain sparsity patterns over others \cite{jenatton2011structured}. Eldar et al.\ \cite{eldar2009robust} consider the union of subspaces framework and cast the model selection problem as a block-sparse model selection one by duplicating the variables that belong to overlaps between the groups, which is the optimization approach proposed also in \cite{jacob2009group}. Moreover, \cite{eldar2009robust} considers a model-based pursuit approach \cite{chen1998atomic} as potential solver for this problem, based on a predefined model $\constraint$.
For these cases, one uses the group lasso norm
\begin{equation}
\label{eq:group_lasso_norm}
\sum_{\G \in \GG} \|{\bf x}_{|_\G}\|_p \; ,
\end{equation} 
where ${\bf x}_{|_\G}$ is the restriction of ${\bf x}$ to only the components indexed by $\G$.

In addition, convex proxies to the group $\ell_0$-norm \eqref{eq:group_0_norm} have been proposed (e.g., \cite{jacob2009group}) for finding  group-sparse approximations of signals. 
Given a group structure $\GG$, an example generalization is defined as
\begin{equation}
\label{eq:atomic_norm}
\|{\bf x}\|_{\GG,\{1,p\}} := \inf\limits_{{\scriptsize \begin{array}{c} \mathbf{v}_1, \ldots, \mathbf{v}_\ngroups \in \R^{\dim}\\\forall i, \supp(\mathbf{v}_i) = \G_i \end{array}}} \left \{ \sum_{i = 1}^\ngroups d_i \|\mathbf{v}_i\|_p ~\big|~ \sum_{i=1}^\ngroups \mathbf{v}_i= {\bf x} \right \} ,
\end{equation}
where $\|{\bf x}\|_p = \left ( \sum_{i=1}^N x_i^p \right )^{1/p}$ is the $\ell_p$-norm, and $d_j$ are positive weights that can be designed to favor certain groups over others \cite{obozinski2011group}. 
This norm can be seen as a weighted instance of the atomic norm described in \cite{chandrasekaran2012convex, rao2012signal}, based on the seminal work of \cite{chen1998atomic} on the sparse synthesis model. There, the authors leverage convex optimization for signal recovery, but not for model selection: Let $\atomic := \left \{ \atom_1, \atom_2, \cdots ~|~ \atom_i \in \R^{\dim}, ~\forall i \right\}$ be the \emph{atomic set} of \emph{group sparse} signals $\x$ that can be synthesized as a $\sparsity$-sparse combination of atoms $\atom_i$ in $\atomic$, i.e., 
\begin{align}
\x =\sum_{i = 1}^{\sparsity} c_i \atom_i, ~\quad c_i \geq 0, ~\atom_i \in \atomic ~\text{ and }~ \|\mathbf{c}\|_0 \leq \sparsity.
\end{align} Here, $\mathbf{a}_i  \in \Real^\dim$, $\text{supp}(\mathbf{a}_i) = \mathcal{G}_i $ and $\|\mathbf{a}_i\|_2 = 1$. We can then define the atomic norm: 
\begin{align}
\label{eq:atomic_norm_2}
\|\x\|_{\mathcal{A}} = \inf \left\{ \sum_{i = 1}^{|\atomic|} c_i ~\big|~ \x = \sum_{i = 1}^{|\atomic|} c_i \atom_i, ~c_i \geq 0, ~ \forall \atom_i \in \atomic \right\}.
\end{align}

\begin{lemma}{\cite{chandrasekaran2012convex, rao2012signal}}
If in \eqref{eq:atomic_norm} the weights are all equal to 1 ($d_i = 1, \forall~i$), we have
$$
\|\x\|_{\mathcal{A}} = \|{\bf x}\|_{\GG,\{1,p\}} \;.
$$
\end{lemma}

The group-norm \eqref{eq:atomic_norm} can also be viewed as the tightest convex relaxation of a particular set function related to the \emph{weighted set-cover} (see Section \ref{sec:submodular} and \cite{obozinski2012convex}).

One can in general use \eqref{eq:atomic_norm} to find a group-sparse approximation under the chosen group norm
\begin{equation}
\label{eq:latent_gl}
\hat{\bf x} \in \argmin\limits_{{\bf w} \in \Real^N} \left \{ \|{\bf w} - {\bf x}\|_2^2 : \|{\bf w}\|_{\GG,\{1,p\}} \leq \lambda\right \}, \;
\end{equation}
where $\lambda > 0$ controls the trade-off between approximation accuracy and group-sparsity. 
However, solving \eqref{eq:latent_gl} does not necessarily yield a group-support for $\hat{\bf x}$:
even though we can recover one through the decomposition $\{ \mathbf{v}^j\}$ used to compute $\|\hat{\bf x}\|_{\GG, \{1,p\}}$, it may not be unique and when it is unique it may not capture the minimal group-cover of ${\bf x}$ \cite{obozinski2011group}. Therefore, the equivalence of $\ell_0$ and $\ell_1$ minimization \cite{donoho2006compressed, candes2006compressive} in the standard compressive sensing setting does not generally hold in the overlapping group-based setting.

The regularized version of problem \eqref{eq:latent_gl} is equivalent to the proximity operator of $\|{\bf x}\|_{\GG,\{1,p\}}$. 
Recently, \cite{mosci2010primaldual, villa2013proximal} proposed an efficient algorithm for this proximity operator in large scale settings with extended overlap among groups. In this case, the proximity operator involves: $(i)$ an active set preprocessing step \cite{wright1999numerical} that restricts the prox operations on a subset of the model---i.e., ``active'' groups and, $(ii)$ a dual optimization step based on Bertsekas' projected Newton method \cite{bertsekas1982projected}; however, its convergence requires the strong regularity of the Hessian of the objective near the optimal solution. 
The authors in \cite{argyriou2011efficient} propose a fixed point method to compute the proximity operator for a wide range of group sparse model variants, given the model-structured mapping of the fixed-point Picard iterations is non-expansive. \cite{yuan2011efficient} develops an efficient primal-dual criterion for the overlapping group sparse proximity operator: the solution of its dual formulation is used to compute the duality gap and tweak the accuracy of the solution and, thus, the convergence of the algorithm.


\subsection{Extensions}

%

In many applications, such as genome-wide association studies \cite{zhou2010association} and multinomial classification \cite{vincent2014sparse}, it is desirable to find approximations that are not only group-sparse, but also sparse in the usual sense (see \cite{simon2013sparse} for an extension of the group lasso). A classic illustrative example is the sparse group lasso approach \cite{friedman2010note}, a regularization method that combines the lasso \cite{tibshirani1996regression} and the group lasso \cite{meier2008group}. 

From a discrete pespective, the original problem \eqref{eq:WMC} can be generalized by introducing a sparsity constraint $\sparsity$ and allowing to individually select variables within a group. 
The generalized integer problem then becomes
\begin{equation}
\label{eq:GWMC}
\max\limits_{\boldsymbol{\omega} \in \mathbb{B}^\ngroups,~{\bf y} \in \mathbb{B}^\dim} \left \{ \sum_{i=1}^\dim y_i x_i^2 ~\big|~ \A \boldsymbol{\omega} \geq {\bf y},~ \sum_{i=1}^\dim y_i \leq \sparsity,~ \sum_{j=1}^\ngroups \omega_j \leq G \right \} \; .
\end{equation}

This problem is in general NP-hard too, but it turns out that it can be solved in polynomial time for the same group structures that allow to solve \eqref{eq:WMC}.

\begin{prop}{\cite{baldassarre2013group}}
\label{prop:DP2}
Given a loopless group structure $\mathfrak{G}$, there exists dynamic programming algorithm that solves \eqref{eq:GWMC} with complexity linear in $\dim$ and $\sparsity$.
\end{prop}

From a convex point of view, Simon et al. \cite{simon2013sparse} are the first to propose the sparse group model regularization in the context of linear regression. Given a group model $\mathfrak{G}$ and constants $\lambda_1, ~\lambda_2 > 0$, they consider the problem
\begin{equation}
	\begin{aligned}
	& \underset{\x \in \R^\dim}{\text{minimize}}
	& & \Big\{f(\x) + \lambda_1 \sum_{\G \in \GG} \|{\bf x}_{|_\G}\|_2 + \lambda_2 \|\x\|_1\Big\}.
	\end{aligned} \nonumber
\end{equation} For this purpose, a generalized gradient descent algorithm is proposed. \cite{vincent2014sparse} provides an efficient and robust sparse group lasso algorithm: in this case, a penalized quadratic approximation of the loss function is optimized via a Newton type algorithm in a coordinate descent framework
\cite{yuan2011efficient}.

%% file: dispersive.tex
\section{Sparse dispersive models}
\label{sec:dispersive}

To describe the \emph{dispersive} structure, we motivate our discussion with an application from neurobiology. Living beings behave and function via transmission of electrical signals between electrically excitable neuronal brain cells. Such chemical ``information'' causes a swift change in the electrical potential of a possibly discharged neuron cell, which results in its electrical excitation. 
Currently, we are far from understanding the grid of neurons in its \emph{entirety}: large-scale brain models are difficult to handle while complex neuronal signal models lead to non-interpretable results. 

\begin{figure}[tb]
\centering \includegraphics[width=0.7\textwidth]{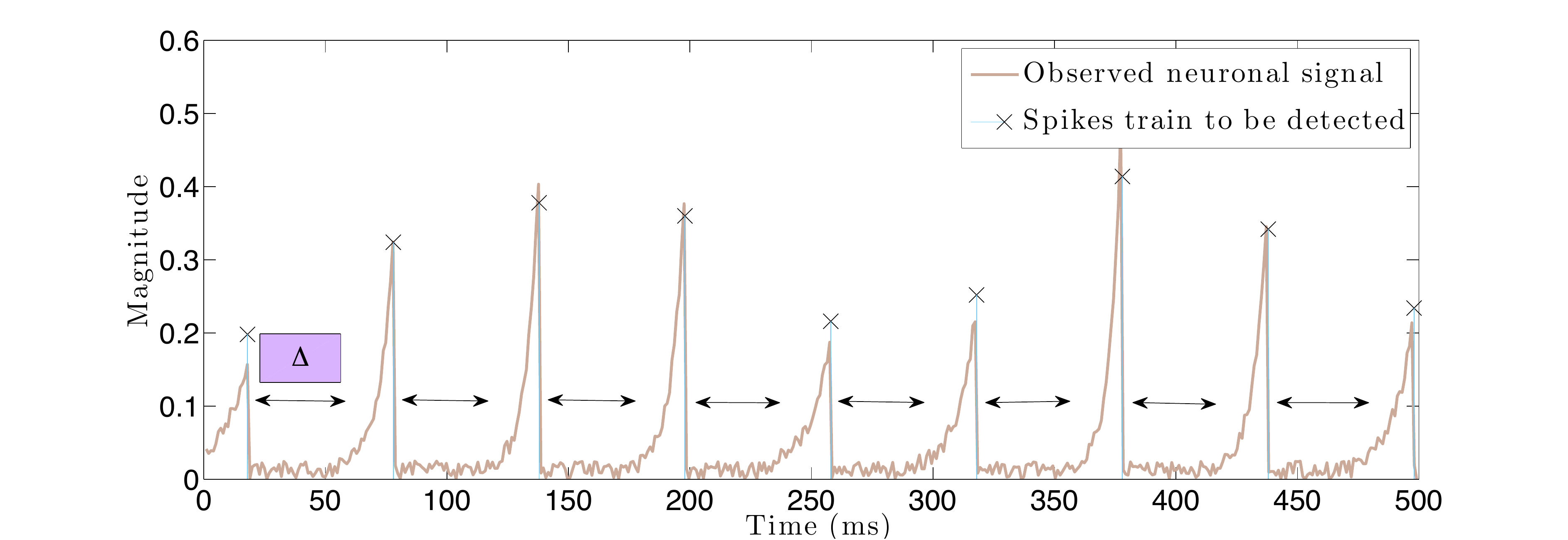}
\caption{Neuronal spike train example.} \label{fig:spike}
\end{figure}

Inspired by the statistical analysis in \cite{gerstner2002spiking}, the authors in \cite{hegde2009compressive} consider a simple one-dimensional model, where the neuronal signal behaves as a train of spike signals with some \emph{refractoriness} period $\Delta > 0$: there is a minimum nonzero time period $\Delta $ where a neuron remains inactive between two consecutive electrical excitations. In statistical terms, neuronal signals are defined by a inter-spike interval distribution that characterizes the probability a new spike to be generated as a function of the inter-arrival time. 
Figure \ref{fig:spike} illustrates how a collection of noisy neuronal spike signals with $\Delta > 0$ might appear in practice. 



\subsection{The discrete model}

We provide next a formal definition of the sparse dispersive model:
\begin{definition}[Dispersive model] \label{def:dispersiveCSM}
We define the dispersive model $\dispersive$ in $n$-dimensions with sparsity level $\sparsity$ and refractory parameter $\Delta \in \mathbb{Z}_{+}$ as:
\begin{align}
\dispersive  = \left\{ {\mathcal{S}}_q ~\big|~ \forall q,~{\mathcal{S}}_q \subseteq  \mathcal{N},~ \abs{{\mathcal{S}}_q} \le \sparsity \text{~~~~and~~~~} |i - j| > \Delta, \forall i \neq j \in \mathcal{S}_{q} \right\},
\end{align} i.e., $\dispersive $ is a collection of index subsets ${\mathcal{S}}_q$ in $\mathcal{N}$ with number of elements no greater than $\sparsity$ and with distance between the indices in $\mathcal{S}_q$ greater than the interval $\Delta$.
\end{definition} We note that if there are no constraints on the interval of consecutive spikes, the dispersive model naturally boils down to the simple sparsity model $\sparse_\sparsity$. 

Given the definition above, the projection \eqref{eq:proj} is:
\begin{align}
\mathcal{P}_{\dispersive}(\signal) \in \argmin_{\w \in \mathbb{R}^{\dim}}\left\{\vectornorm{\w - \signal}_2^2 ~\big|~ {\rm  supp}(\w) \in \dispersive \right\}.
\label{eq:disp_proj} 
\end{align} Let $\boldsymbol{\omega} \in \mathbb{B}^{n}$ be a \emph{support indicator} binary vector, i.e., $\boldsymbol{\omega}$ represents the support set of a sparse vector $\x$ such that $\text{supp}(\boldsymbol{\omega}) = \text{supp}(\x)$. Moreover, let $\D \in \mathbb{B}^{(n - \Delta + 1) \times n} $ such that:
\begin{align}
\D = \begin{bmatrix}
1 & 1 & \cdots & 1 & 1 & 0 & 0 & \cdots & 0 \vspace{0.1cm} \\
0 & 1 & 1& \cdots & 1 & 1 & 0 & \cdots & 0 \vspace{0.1cm} \\
& & & & \ddots  & & & \vspace{0.1cm} \\
0& \cdots & 0 & 0 & 1 & 1 & \cdots & 1 & 1
\end{bmatrix}_{(n - \Delta + 1) \times n}
\end{align} Here, per row, there are $\Delta$ consecutive ones that denote the time interval between two potential consecutive spikes. Finally, let $\mathbf{b} \in \R^{n - \Delta + 2}$ such that $\mathbf{b} := \begin{bmatrix}
\sparsity & 1 & 1 & \cdots & 1 & 1
\end{bmatrix}^T.$ 

According to \cite{hegde2009compressive}, the following linear support constraints encodes the definition of the dispersive model $\dispersive$:
\begin{align}
\A := 
\begin{bmatrix}
\mathbf{1} \\
\D
\end{bmatrix} \boldsymbol{\omega} \leq \mathbf{b}.
\end{align}
One can observe that 
$\dispersive \equiv \left\{ \bigcup_{\forall \boldsymbol{\omega} \in \mathfrak{Z}} \text{supp}\left(\boldsymbol{\omega}\right) ~\big|~ \mathfrak{Z} := \left\{\boldsymbol{\omega} \in \mathbb{B}^\dim:  \A \boldsymbol{\omega} \leq \mathbf{b} \right\} \right\}.$
To this end, \eqref{eq:disp_proj} becomes:
\begin{align}
\mathcal{P}_{\dispersive}(\signal) \in \argmin_{\w \in \mathbb{R}^{\dim}}\left\{\vectornorm{\w - \signal}_2^2 ~\big|~ \A \cdot \text{supp}(\w) \leq \mathbf{b} \right\}.
\label{eq:disp_proj2} 
\end{align}
A key observation is given in the next lemma.
\begin{lemma}{\cite{hegde2009compressive}}\label{lem:dispersive}
Given the problem setting above, it is easy to observe that \eqref{eq:disp_proj2} has solution $\mathcal{P}_{\dispersive}(\signal)$ such that $\mathcal{S} := \text{supp}\left(\mathcal{P}_{\dispersive}(\signal)\right)$ and $\left(\mathcal{P}_{\dispersive}(\signal)\right)_{\mathcal{S}} = \signal_{\mathcal{S}}$ where: 
\begin{align}
\mathcal{S} \in \text{supp}\left(\argmax_{\boldsymbol{\omega} \in \mathbb{B}^{\dim}:~\A \boldsymbol{\omega} \leq \mathbf{b} }\left\{\mathbf{c}^T \boldsymbol{\omega}\right\}\right), \quad \text{ where } \mathbf{c} := \left[x_1^2\quad x_2^2\quad \cdots \quad x_n^2\right]^T,
\label{eq:disp_proj3} 
\end{align} i.e., we target to capture most of the signal's $\signal$ energy, given structure $\dispersive$. To solve \eqref{eq:disp_proj3}, the authors in \cite{hegde2009compressive} identify that the binary integer program \eqref{eq:disp_proj3} is identical to the solution of the linear program, obtained by relaxing the integer constraints into continuous constraints.
\end{lemma}

To this end, Lemma \ref{lem:dispersive} indicates that \eqref{eq:disp_proj2} can be efficiently performed using linear programming tools \cite{boyd2004convex}. Once \eqref{eq:disp_proj2} is relaxed to a convex problem, decades of knowledge on convex analysis and optimization can be leveraged. Interior point methods find a solution with fixed precision in polynomial time but their complexity might be prohibitive even for moderate-sized problems.

\subsection{Convex approaches}
The constraint matrix $\D$ describes a \emph{collection of groups}, where each group is assumed to have at most one non-zero entry to model the refractoriness property.\footnote{Other convex structured models that can be described as the composition of a simple function over a linear transformation $\D$ can be found in \cite{argyriou2011efficient}.} Moreover, these groups are \emph{overlapping} which aggrandizes the ``clash'' between neighboring groups: a non-zero entry in a group discourages every other overlapping group to have a distinct non-zero entry.

In mathematical terms, each row $i$ of $\D$ defines a group $\mathcal{G}_i$ such that $\mathcal{G}_i = \text{supp}(\mathbf{d}_i) \subseteq \N$ where $\mathbf{d}_i$ denotes the $i$-th row of $\D$, $\forall i \in \lbrace 1, \dots, M:= \dim - \Delta + 1 \rbrace$: 
\begin{center}
\includegraphics[width = 0.3\textwidth]{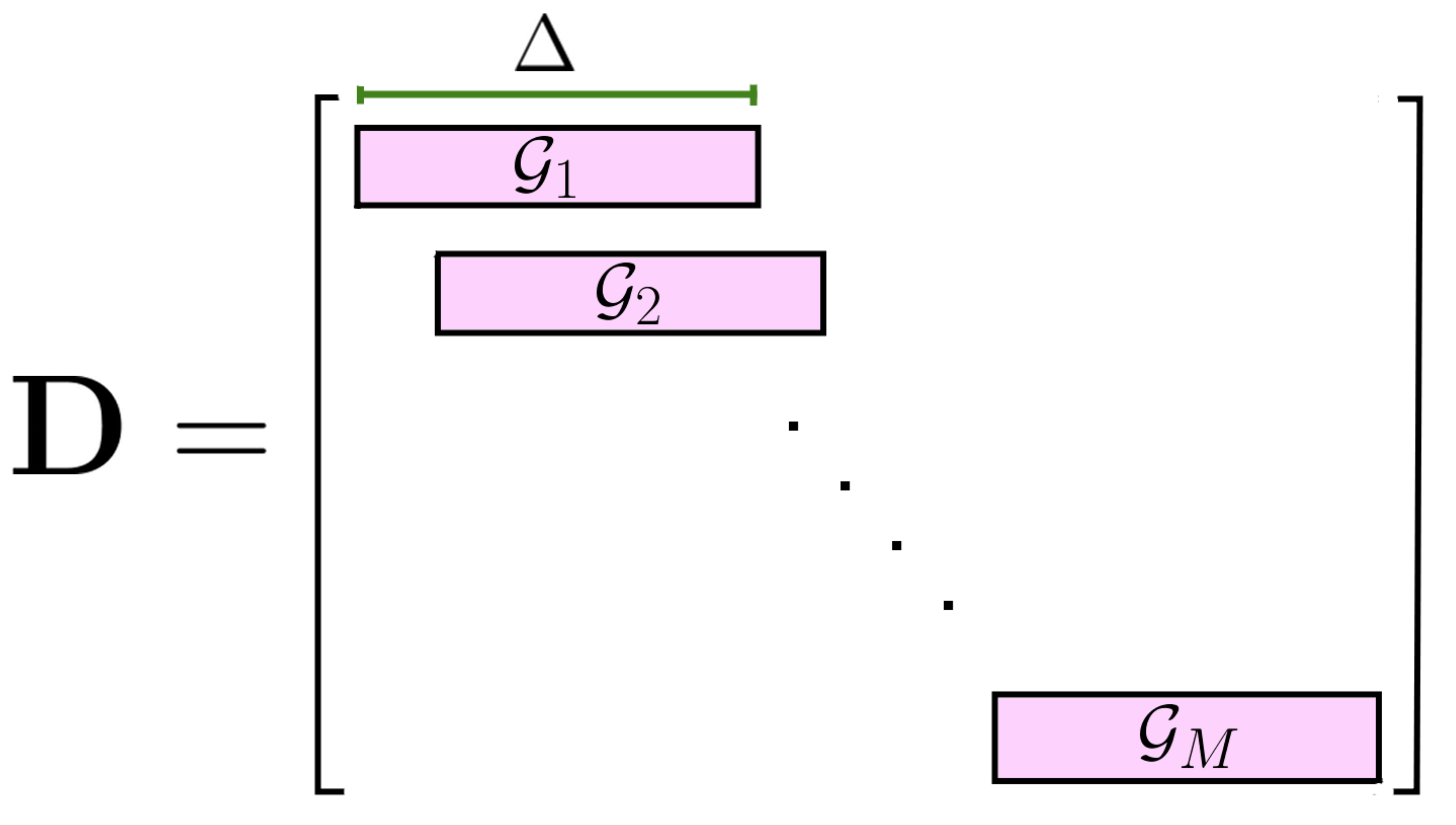}
\end{center} Given such group structure, the dispersive model is characterized both by \emph{inter-group} and \emph{intra-group} properties: 
\begin{itemize}
\item \emph{Intra-group sparity}: we desire $\|\D \boldsymbol{\omega}\|_{\infty} \leq 1$, i.e., per refractoriness period of length $\Delta$, we require only one ``active'' spike.
\item \emph{Inter-group exclusion}: due to the refractoriness property, the activation of a group implies the deactivation of its closely neighboring groups.
\end{itemize}

While the sparsity level within a group can be easily ``convexified'' using standard $\ell_1$-norm regularization, the dispersive model further introduces the notion of \emph{inter-group exclusion}, which is highly \emph{combinatorial}. 
However, one can relax it by introducing \emph{competitions} among variables in overlapping groups: 
variables that have a ``large'' neighbor should be penalized more than variables with ``smaller'' neighbors.

In this premise and based on \cite{zhou2010exclusive}, we identify the following family of norms\footnote{The proposed norm originates from the composite absolute penalties (CAP) convex norm, proposed in \cite{zhao2009composite}, according to which:
\begin{align}
g(\x) = \sum_{\mathcal{G}_i	} \left(\sum_{j \in \mathcal{G}_i} |x_j|^{\gamma}\right)^{p},
\end{align} for various values of $\gamma$ and $p$. Observe that this model also includes the famous group sparse model where $g(\x) = \sum_{\mathcal{G}_i} \|\x_{\mathcal{G}_i}\|_2$, described in Section \ref{sec:groups}, for $p = 1/2$ and $\gamma = 2$.}: 
\begin{align}{\label{eq:exclusive}}
\Omega_{\text{exclusive}}(\x) = \sum_{\mathcal{G}_i} \left(\sum_{j \in \mathcal{G}_i} |x_j|\right)^{p},\quad p = 2, 3, \dots,
\end{align} as convex regularizers that imitate the dispersive model. In \eqref{eq:exclusive}, $\left(\sum_{j \in \mathcal{G}_i} |x_j|\right) := \|\x_{\mathcal{G}_i}\|_1$ promotes sparsity within each group $\mathcal{G}_i$, while the outer sum over groups $\sum_{\mathcal{G}_i} \|\x_{\mathcal{G}_i}\|_1^p$ imposes sparsity over the number of groups that are activated. Observe that for $p = 1$, \eqref{eq:exclusive} becomes the standard $\ell_1$-norm over $\mathcal{N}$. Notice that the definition of the overlapping groups (instead of non-overlapping) is a key property for capturing the discrete structure:
variables belonging to overlapping groups are weighted differently when considered parts of different groups. This leads to variable ``suppression'' (i.e., thresholding) of elements, depending on the ``weight'' of their neighborhood within the groups they belong to. 

%% file: hierarchical.tex
\section{Hierarchical sparse models}
\label{sec:hierarchical}

\begin{figure}[tb]
\captionsetup{width=0.8\textwidth}
\centering
\begin{tabular}{ccc}
\includegraphics[width=0.25\columnwidth]{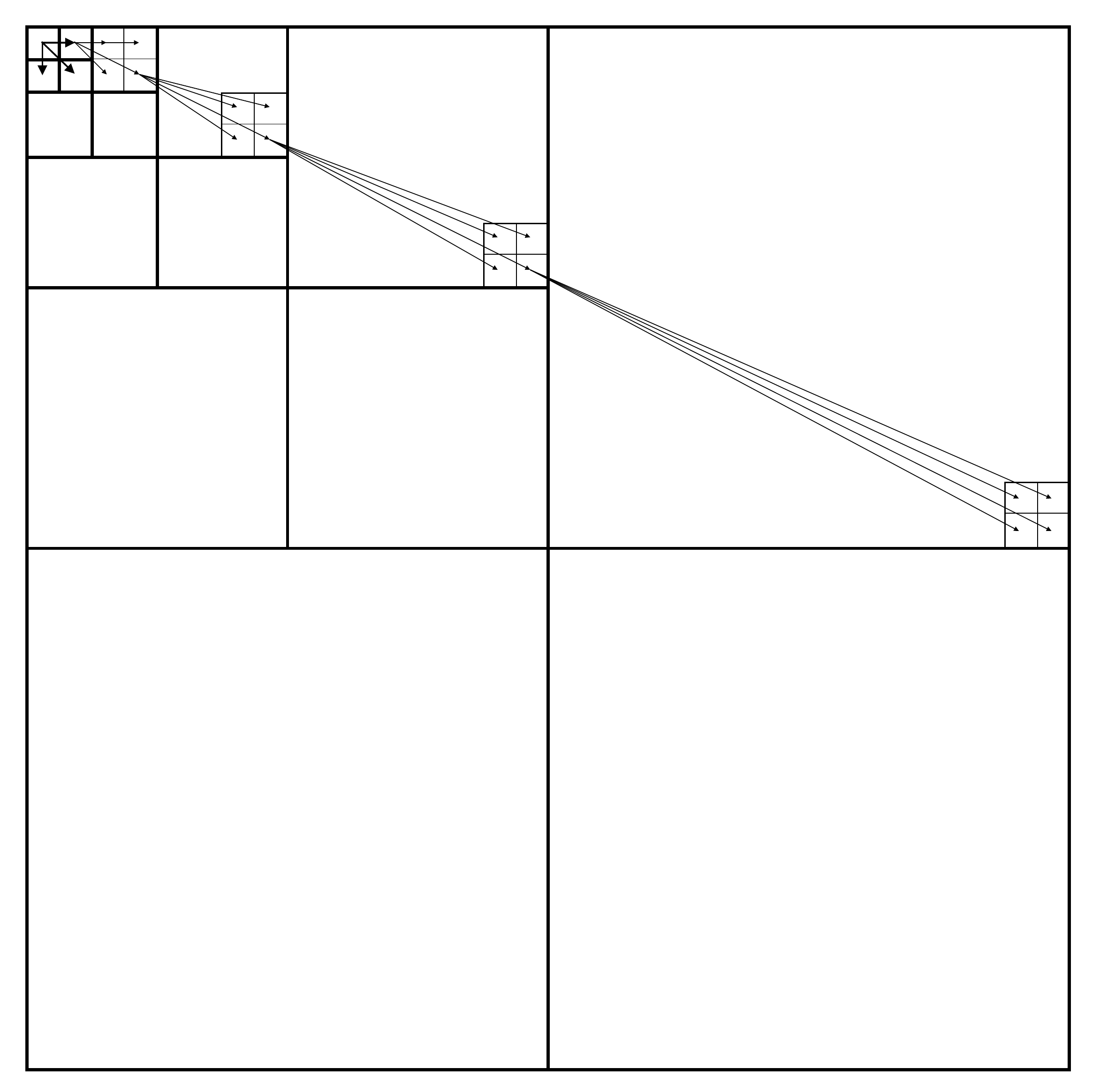} &
\includegraphics[width=0.25\columnwidth]{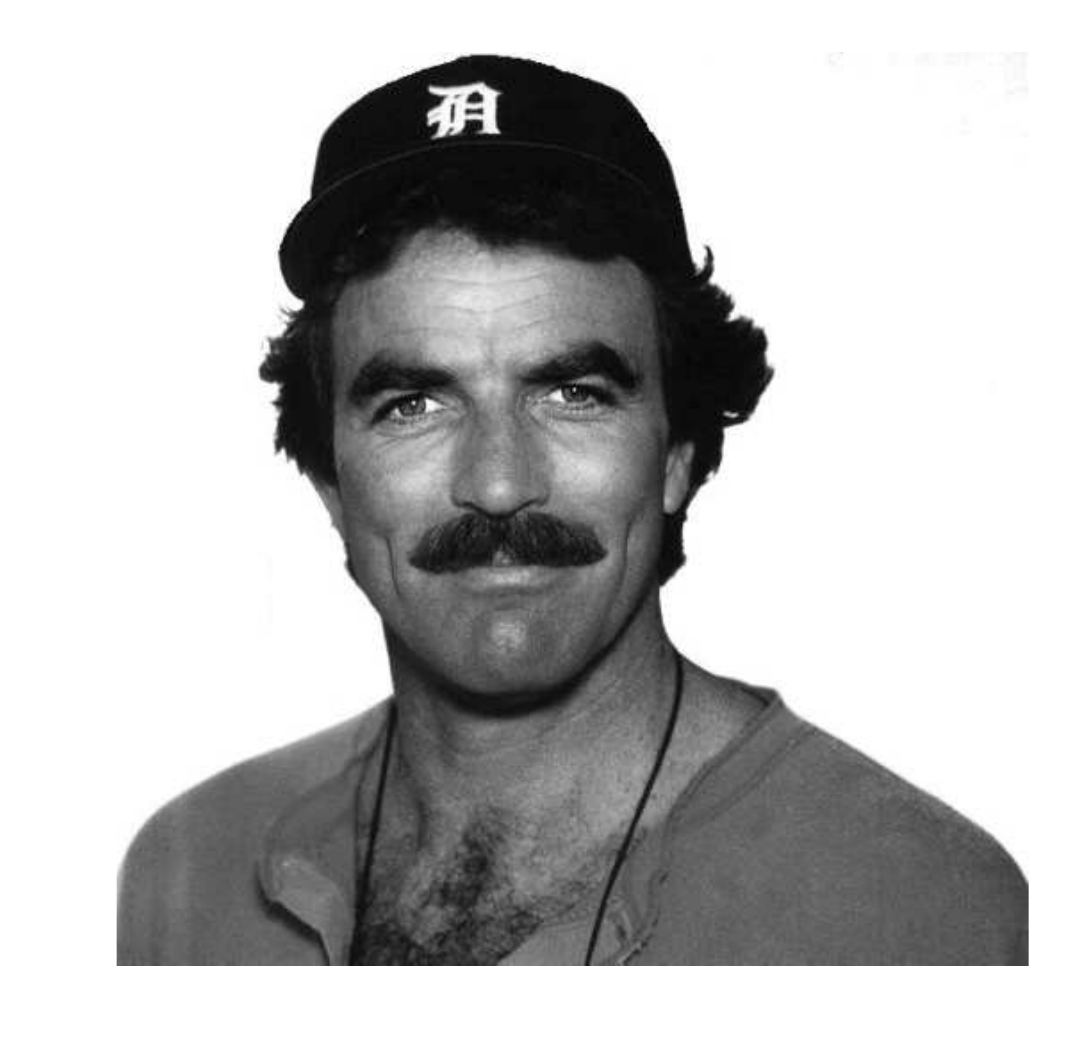} &\includegraphics[width=0.25\columnwidth]{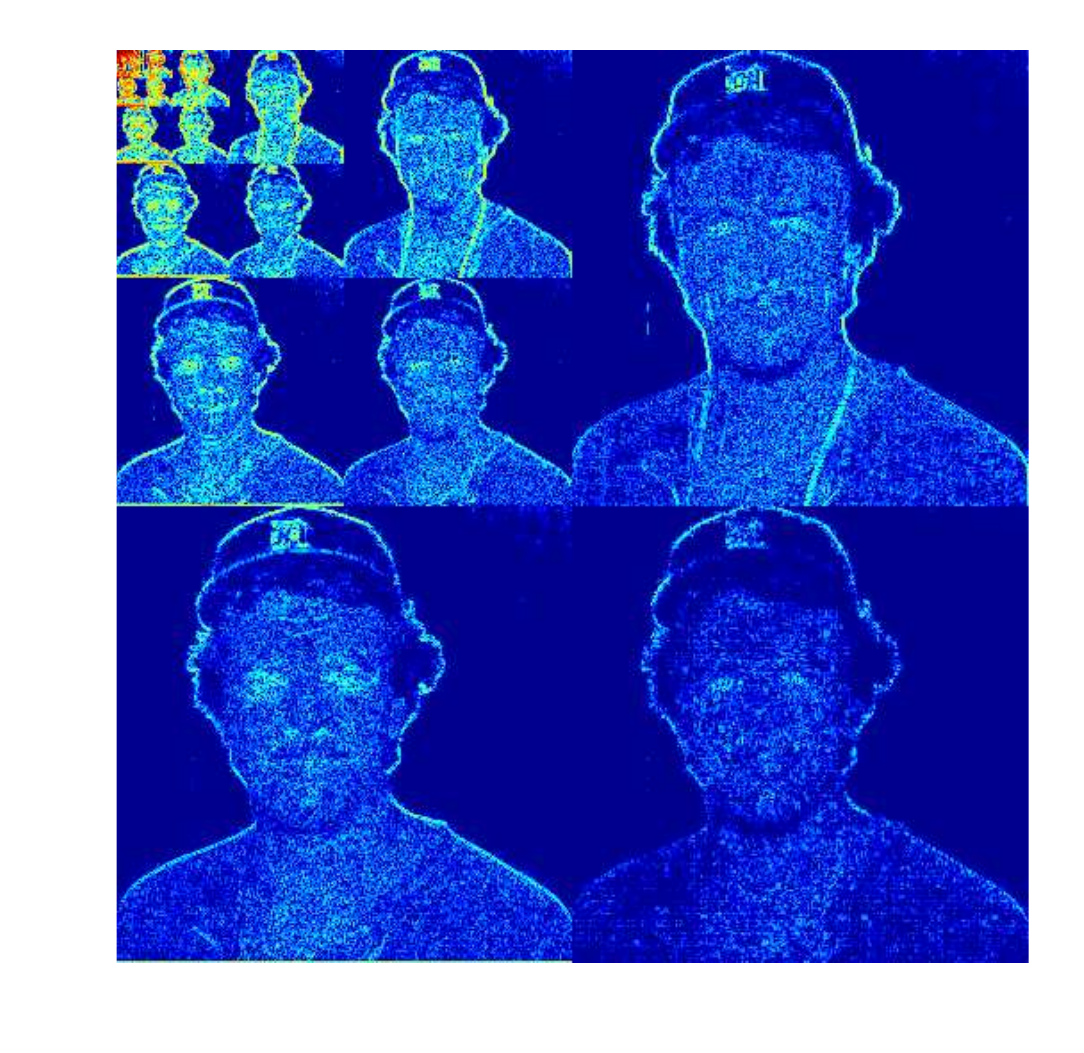}
\end{tabular}
\caption{\label{fig:wavelet_coeff} Wavelet coefficients naturally cluster along a rooted connected subtree of a regular tree and tend to decay towards the leaves. (Left) Example of wavelet tree for a $32$ x $32$ image. The root of the tree is the top-left corner and there are three regular subtrees related to horizontal, vertical and diagonal details. Each node is connected to four children representing detail at a finer scale. (Centre) Grayscale $512$ x $512$ image. (Right) Wavelet coefficients for the image at center. Best viewed in color, dark blue represents values closer to zero.}
\end{figure}

Hierarchical structures are found in many signals and applications. 
For example, the wavelet coefficients of images are naturally organized on regular quad-trees to reflect their multi-scale structure, Figure \ref{fig:wavelet_coeff} and \cite{shapiro1993embedded, crouse1998wavelet, mallat1999wavelet, baraniuk1999optimal, baraniuk2002near, he2009exploiting, zhao2009composite, baraniuk2010model, huang2011learning}; gene networks are described by a hierarchical structure that can be leveraged for multi-task regression \cite{kim2010tree}; hierarchies of latent variables are typically used for deep learning \cite{bengio2009learning}.

In essence, a hierarchical structure defines an ordering of importance among the elements (either individual variables or groups of them) of a signal with the rule that an element can be selected only after its  ancestors. Such structured models result into more robust solutions and allow recovery with far fewer samples. In compressive sensing, assuming that the signal possesses a hierarchical structure with sparsity $k$ leads to improved sample complexity bounds of the order of $O(k)$ for dense measurement matrices \cite{baraniuk2010model}, compared to the bound of $O(k\log(\dim/k))$ for standard sparsity. Also in the case of sparse measurement matrices, e.g. expanders, hierarchical structures yield improved sample complexity bounds \cite{indyk2013model, bah2014model}.


\subsection{The discrete model}

The discrete model underlying the hierarchical structure is given by the next definition. 
\begin{definition}[Hierarchical model]
Let $\T$ denote an arbitrary \emph{tree} or \emph{forest} representation over the variables in a set $\N$. We define a $\sparsity$ rooted connected (RC) subtree $\mathcal{S}$ with respect to $\mathcal{T}$ as a collection of $\sparsity$ variables in $\N$ such that $v \in \mathcal{S}$ implies $\mathcal{A}(v) \in \mathcal{S}$, where $\mathcal{A}(v)$ is the set that contains all the ancestors of the node $v$. 

The hierarchical model of budget $\sparsity$, $\mathcal{T}_\sparsity$ is the set of all $\sparsity$ rooted-connected subtrees of $\mathcal{T}$.
\end{definition}
An example of rooted connected subtree over $|\N| = 9$ variables is given in Figure \ref{fig:hier}. 

\tikzstyle{notsel}=[circle,draw=black,fill=white,thick, minimum size=6pt, inner sep=0pt]
\tikzstyle{sel}=[circle,draw=black,fill=black,thick, minimum size=6pt, inner sep=0pt]
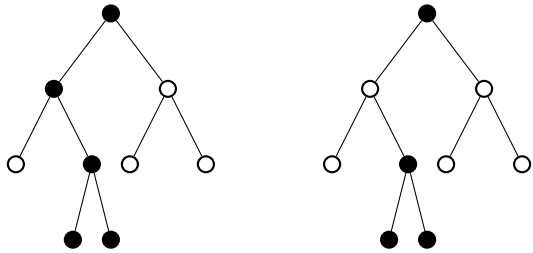
\begin{figure}
\captionsetup{width=0.8\textwidth}
\centering
\begin{tabular}{cc}
\begin{tikzpicture}[level distance=10mm]
  \tikzstyle{level 1}=[sibling distance=15mm]
  \tikzstyle{level 2}=[sibling distance=10mm]
  \tikzstyle{level 3}=[sibling distance=5mm]
  \node[sel] {}
    child {node[sel] {} 
  		child{node[notsel] {}} 
		child{node[sel] {} 
			child{node[sel] {}} 
			child{node[sel] {}}}}
    child {node[notsel] {} 
		child{node[notsel] {}} 
		child{node[notsel] {}}};
\end{tikzpicture}
& \hskip 1cm
\begin{tikzpicture}[level distance=10mm]
  \tikzstyle{level 1}=[sibling distance=15mm]
  \tikzstyle{level 2}=[sibling distance=10mm]
  \tikzstyle{level 3}=[sibling distance=5mm]
  \node[sel] {}
    child {node[notsel] {} 
  		child{node[notsel] {}} 
		child{node[sel] {} 
			child{node[sel] {}} 
			child{node[sel] {}}}}
    child {node[notsel] {} 
		child{node[notsel] {}} 
		child{node[notsel] {}}};
\end{tikzpicture}
\end{tabular}	
\caption{\label{fig:hier} Hierarchical constraints. Each node represent a variable. (Left) A valid selection of nodes. (Right) An {\em invalid} selection of nodes.}
\end{figure}


Given a tree $\mathcal{T}$, the rooted connected approximation is the solution of the following discrete problem
\begin{equation}
\label{eq:tree_proj}
\mathcal{P}_{\mathcal{T}_\sparsity}(\signal) \in \argmin\limits_{{\bf z} \in \Real^\dim} \left \{ \|{\bf x} - {\bf z}\|_2^2 ~\big|~ \supp({\bf z}) \in \mathcal{T}_\sparsity \right \}.
\end{equation}
which can be reformulated as follows
\begin{equation}
\label{eq:hier}
\widehat{\mathbf{y}} \in \argmax_{{\bf y} \in \BB^\dim} \left \{ \sum_{i=1}^N y_ix_i^2: {\bf y} \in \mathcal{T}_\sparsity \right \} \; ,
\end{equation}
where ${\bf y}$ is a binary vector with $\sparsity$ non-zero components that indicates which components of ${\bf x}$ are selected.
Given a solution $\widehat{\bf y}$ of the above problem, a solution $\widehat{\bf z}$ of \eqref{eq:tree_proj} is then obtained as $\widehat{\bf z}_{|_\mathcal{S}} = {\bf x}_{|_\mathcal{S}}$ and $\widehat{\bf z}_{|_{\mathcal{N}\setminus \mathcal{S}}} = 0$, where $\mathcal{S} = \supp(\widehat{\bf y})$.

This type of constraint can be represented by a group structure with an overall sparsity constraint $\sparsity$, where for each node in the tree we define a group consisting of that node and all its ancestors.  When a group is selected, we require that all its elements are selected as well. Problem \eqref{eq:hier} can then be cast as a special case of the Weighted Maximum Coverage problem \eqref{eq:WMC}. Fortunately, this particular group structure leads to tractable solutions.

Indeed \eqref{eq:hier} can be solved exactly via a dynamic program that runs in polynomial time \cite{cartis2013exact, baldassarre2013group}. For $d$-regular trees, that is trees for which each node has $d$ children, the algorithm in \cite{baldassarre2013group} has complexity $\mathcal{O}(\dim \sparsity d)$.

%

\subsection{Convex approaches}

The hierarchical structure can also be enforced by convex penalties, based on groups of variables. 
Given a tree structure $\mathcal{T}$, define groups consisting of a node and all its descendants and let $\mathfrak{G}_\mathcal{T}$ represent the set of all these groups.
Based on this construction, the hierarchical group lasso penalty \cite{zhao2009composite, kim2010tree, jenatton2011proximal} imitates the hierarchical sparse model and is defined as follows
\begin{equation}
\Omega({\bf x})_\text{HGL} = \sum_{\mathcal{G} \in \mathfrak{G}_\mathcal{T}} w_\mathcal{G} \|{\bf x}_{|\mathcal{G}}\|_p
\end{equation}
where $p \geq 1$, $w_\mathcal{G}$ are positive weights and ${\bf x}_{|\mathcal{G}}$ is the restriction of ${\bf x}$ to the elements contained in $\mathcal{G}$. 
Since the nodes lower down in the tree appear in more groups than their ancestors, they will contribute more to $\Omega({\bf x})_\text{HGL}$ and therefore will be more easily encouraged to be zero.
The proximity operator of $\Omega_\text{HGL}$ can be computed exactly for $p = 2$ and $p = \infty$ via an active set algorithm \cite{jenatton2011proximal}.

\begin{figure}[tb]
\captionsetup{width=0.8\textwidth}
\centering
\includegraphics[width=0.75\columnwidth]{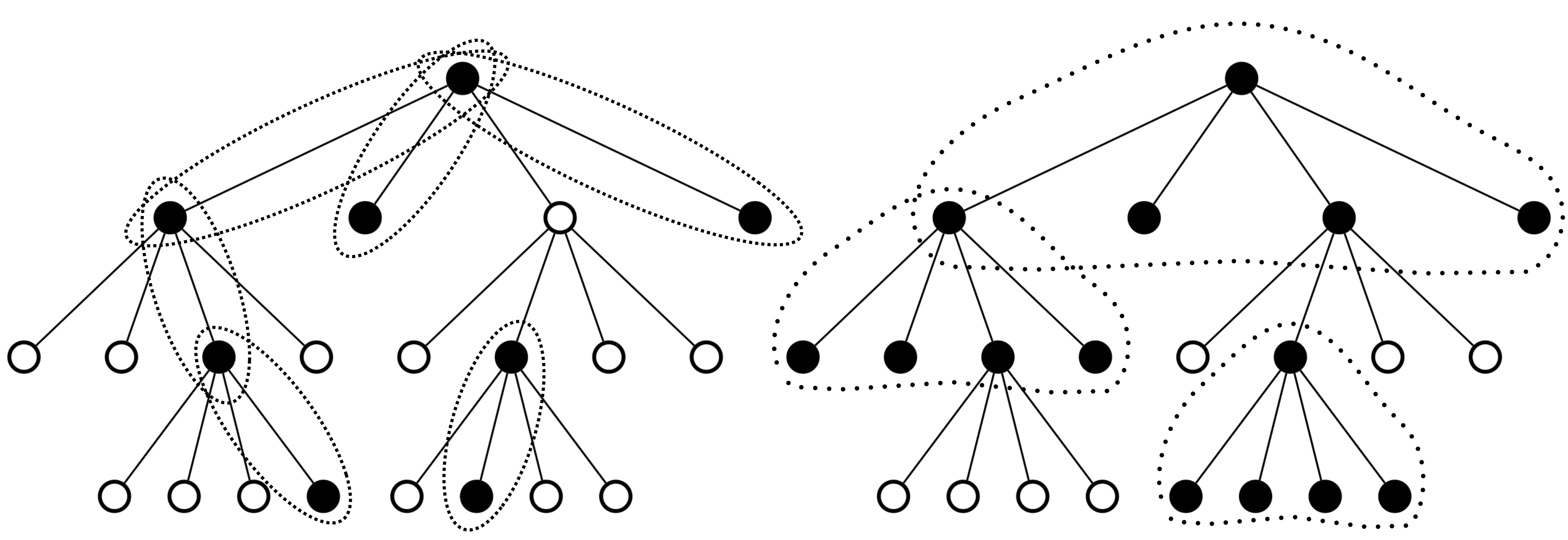}
\caption{\label{fig:family_models} Examples of parent-child and family models. Active groups are indicated by dotted ellipses. The support (black nodes) is given by the union of the active groups. (Left) Parent-child model. (Right) Family model.}
\end{figure}

Other convex penalties have been recently proposed in order to favor hierarchical structures, but also allowing for a certain degree of flexibility in deviations from the discrete model.
One approach considers groups consisting of all parent-child pairs and uses the latent group lasso penalty (see Section \ref{sec:group_convex}) in order to obtain solutions whose support is the union of few such pairs \cite{rao2011convex}, see Figure \ref{fig:family_models} (left).

An interesting extension is given by the {\em family} model \cite{bhan2013tractability, zhang2013mr}, where the groups consist of a node and all its children, see Figure \ref{fig:family_models} (right). Again the latent groups lasso penalty is used. This model is better suited for wavelet decomposition of images because it better reflects the fact that a large coefficient value implies large coefficients values for all its children at a finer scale.

For both these cases, one can use the duplication strategy to transform the overlapping proximity problem into a block one, which can be efficiently solved in closed-form \cite{jacob2009group}.

%% file: submodular.tex
\section{Submodular models}
\label{sec:submodular}

Most of the structures described so far are naturally combinatorial, but the models presented in the previous sections are incapable of capturing more complex structures. 
A more general approach is to use a set function $R$ to quantify how much a given support set deviates from the desired structure.
For instance, we can describe all previous structures by using an indicator function that assigns an infinite value to sets that do not belong to the chosen structure $\mathcal{M}_k$. Next, we provide a definition of such set functions and the sets that they describe:

\begin{definition}
Given a set function $R: 2^\N \rightarrow \R$, we can define a model $\mathcal{R}_\tau$ consisting of all sets $\mathcal{S}$ for which $R(\mathcal{S}) \leq \tau$:\begin{equation}\label{eq:SubModel}
\mathcal{R}_\tau := \left\{ \mathcal{S} ~\big|~ R ( \mathcal{S}) \leq \tau\right\}.
\end{equation}
\end{definition}

Unfortunately, computing the projection of a given signal onto this set is not generally feasible for any set function, due to their intractable combinatorial nature.

Therefore, it is necessary to restrict our attention to set functions with specific properties that lead to tractable problems. 
It turns out that in many applications (See for e.g, \cite{bach2011learning, bach2010structured, cevher2009sparse, kolmogorov2004energy, seeger2009submodularity, heckerman1995learning, narasimhan2005q,baraniuk1999optimal}) the combinatorial penalties have a convenient ``diminishing returns" property that we can exploit. These are the so-called \textit{submodular} set functions.

\begin{definition}\label{def:Sub}
A set function $R: 2^{\N} \to \reals$ is submodular iff $R(\sss \cup \{v\}) - R(\sss) \geq R(\mathcal{U} \cup \{v\}) - R(\mathcal{U})$, $\forall \sss \subseteq \mathcal{U} \subseteq \N$, and $v \in \N \setminus \mathcal{U}$.
\end{definition}

Given a combinatorial penalty that encodes the desired structure, we want to solve an inverse problem of the form:
\begin{equation}\label{eq:SubModel2}
	\underset{\x\in\mathbb{R}^n}{\text{minimize}} ~f(\x) ~ \text{ subject to } \supp(\x) \in \mathcal{R}_\tau .
\end{equation}
To simplify things, one can relax the hard constraint required in \eqref{eq:SubModel2}, by using the combinatorial penalty as a regularizer to discourage unwanted patterns of the support:
\begin{equation}
\label{eq: RegProbl}
 \underset{\x\in\mathbb{R}^n}{\text{minimize}}  ~ f(\x)  + \lambda \cdot R\left(\supp(\x)\right),
\end{equation} 
for some regularization parameter $\lambda > 0$. \\

Unfortunately, the composite minimization of a continuous and discrete function required in \eqref{eq: RegProbl} leads to an NP-hard problem, even for submodular set functions. In fact, even in the simple sparsity case in CS setting, where $f(\signal) := \frac{1}{2}\vectornorm{\obs - \sensing \signal}_2^2$, and $R(\sss) = |\sss|$, the problem \eqref{eq: RegProbl} is NP- hard. 

When submodular functions are paired with continuous functions like in \eqref{eq: RegProbl}, the minimization becomes very difficult. However when considered alone, submodular functions can be efficiently minimized, see Section \ref{sec:SFMalgo}. Submodular function minimization constitutes a key component in both the convex and discrete approaches to solving \eqref{eq: RegProbl}.

\subsection{The discrete model} \label{sec:sub_discrete}


In the discrete setting, \eqref{eq: RegProbl} is preserved in an attempt to faithfully encode the discrete model at hand. While such criterion seems cumbersome to solve, it has favorable properties that lead to polynomial solvability, irrespective of its combinatorial nature: there are efficient combinatorial algorithms that provide practical solutions to overcome this bottleneck and are guaranteed to converge in polynomial time \cite{el2013convexify}; see Section \ref{sec:algo_nonconvex}. 

Within this context, our intention here is to present fundamental algorithmic steps that reveal the underlying structure of \eqref{eq: RegProbl}. Here, we assume that $f$ has an $L$-Lipschitz continuous gradient and, thus, it admits the following upper bound: 
\begin{equation} \nonumber
f(\x) \leq f(\x^i) + \langle \nabla f(\x^i) , \x - \x^i \rangle + \frac{L}{2} \|\x - \x^i\|_2^2 := Q(\x,\x^i) \; \quad \forall \x, \x^i \in \text{dom}(f).
\end{equation}
Then, we perform the following iterative Majorization-Minimization (MM) scheme: 
\begin{equation}
\x^{i+1} \in \argmin_{\x \in \reals^n} \left\{Q(\x,\x^i) + \lambda R(\supp(\x))\right\} = \argmin_{\sss \subseteq \N} \left\{\min_{\x: \supp(\x)=\sss} Q(\x,\x^i) + \lambda R(\sss)\right\} \label{eq:MMiter}
\end{equation} 
If we focus on in the inner minimization above and for a given support $\sss$, one can compute the optimal minimizer $\widehat{\x}^i$ as:  $\widehat{\x}^i_{\sss^c} =0$ and
$\widehat{\x}^i_\sss = \x^i_\sss - (1/L)\nabla f(\x^i)_\sss.$
By substituting $\widehat{\x}^i_\sss$ into the upper bound $Q(\cdot,\cdot)$, we obtain a modular majorizer $M^i$:
\begin{align}
f(\x) + \lambda R(\supp(\x))  & \le C - \frac{L}{2} \sum_{j \in \supp(\x)} \left(x^i_j - \frac{1}{L} \nabla f(\x^i)_j\right)^2 + \lambda R(\supp(\x)) \nonumber\\
 & := M^i(\supp(\x)) +\lambda R(\supp(\x)), \nonumber
\end{align}
where $C$ is a constant. 
Finally, the update \eqref{eq:MMiter} then becomes:
\begin{equation} \label{eq:MMSFM}
\x^{i+1}= \widehat{\x}^i_{\sss} ~~~~ \quad \text{ subject to }\quad \quad  ~~~~\sss \in \arg \min_{\sss \subseteq \N} M^i(\sss) + R(\sss)
\end{equation}

The minimization in \eqref{eq:MMSFM} is a \emph{submodular function minimization} (SFM) problem, since the majorizer $M^i+R$ is submodular, and thus can be done efficiently.

\begin{prop}[\cite{el2013convexify}]
At each iteration, the new estimate $\x^{i+1}$ produced by the Majorization-Minimization algorithm satisfies $$f(\x^{i+1}) + R(\supp(\x^{i+1})) \leq  f(\x^{i}) + R(\supp(\x^{i})),$$ which implies convergence.
\end{prop}

Note that this scheme is a type of proximal gradient method (cf., Sect. \ref{sec:ProxGrad}), applied to submodular regularizers, where the proximal operation in \eqref{opt:05} corresponds to the SFM step in \eqref{eq:MMSFM}. We refer the reader to Section \ref{sec:algo_convex} for more information.

\subsection{Convex approaches} \label{sec:SubmodularConvex}

The non-convex problem \eqref{eq: RegProbl} can be converted to a closely related convex problem by replacing the discrete regularizer $g(\x)=R(\supp(\x))$ by its largest convex lower bound. This is also called convex envelope and is given by the biconjugate $g^{**}$, i.e. by applying the Fenchel-Legendre conjugate twice \cite{borwein2006convex}. We call this approach a convex relaxation of \eqref{eq: RegProbl}. However, for some functions $g$, there is no meaningful convex envelope. For these cases, one can use the convex closure $R^-: [0,1]^n \to \reals$, of $R(\sss)$, which is the point-wise largest lower bound of $R$ \cite{dughmi2009submodular}. Note that both the convex envelope and the convex closure are ``tight'' relaxations, but one in the continuous domain, the other in the discrete domain.

Both notions of convex relaxation for submodular functions use the Lov\'asz extension (LE), introduced in \cite{lovasz1983submodular}. We give here only one of the equivalent definitions:
\begin{definition}
Given a set function $R$ such that $R(\emptyset) = 0$, we define its Lov\'asz extension as follows: $\forall x \in \reals^N$, 
\begin{equation}
\label{eq:lovasz1}
r(x)= \sum_{k=1}^{N} x_{j_k} \Big (R(\{j_1, \ldots, j_k\}) - R(\{j_1, \ldots, j_{k-1}\} ) \Big )
\end{equation}
where $x_{j_1}\geq \cdots \geq x_{j_N}$.
\end{definition} 
The Lov\'asz extension is the convex closure of its corresponding submodular function on the unit hypercube, i.e $R^- = r$ \cite{dughmi2009submodular}. In this sense, this extension already gives a convex relaxation for any submodular function. However, it turns out that using the Lov\'asz extension as a convexification might not always fully capture the structure of the discrete penalty. We elaborate more on this in Section \ref{sec:sub_ex}.

For a monotone\footnote{A monotone function is a function that satisfies: $\forall \S \subseteq \T \subseteq \N, R(\S) \leq R(\T)$} submodular function its convex envelope is also obtained through the LE.

\begin{prop}[\cite{bach2010structured}]\label{prop:relSubMonot}
Given a monotone submodular function $R$, s.t $R(\emptyset)=0$, and its Lov\'asz extension $r$, the convex envelope of $g(\x) =R(\supp(\x))$ on the unit $\ell_\infty$-ball is given by the norm $\Omega_{sub}(\x) := g^{**}(\x) = r(|\x|)$. 
\end{prop}

The proximity operator \eqref{eq:prox} of $\Omega_{sub}$ can be efficiently computed. In fact, computing $\text{prox}_{\lambda}^{\Omega_{sub}}(\x)$ is shown in \cite{bach2010structured} to be equivalent to minimizing the submodular function $\lambda R(\sss) - \sum_{i \in \sss} |x_i|$ using the minimum-norm point algorithm. Other SFM algorithms can also be used, but then solving a sequence of SFMs is needed. Note that simpler subcases of submodular functions may admit more efficient proximity operator than this general one.\\
As a result, the convex relaxation of \eqref{eq: RegProbl}, where $R(\supp(\x))$ is replaced by $\Omega_{sub}(\x)$, can then be efficiently solved by a proximal gradient method (cf., Sect. \ref{sec:ProxGrad}).\\

The monotonicity requirement in \eqref{prop:relSubMonot} is necessary to guarantee the convexity of $r(|\x|)$. Unfortunately, all symmetric\footnote{A symmetric function is a function that satisfies: $\forall \S \subseteq \N, R(\S)= R(\N \setminus \S)$} submodular functions. among which undirected cut functions (cf., Sect. \ref{sec:sub_ex}), are not monotone. 
In fact, the convex envelope of $R(\supp(\x))$ on the unit $\ell_{\infty}$-ball, for $R$ any submodular function, is given by \cite{bach2010structured}: 
\begin{equation}
\label{eq:Bidual} 
g^{**}(\x)= \min_{\delta \in [0,1]^n, \bf \delta \geq |\x|} r(\delta). 
\end{equation} 
Thus, when $R$ is non-decreasing, the convex envelope is $r(|\x|)$ as stated in Proposition \ref{prop:relSubMonot}. When $R$ is symmetric and $R(\N)=R(\emptyset)=0$ (which is assumed w.l.o.g since addition of constants doesn't affect the regularization), $g^{**}(\x)=0, \forall \x \in \reals^n$ and the minimum in \eqref{eq:Bidual} is attained at any constant value vector $\delta$. 

So the convexification of symmetric submodular functions consists of the Lov\'asz extension alone. However the LE is a poor convexification that can significantly modify the intended structure. This can already be seen by the fact that the LE is tight only on the unit hypercube, while most penalties $R(\supp(\x))$ are not constrained there, and that $R(\supp(\x))$ is symmetric around the origin which is not the case for $r(\x)$. Some artifacts of using the LE as a convexification for symmetric functions are illustrated in \cite{el2013convexify}.


 For some problems, the submodularity requirement can be relaxed. Convex relaxations of general set functions, when paired with the $\ell_p$-norm $(p>1)$ as a continuous prior are studied in \cite{obozinski2012convex}.
 
 \begin{prop}[\cite{obozinski2012convex}] \label{prop:lpRel}
 Define the norm
\begin{equation}
\Omega_p(\x) := \min_{\mathbf{v} \in \mathcal{V}} \{ \sum_{\sss \subseteq \N} R(\sss)^{\frac{1}{q}} \|\mathbf{v}^\sss\|_p ~\big|~ \sum_{\sss \subseteq \N} \mathbf{v}^\sss= {\bf x}  \} ,
\end{equation}
where $\frac{1}{p} + \frac{1}{q}=1$ and $\mathcal{V}=\{ \mathbf{v}=(\mathbf{v}^\sss)_{\sss \subseteq \N} \in (\reals^n)^{2^\N} \text{ s.t } \supp(\mathbf{v}^\sss) \subseteq \sss\}$.\\ \\
$\Omega_p$ is the convex positively homogeneous envelope of the function $\lambda R(\supp(\x)) + \mu \Vert \x \Vert_p^p$ (up to a constant factor), and the convex envelope of $R(\supp(\x))^{\frac{1}{q}} \Vert \x \Vert_p$.
 \end{prop}
 Note that this is the same norm as \eqref{eq:atomic_norm}, with weights $d_j$ given by a set function $R$, such that $d_j = R(\G_j)^{1/q}$, and the group structure $\GG$ is the entire power set of $\N$. 
 
The convex relaxation proposed in Proposition \ref{prop:lpRel}, when applied to a submodular function, generalizes the relaxation of Proposition \ref{prop:relSubMonot} to unit balls with respect to the $\ell_p$-norm $(p > 1)$ and not only to the $\ell_\infty$ one.



\subsection{Examples}\label{sec:sub_ex}

We now offer some examples of structures that can be described via submodular functions.

\begin{figure}[ht]
\centering
\begin{minipage}[b]{0.35\linewidth}
\centering
\begin{tabular}{cc}
\includegraphics[width=0.5\columnwidth]{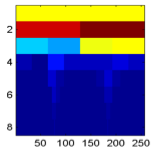}&
\includegraphics[width=0.5\columnwidth, height=0.13\textheight]{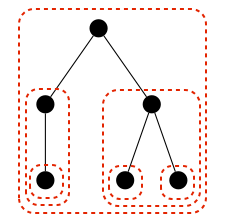} 
\end{tabular}
\caption{Wavelet coefficients and underlying hierarchical group model.}
\label{fig:wavelet}
\end{minipage}
\hspace{0.5cm}
\begin{minipage}[b]{0.35\linewidth}
\centering
\begin{tabular}{cc}
\includegraphics[width=0.5\columnwidth]{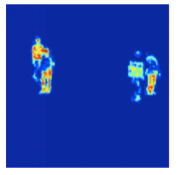}&
\includegraphics[width=0.5\columnwidth]{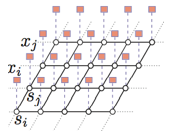} 
\end{tabular}
\caption{Background subtraction and underlying Ising model.}
\label{fig:Ising}
\end{minipage}
\end{figure}

\medskip
\textbf{Simple sparsity:}
The most ubiquitous structure in practice is simple sparsity, which corresponds to the set function $R(\sss) = |\sss|$ that measures the cardinality of $\sss$. 
The convexification obtained by Proposition \ref{prop:relSubMonot} is the usual $\ell_1$-norm. 

\medskip
\textbf{Group sparsity:} Given a group structure $\GG$, a group sparse model can be enforced by penalizing the number of groups that intersects with the support: $R_{\cap}(\sss) = \sum_{\sss \cap \G_i \ne \emptyset} ~ d_i$ where $d_i\geq 0$ is the weight associated with the group $\G_i \in \GG$. For example, defining the groups to be each node and all of its descendants on a tree (see Figure \ref{fig:wavelet}), enforces the support to form a rooted connected tree, which is characteristic of wavelet coefficients (cf., Sect. \ref{sec:hierarchical}). The way the groups and their weights are defined leads to different sparsity patterns \cite{bach2011learning}.

 The convexification of $R_{\cap}(\S)$ is the $\ell_1/\ell_\infty$-norm\footnote{Actually, it is a norm iff $\N = \cup_{d_i > 0} ~G_i $} over groups $\sum_{\G_i \in \GG} d_i \Vert \x \Vert_\infty$  (even for overlapping groups) \cite{bach2010structured, zhao2009composite}. 
Note that for groups that form a partition of $\N$,  $R_{\cap}(\S)$ is equivalent to the minimum weight set cover, defined as:
\begin{equation}\label{eq:SetCover}
R_{sc}(\sss) = \min_{s \in \BB^M} ~\sum_{i=1}^M d_i s_i ~~\text{ s.t } \sum_{i=1}^M s_i \1_{\dim,\G_i} \geq \1_{\dim,\sss}
\end{equation}
with the same set of groups $\GG$ and weights $d$. $R_{sc}$ computes the total weight of the lightest cover for $\S$ using the groups in $\GG$. Note that $R_{sc}(\sss)$ is not a submodular function\footnote{Consider this example: let $\N=\{1,2,3,4\}$, and $\GG= \{ \G_1=\{1\}, \G_2=\{2,3\}, \G_3=\{1,2,4\} \}$, with weights defined as $d_i = |\G_i|$. Then the inequality in definition \ref{def:Sub} is not satisfied for the sets $\S=\{1,2\}$ for which $R_{sc}(\S)=3$, and $\T=\{1,2,4\}$ for which $R_{sc}(\T)=4$ ,with the addition of the element $\{3\}$.}, unless the groups form a partition. 

\medskip
\textbf{Ising model:}
The Ising model  \cite{mccoy1973two}  is a model that associates with each coefficient $x_i$ of a signal $\x \in \reals^n$ a discrete variable $s_i\in \{-1,1\}$ to represent the state (zero/non-zero) of the coefficient. The Ising penalty enforces clustering over the coefficients, which is a desired structure for example in background subtraction in images or videos \cite{cevher2009sparse}. 
It can be naturally encoded on a graph $\mathcal{G}=(\mathcal{V,E})$ where the vertices are the coefficients of the signal, i.e $\mathcal{V}= \N$, and the edges connects neighboring coefficients. For example, for images, the coefficients of the signal are the pixels of the image, and edges connect pixels next to each other, forming a lattice (see Figure \ref{fig:Ising}). This is called the two dimensional lattice Ising model. The Ising penalty is then expressed via the following symmetric submodular function: 
\begin{equation}
\label{eq:ising}
R_{\text{\sc Ising}}(\sss) =  \frac{1}{2} \bigg ( |\mathcal{E}|\ - \sum_{(i,j) \in \mathcal{E}} s_i s_j  \bigg ), \; 
\end{equation}
where ${\bf s} \in \reals^\dim$ is an indicator vector for the set $\sss$ such that $s_i = 1$ if $i \in \sss$ and $s_i = -1$, otherwise. When $\sss=\supp(\x)$ for a signal $\x \in \reals^\dim$, $s_i$ encodes the state of the corresponding coefficient $x_i$.
We can also view the Ising penalty as a cut function: $R_{\text{\sc Ising}}(\sss)$  indeed just counts the number of edges that are cut by the set $\sss$. 
Cut functions with appropriate graphs and weights actually capture a large subset of submodular functions \cite{jegelka2011fast, kolmogorov2004energy}. 
Since $R_{\text{\sc Ising}}$ is symmetric, its convexification is just its Lov\'asz extension, as discussed above, which is shown \cite{el2013convexify} to be the anisotropic discrete Total Variation semi-norm 
\begin{equation}
\|\x\|_{TV} = \sum_{(i,j) \in \mathcal{E}} |x_i - x_j| \label{def:TV}
\end{equation}
 While the Ising model enforces the clustering of non-zero coefficients, its convexification encodes a different structure by favoring piece-wise constant signals. Furthermore, this penalty cannot be described by any of the structures introduced in the previous section.
 
 \medskip
\textbf{Entropy:}
Given $n$ random variables $X_1, \cdots, X_n$, the joint entropy of $X_\sss= (X_i)_{i\in \sss}$ defines a submodular function, $R(\sss)=H(X_\sss)$. Moreover, the mutual information also defines a symmetric submodular function, $R(\sss)= I(X_\sss, X_{\N \setminus \sss})$. Such functions occur, for example, in experimental design \cite{seeger2009submodularity}, in learning of Bayesian networks \cite{heckerman1995learning}, in semi-supervised clustering \cite{narasimhan2005q}, and in diverse feature selection \cite{das2012selecting}. Both of these set functions encourage structures that cannot be defined by the models presented in the previous sections.

\medskip
\textbf{Supermodular functions:}
Another variant of set function regularizers used in practice, for example in diverse feature selection, are supermodular functions, i.e. the negative of a submodular function. The problem \eqref{eq: RegProbl} with supermodular regularizers can then be cast as a maximization problem with a submodular regularizer. The notion of ``approximate" submodularity introduced in \cite{das2011submodular, krause2010submodular} is used to reduce the problem to an ``approximate" submodular maximization problem, where several efficient algorithms for constrained/unconstrained submodular maximization \cite{buchbinder2012tight, buchbinder2014submodular} with provable approximation guarantees can be employed. For more information, we refer the reader to \cite{das2012selecting}.\\ \\


%% file: algorithms.tex
\section{Optimization}
\label{sec:optimization}

In order to use the previous structures in practice, one needs efficient optimization solutions for structured sparsity problems that scale up in high-dimensional settings. From our discussions above, it is apparent that the key actors for this purpose are projection and proximity operations over structured sets that go beyond simple selection heuristics and towards provable solution quality as well as runtime/space bounds. 

Projection operations faithfully follow the underlying combinatorial model but, in most cases, result in hard-to-solve or even combinatorial optimization problems. Furthermore, model misspecification often results in wildly inaccurate solutions. 

Proximity operators of convex sparsity-inducing norms often can only partially describe the underlying discrete model and might lead to ``rules-of-thumb'' in problem solving (e.g., how to set up the regularization parameter). However, such approaches work quite well in practice and are more robust to deviations from the model, leading to satisfactory solutions.

Here, our intention is to present an overview of the dominant approaches followed in practice. We consider the following three general optimization formulations:\footnote{We acknowledge that there are other criteria that can be considered in practice; for completeness, in the simple sparsity case, we refer the reader to the $\ell_1$-norm constrained linear regression (a.k.a. Lasso \cite{tibshirani1996regression})---similarly, there are alternative optimization approaches for the discrete case \cite{wright2009sparse}. However, our intention in this chapter is to use the most prevalent formulations used in practice.}
\begin{itemize}
\item {\it Discrete projection formulation:} Given a signal model $\constraint$, let $f:~\R^\dim \rightarrow \R$ be a closed convex data fidelity$/$loss function. Here, we focus on the {\it projected} non-convex minimization problem: \vspace{-0.3cm}
\begin{equation}
	\begin{aligned}
	& \underset{\x \in \R^\dim}{\text{minimize}}
	& & f(\x)
	& \text{subject to} 
	&& \x \in \constraint.
	\end{aligned} \label{opt:01}
\end{equation}  \vspace{-0.3cm}
\item {\it Convex proximity formulation:} Given a signal model $\constraint$, let $f:~\R^\dim \rightarrow \R$ be a closed convex data fidelity$/$loss function, $g:~\R^{\dim} \rightarrow \R$ a closed convex regularization term, possibly non-smooth, that faithfully models $\constraint$ and $\lambda > 0$. In this chapter, we focus on the convex composite minimization problem:  \vspace{-0.3cm}
\begin{equation}
	\begin{aligned}
	& \underset{\x \in \R^\dim}{\text{minimize}}
	& & f(\x) + \lambda\cdot g(\x).
	\end{aligned} \label{opt:00}
\end{equation}   \vspace{-0.3cm}
\item {\it Convex structured-norm minimization:} Given a signal model $\constraint$, let $g:~\R^{\dim} \rightarrow \R$ be a closed convex regularization term, possibly non-smooth, that faithfully models $\constraint$. Moreover, let $f:~\R^\dim \rightarrow \R$ be a closed convex data fidelity$/$loss function and $\sigma > 0$. We consider the following minimization problem:  \vspace{-0.3cm}
\begin{equation}
	\begin{aligned}
	& \underset{\x \in \R^\dim}{\text{minimize}}
	& & g(\x)
	& \text{subject to} 
	&& f(\x) \leq \sigma.
	\end{aligned} \label{opt:02}
\end{equation}  \vspace{-0.3cm}
\end{itemize}   \vspace{-0.3cm}

\noindent In addition, we also briefly mention submodular function minimization (SFM)  \vspace{-0.3cm}
$$
	\begin{aligned}
	& \underset{\mathcal{S} \in 2^{\N}}{\text{minimize}}
	& & R(\mathcal{S})
	\end{aligned} \label{opt:sub}  \vspace{-0.3cm}
$$ 
where $R: 2^{\N} \to \reals$ is a submodular function; see Definition \ref{def:Sub}.
SFM is at the core of both discrete and convex approaches for solving estimation problems, where the structure is imposed via a submodular set function; see Section \ref{sec:submodular}.

Some representative problem instances are given next.

\medskip
\noindent \textbf{Compressive sensing (CS): }In classic CS, we are interested in recovering a high-dimensional signal $ \xtrue \in \constraint $ from an underdetermined set of linear observations $ \obs \in \mathbb{R}^\numsam ~(m < n)$, generated via a known $ \sensing \in \R^{\numsam \times \dim}$:\vspace{-2mm}
\begin{align*}
\obs = \sensing \xtrue + \noise. \vspace{-2mm}
\end{align*} Here, $ \noise \in \mathbb{R}^{\numsam} $ denotes an additive noise term with $ \vectornorm{\noise}_2 \leq \sigma $. 

For $ f(\signal) := \frac{1}{2}\vectornorm{\obs - \sensing \signal}_2^2 $ and using the fact that $ \xtrue \in \constraint $, one can formulate the problem at hand as in \eqref{opt:01}; observe that, in the case where $\constraint \equiv \sparse_\sparsity$, \eqref{opt:01} corresponds to the $\ell_0$ pseudo-norm optimization problem \cite{natarajan1995sparse}. Alternatively, given a function $g$ that encourages solutions to be in $\constraint$, one can solve the CS problem with convex machinery as in \eqref{opt:00}-\eqref{opt:02}; when $\constraint \equiv \sparse_\sparsity$, \eqref{opt:00}-\eqref{opt:02} corresponds to the Basis Pursuit DeNoising (BPDN) criterion \cite{chen1998atomic}.


\medskip
\noindent \textbf{Sparse graph modeling:} 
In Gaussian graph model selection \cite{dahl2008covariance, tran2013proximal, kyrillidis2013fast, kyrillidis2014scalable, dinh2013inexact, hsieh2013big, hsieh2011sparse}, the sparse-regularized maximum likelihood estimation yields the criterion: \vspace{-0.2cm}
\begin{equation}
	\begin{aligned}
		\hat{\x} ~=~& \underset{\x \in \R^{n^2}}{\text{argmin}}
		& \Big\{-\log\det(\text{mat}(\x)) + \tr(\text{mat}(\x) \cdot \widehat{\mathbf{C}}) + \lambda \cdot g(\x)\Big\}, \label{eq:1}
	\end{aligned}  \vspace{-0.15cm}
\end{equation} where $\x \in \R^{n^2}$ is the vectorized version of the inverse covariance estimate, $\text{mat}(\cdot)$ is the linear matricization operation and $\widehat{\mathbf{C}}$ is the sample covariance. Developments in random matrix theory \cite{johnstone2001distribution} divulge the poor performance of solving \eqref{eq:1} without regularization: in that case, the solution is usually fully dense and no inference about the dependence graph structure is possible. Choosing a suitable regularizer $g$ \cite{jalali2011learning, lee2013structure, loh2013structure}, that models the connections between the variables, reduces the degrees of freedom, leading to robust and more interpretable results.

\medskip
\noindent \textbf{Other applications:} Similar sparsity regularization in discrete or convex settings is found many diverse disciplines: low-light imaging problem in signal processing \cite{harmany2012spiral}, heteroschedastic LASSO \cite{dalalyan2013learning}, dictionary learning, \cite{szabo2011online}, etc.

\subsection{Greedy and discrete approaches}{\label{sec:algo_nonconvex}}

\subsubsection{Projected gradient descent and matching pursuit variants}
Iterative greedy algorithms maintain the combinatorial nature of (\ref{opt:01}), but instead of tackling it directly, which would be intractable, the algorithms of this class greedily refine a $ k $-model-sparse solution using only ``local'' information available at the current iteration. Within this context, matching pursuit approaches \cite{mallat1993matching, tropp2007signal} gradually construct the sparse estimate by greedily choosing the non-zero coefficients that best explain the current residual, e.g. ${\bf y} - {\bf A}{\bf x}^i$, where ${\bf x}^i$ is the current iterate. 
Extensions of such greedy approaches have been recently proposed to accommodate structured models in the selection process \cite{huang2011learning}.

Most of the algorithmic solutions so far concentrate on the non-convex projected gradient descent algorithm: a popular method, known for its simplicity and ease of implementation. Per iteration, the total computational complexity is determined by the calculation of the gradient and the projection operation on $\constraint$ as in \eqref{eq:proj}. Most algorithms in the discrete model can be described or easily deduced by the following simple recursion: 
\begin{align}
\x^{i+1} = \mathcal{P}_{\constraint}\left(\x^i - \frac{\mu}{2} \nabla f(\x^i)\right), \label{eqq:00}
\end{align} where $ \mu $ is a step size and $ \mathcal{P}_{\constraint}(\cdot) $ is the projection onto $ k $-model-sparse signals.

Representative examples for the CS application are hard thresholding methods over simple sparse sets \cite{blumensath2009iterative, needell2009cosamp, foucart2011hard, kyrillidis2011recipes, kyrillidis2012hard, cevher2014linear}. \cite{baraniuk2010model} further extends these ideas to \emph{model-based CS}, where non-overlapping group structures and tree structures are used to perform the projection \eqref{eq:proj}. From a theoretical computer science perspective, \cite{indyk2013model, bah2014model} address the CS problem using \emph{sparse} sensing matrices. Exploiting model-based sparsity in recovery provably reduces the number of measurements $\numsam$, without sacrificing accuracy. The resulting algorithm reduces the main computational cost
of the proposed scheme on the difficulty of projecting onto the model-sparse set, which, as mentioned in Section \ref{sec:groups}, in most relevant cases, can be computed in linear time using dynamic programming.\footnote{In the case of CS, an important modification of \eqref{eqq:00} to achieve linear computational time per iteration is the substitution of the gradient with the \emph{median} operator, which is nonlinear and defined component-wise on a vector; for more information, we refer to \cite{foucart2011hard, gilbert2010sparse}.} 

\subsubsection{Submodular function minimization} \label{sec:SFMalgo}

Submodularity is considered the discrete equivalent of convexity in the sense that it allows efficient minimization. The best combinatorial algorithm for submodular function minimization (SFM) has a  proven complexity of $O(n^5 T + n^6)$, where $T$ is the function evaluation complexity \cite{orlin2009faster}. For practical purposes however, another algorithm called minimum-norm point algorithm \cite{fujishige2011submodular} is usually used. This algorithm has no known worst case complexity but in practice it usually runs in $O(n^2)$. Furthermore, for certain functions which are ``graph representable" \cite{jegelka2011fast, fujishige2001realization}, submodular minimization becomes equivalent to computing the  minimum s-t cut on the corresponding graph $G(\mathcal{V},\mathcal{E})$, which has time complexity\footnote{the notation $\tilde{O}(\cdot) $ ignores $\log$ terms} $\tilde{O}(|\mathcal{E}| \min\{|\mathcal{V}|^{2/3},|\mathcal{E}|^{1/2}\})$ \cite{Goldberg:1998:BFD:290179.290181}.


\subsection{Convex approaches}\label{sec:algo_convex}

\subsubsection{Proximity methods}\label{sec:ProxGrad}
Proximity gradient methods are iterative processes that rely on two key structural assumptions: i) $f$ has Lipschitz continuous gradient\footnote{In \cite{tran2013composite}, the authors consider a more general class of functions with no \emph{global} Lipschitz constant $L$ over their domain. The description of this material is out of the scope of this chapter and is left to the reader who is interested in deeper convex analysis and optimization.} (see Definition \ref{def:lipschitz}) and ii) the regularizing term $g$ is endowed with a \emph{tractable} proximity operator.

By the Lipschitz gradient continuity and given a putative solution $\x^i \in \text{dom}(f + g)$, one can locally approximate $f$ around $\x^i$ using a quadratic function as:
\begin{align*}
f(\x) \leq Q(\x, \x^i) := f(\x^i) + \nabla f(\x^i)^T (\x - \x^i) + \frac{L}{2}\|\x - \x^i\|_2^2, \quad \forall \x \in \text{dom}(F).
\end{align*} 
The special structure of this upper-bound allows us to consider a majorization-minimization approach: instead of solving \eqref{opt:00} directly, we solve a sequence of simpler composite quadratic problems:
\begin{align}
\x^{i+1} \in \argmin_{\x \in \R^n} \left \{ Q(\x, \x^i) + g(\x) \right \}. \label{opt:04}
\end{align}
In particular, we observe that \eqref{opt:04} is equivalent to the following iterative \emph{proximity} operation, similar to \eqref{eq:prox}:
\begin{align}{\label{opt:05}}
\x^{i+1} \in \argmin_{\x \in \R^n} \left \{ \frac{1}{2}\left\|\x - \left(\x^i - \frac{1}{L}\nabla f(\x^i) \right) \right\|_2^2 + \frac{1}{2L} g(\x) \right \}.
\end{align} Here, the anchor point $\w$ in \eqref{eq:prox} is the gradient descent step: $\w := \x^i - \frac{1}{L}\nabla f(\x^i)$. 

In the case where $g(\x) := \|\x\|_1$, the proximity algorithm in \eqref{opt:05} is known as the Iterative Soft-Thresholding Algorithm (ISTA) \cite{combettes2005signal, chambolle1998nonlinear, daubechies2004iterative}. 
Iterative algorithms can use memory to provide momentum in convergence. Based on Nesterov's optimal gradient methods \cite{nesterov1983method}, \cite{beck2009fast} proves the universality of such acceleration in the composite convex minimization case of \eqref{opt:00}, where $g(\x)$ can be any convex norm with tractable proximity operator.
However, the resulting optimization criterion in \eqref{opt:05} is more challenging when $g$ stands for more elaborate sparsity structures. 

Within this context, \cite{Schmidt2011, villa2013accelerated} present a new convergence analysis for proximity (accelerated) gradient problems, under the assumption of \emph{inexact proximity evaluations} and study how these errors propagate into the convergence rate. 


An emerging direction for solving composite minimization problems of the form \eqref{opt:00} is based on the proximity-Newton method \cite{fukushima1981generalized}. The origins of this method can be traced back to the work of \cite{fukushima1981generalized, Bonnans1994}, which relies on the concept of \textit{strong regularity} introduced by \cite{Robinson1980} for generalized equations. 
In this case, we identify that the basic optimization framework above can be easily adjusted to second-order Newton gradient and quasi-Newton approaches:
\begin{align}{\label{opt:06}}
\x^{i+1} \in \argmin_{\x \in \R^n} \left \{ \frac{1}{2}\left\|\x - \left(\x^i - \mathbf{H}_i^{-1}\nabla f(\x^i) \right) \right\|_{\mathbf{H}_i}^2 + \frac{1}{2} g(\x) \right \}.
\end{align}
where $\mathbf{H}_i $ represents either the actual Hessian of $f$ at $\x^i$ (i.e., $\nabla^2 f(\x_i)$) or a symmetric positive definite matrix approximating $\nabla^2f(\x^i)$. Given a computationally efficient Newton direction, one can re-use the model-based proximity solutions presented in the previous subsection along with a second order \emph{variable metric} gradient descent scheme, as presented in \eqref{opt:06} \cite{tran2013composite}.

\subsubsection{Primal-dual and alternating minimization approaches}

Several model-based problems can be solved using the convex structured-norm minimization in \eqref{opt:02}. As a stylized example, consider the noiseless CS problem formulation
\begin{equation}\label{eq:cvx_prob}
	\begin{aligned}
	& \underset{\mathbf{x} \in \mathbb{R}^n}{\text{minimize}}
	& & g(\mathbf{x})
	& \text{subject to} 
	&& \mathbf{u} = \boldsymbol{\Phi} \mathbf{x},
	\end{aligned} 
\end{equation} where $g$ is a convex structured norm, mimicking the model at hand.

Within this context, primal-dual convex optimization methods provide attractive approaches, by exploiting both the primal and the dual formulations at the same time (via Lagrangian dualization). More precisely, primal-dual methods solve the following minimax problem:
\begin{equation}\label{eq:minmax}
\min_{\mathbf{x}\in\mathbb{R}^n} \max_{\mathbf{y}\in\mathbb{R}^m}\left\{  g(\mathbf{x}) + \langle \boldsymbol{\Phi}\mathbf{x} - \mathbf{u}, \mathbf{y}\rangle\right\},
\end{equation} 
where $\mathbf{y}$ is the Lagrange multiplier associated to the equality constraint $\mathbf{u} = \boldsymbol{\Phi}\mathbf{x}$. 
The minimax formulation \eqref{eq:minmax} can be considered as a special case of the general formulation in \cite{Chambolle2011,Nesterov2005d}.

If the function $g$ is nonsmooth and possesses a closed form proximity operator,  then general primal-dual methods such as \cite{Chambolle2011,He2012} have been known as powerful tools to tackle problem \eqref{eq:minmax}. Since $g$ is nonsmooth, one can also use the primal-dual subgradient method \cite{Nesterov2009b} as well as the prox-method proposed in \cite{Nemirovskii2004} for solving \eqref{eq:cvx_prob}.
An alternative approach is to combine primal-dual optimization methods and smoothing technique as done in  \cite{Nesterov2005d,Nesterov2005c} for solving \eqref{eq:minmax}.

When \eqref{eq:minmax} possesses a separable structure $g(\mathbf{x}) := \sum_{i=1}^p g_i(\mathbf{x}_i)$, distributed approaches can be applied. Recently, the authors in \cite{Tran-Dinh2014a} developed a unified primal-dual decomposition framework for solving \eqref{eq:cvx_prob}, using also ideas from the alternating direction methods of multipliers \cite{Tseng1991a, boyd2011distributed, combettes2005signal}. 
Such alternating optimization methods offer a unifying computational perspective on a broad set of large-scale convex estimation problems in machine learning \cite{jenatton2011structured} and signal processing \cite{combettes2005signal}, including sparse recovery, deconvolution, and de-mixing \cite{goldstein2012fast}. They have numerous universality and scalability benefits and, in most cases, they are well-suited to distributed convex optimization. In this case, one can borrow ideas and tools from gradient descent and Newton schemes in solving the subproblems of the alternating minimization. While their theoretical convergence guarantees are not optimal, they usually work well in practice. 

%% file: experiments.tex
\section{Applications}
\label{sec:applications}

\subsection{Compressive Imaging}

Natural images are usually sparse in wavelet basis. In this experiment, we study the image reconstruction problem from compressive measurements, where structured sparsity ideas are applied in practice. 

For this purpose and given a $p \times p$ natural grayscale image ${\bf x} \in \reals^{p^2}$, we use the Discrete Wavelet Transform (DWT) with $\log_2(p)$ levels, based on the Daubechies $4$ wavelet, to represent $\x$; see the Wavelet representation of two images in Figures \ref{fig:cs_image_results_woman}-\ref{fig:cs_image_results_mountains}. In math terms, the DWT can be represented by an operator matrix $\mathbf{W}^\top$, so that $\x$ can be sparsely represented (or well-approximated) as $\x = \mathbf{W}\mathbf{c}$, where $\mathbf{c} \in \reals^{n}$, $n := p^2$ are the wavelet coefficients for ${\bf x}$.

To exploit this fact in practice, we consider the problem of recovering ${\bf x} \in \reals^{n}$ from $\numsam$ compressive measurements $\obs \in \reals^\numsam$. 
The measurements are obtained by applying a sparse matrix ${\bf A} \in \reals^{\numsam \times p^2}$ to the vectorized image such that:
$$\obs = {\bf A} \x. $$ 
Here, ${\bf A}$ is the adjacency matrix of an expander graph of degree $d = 8$, so that $\|{\bf A}\|_0 = dn$. 
Thus, the overall measurement operator on the wavelet coefficients is then given by the concatenation of the expander matrix with the DWT: $\obs = \A \mathbf{W}\mathbf{c}$, with $\mathbf{c} \approx \mathbf{\widehat{c}}$ with $\|\mathbf{\widehat{c}}\|_0 \ll m^2$, i.e., $\x$ can be well-approximated by using only a limited number of wavelet coefficients.

We use the following methods for recovering $\mathbf{c}$ from the measurements $\obs$:
\begin{equation}
\begin{aligned}
& \underset{\mathbf{c} \in \R^{n}}{\text{minimize}} && 
 \|\obs - \A\mathbf{W}\mathbf{c} \|_2^2 \\ & \text{subject to} &&  \supp(\mathbf{c}) \in \mathcal{T}_k. 
\end{aligned}   ~~~~\quad \quad  \tag{Rooted Connected Tree model (RC)}
\end{equation}
\medskip 
\begin{equation}
\begin{aligned}
& \underset{\mathbf{c} \in \R^{n}}{\text{minimize}} && 
\|\mathbf{c}\|_1 \\ & \text{subject to} && \obs = \A\mathbf{W}\mathbf{c}. 
\end{aligned}  ~~~~\quad \quad \quad  \quad \quad \quad \quad \quad \quad \quad \quad  \tag{Basis Pursuit (BP)}
\end{equation}
\medskip 
\begin{equation}
\begin{aligned}
& \underset{\mathbf{c} \in \R^{n}}{\text{minimize}} && 
\|\mathbf{c}\|_\text{HGL} \\ & \text{subject to} && \obs = \A\mathbf{W}\mathbf{c}.
\end{aligned}  ~~~\quad \quad  \tag{Hierarchical Group Lasso (HGL) pursuit}
\end{equation}
\begin{equation}
\medskip
\begin{aligned}
& \underset{\mathbf{c} \in \R^{n}}{\text{minimize}} && 
\|\mathbf{c}\|_\text{PC} \\ & \text{subject to} && \obs = \A\mathbf{W}\mathbf{c}.
\end{aligned} \tag{Parent-Child Latent Group Lasso (PC) pursuit}
\end{equation}
\medskip 
\begin{equation}
\begin{aligned}
& \underset{\mathbf{c} \in \R^{n}}{\text{minimize}} && 
\|\mathbf{c}\|_\text{FAM} \\ & \text{subject to} && \obs = \A\mathbf{W}\mathbf{c}.
\end{aligned} ~~\quad \quad ~~~ \tag{Family Latent Group Lasso (FAM) pursuit}
\end{equation}

The RC model is solved via the improved projected gradient descent given in \cite{kyrillidis2011recipes} with the projections computed via the dynamic program proposed in \cite{baldassarre2013group}. All the remaining methods are solved using the primal-dual method described in \cite{trandinh2014optimal} which relies on the proximity operator of the associated structure-sparsity inducing penalties. For BP the proximity operator is given by the standard soft-thresholding function. For HGL, we use the algorithm and code given by \cite{jenatton2011proximal}. For the latent group lasso approaches, PC and FAM, we adopt the duplication strategy proposed in \cite{jacob2009group, obozinski2011group}, for which the proximity operator reduces to the standard block-wise soft-thresholding on the duplicated variables. All algorithms are written in \texttt{Matlab}, except for the HGL proximity operator and the RC projection that are in \texttt{C}.

The duplication approach consists in creating a latent vector that contains copies of the original variables. 
The number of copies is determined by the number of groups that a given variable belongs to.
For the Parent-Child model, each node belongs to the four groups that contain each of its children and the group that contains its father. 
The root has only three children, corresponding to the roots of the horizontal, vertical and diagonal wavelet trees. 
The leaves belong only to the group that contains their fathers. 
A simple calculation shows that the latent vector for the PC model contains $2(\dim-1)$ variables. 

For the Family model, instead, each node belongs to only two groups: the group containing its children and the group containing its siblings and its father. Each leaf belongs to only the group that contains its siblings and its father.
Overall, the number of variables in the latent vector for the Family models is equal to $\frac{5\dim}{4}-1$.

The duplication approach does not significantly increase the problem size, while it allows an efficient implementation of the proximity operator.
Indeed, given that the proximity operator can be computed in closed form over the duplicated variables, this approach is as fast as the hierarchical group lasso one, where the proximity operator is computed via \texttt{C} code.

In order to obtain a good performance, both the parent-child and the family model require a proper weighting scheme to penalize groups lower down in the tree, where smaller wavelet coefficients are expected, compared to nodes closer to the root, which normally carry most of the energy of the signals and should be penalized less.
We have observed that setting the group weights proportional to the level $L$ of the node of the group closest to the root gives good results. In particular, we set the weights equal to $L^2$, with $0$ being the root level.

\subsubsection{Results}

We performed the compressive imaging experiments on both a $256 \times 256$ portrait of a woman and a $2048 \times 2048$ mountain landscape.
Apart from conversion to grayscale and resizing through the Matlab function \texttt{imresize}, no preprocessing has been carried out. 
The primal-dual algorithm of \cite{trandinh2014optimal} has been run up to precision $10^{-5}$.
We measure the recovery performance in terms of Power Signal to Noise Ratio (PSNR) and relative recovery error $\ell_2$ norm as $\frac{\|\widehat{\bf x} - {\bf x}\|_2}{\|{\bf x}\|}$, where $\widehat{\bf x}$ is the estimated image and ${\bf x}$ is the true image.

\begin{figure}
\centering
\captionsetup{width=0.8\textwidth}
\includegraphics[width=0.7\textwidth]{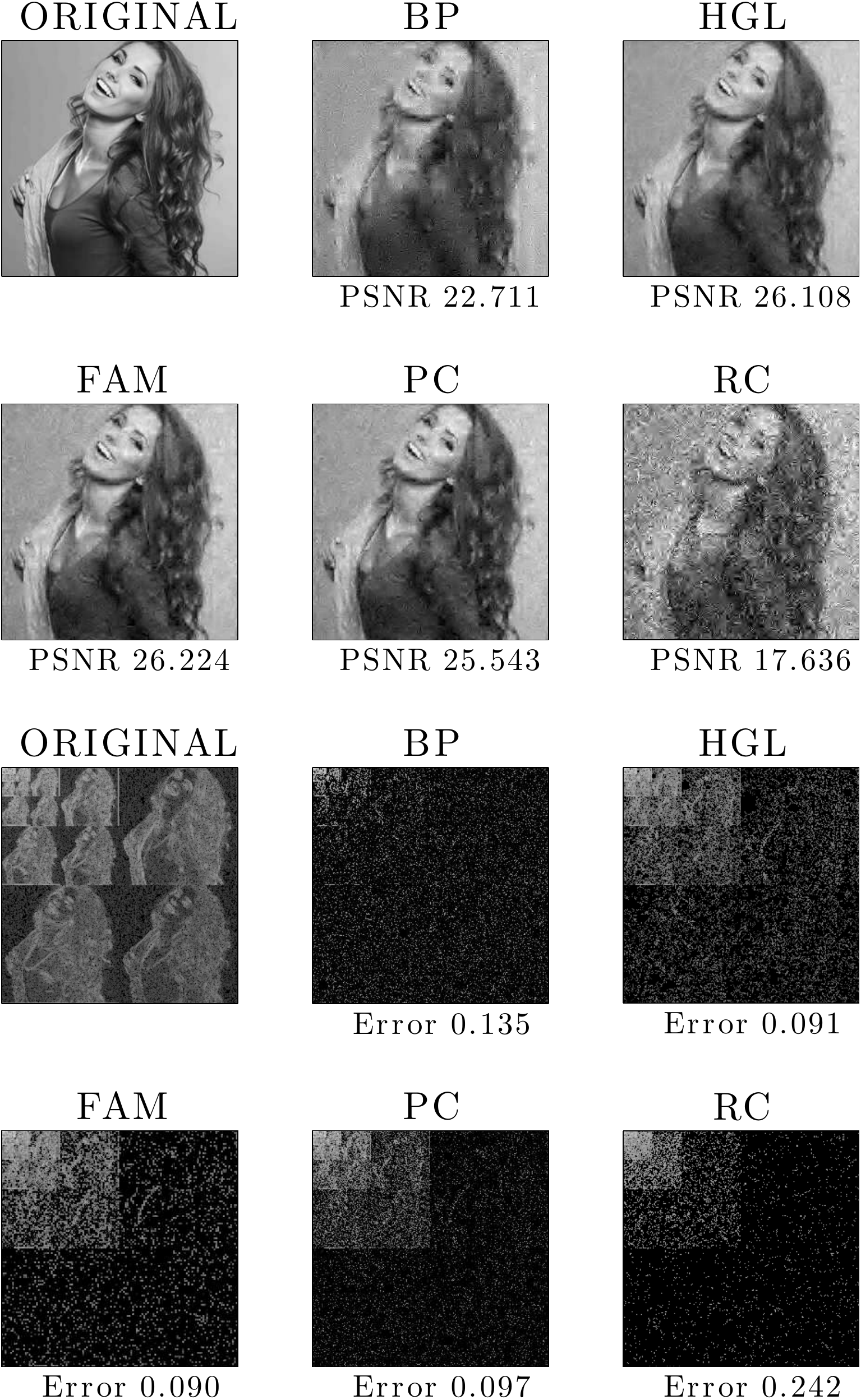}
\caption{\label{fig:cs_image_results_woman} Woman image recovery performance from compressive measurements. Here, $p = 256$. The top two rows show the reconstruction performance in the original domain, along with the PSNR levels achieved. The bottom two rows show the corresponding representations into the wavelet domain.}
\end{figure}

\begin{figure}
\centering
\captionsetup{width=0.8\textwidth}
\includegraphics[width=0.7\textwidth]{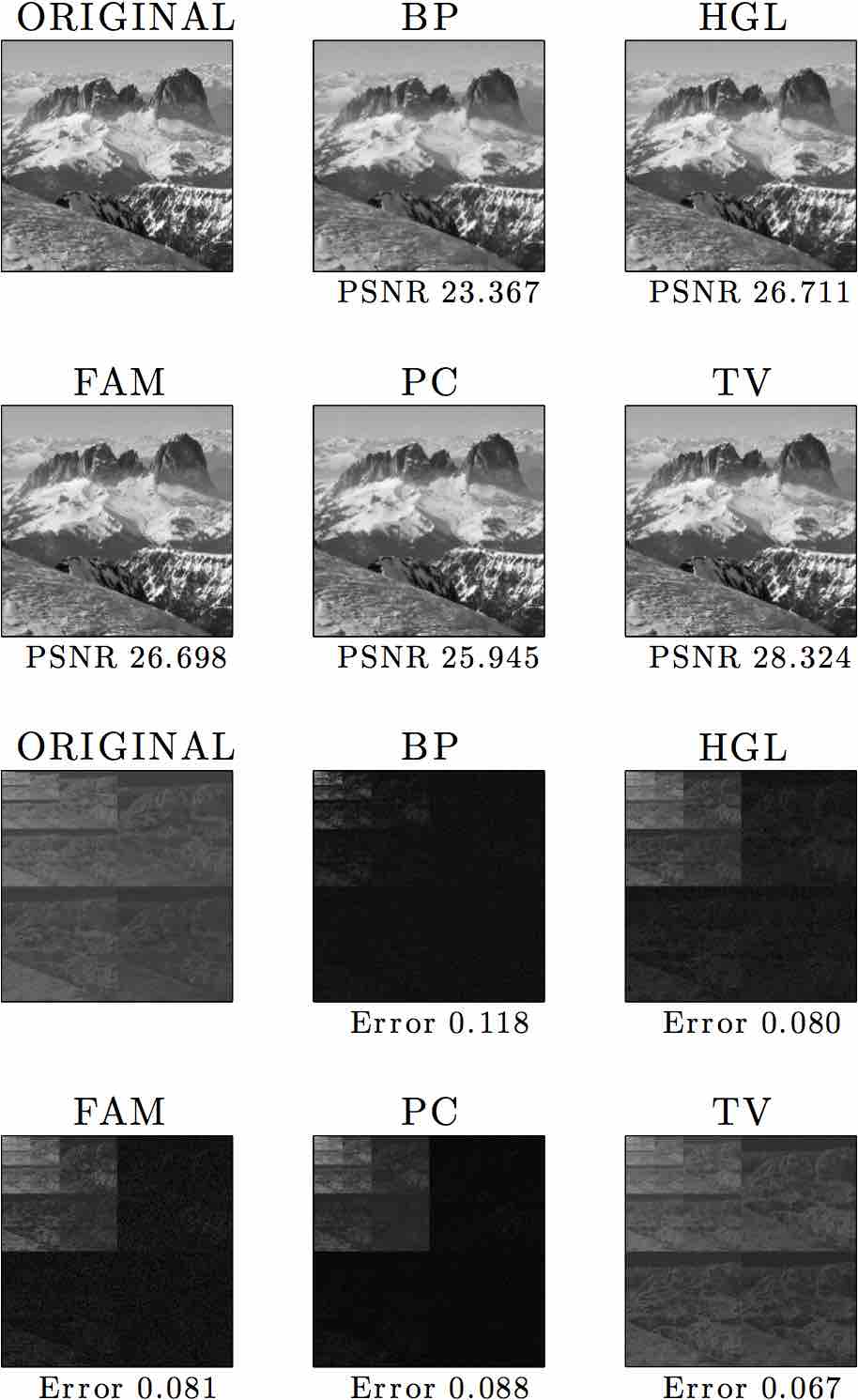}
\caption{\label{fig:cs_image_results_mountains} Mountains image recovery performance from compressive measurements. Here, $p = 2048$.  The top two rows show the reconstruction performance in the original domain, along with the PSNR levels achieved. The bottom two rows show the corresponding representations into the wavelet domain.}
\end{figure}

Figures~\ref{fig:cs_image_results_woman}-\ref{fig:cs_image_results_mountains} report the recovery results, using $\numsam = \frac{\dim}{8}$, that is using only $12.4\%$ samples compared to the ambient dimension.
The estimated images are on the top two rows, while the third and fourth rows show the estimated wavelet coefficients. 

The effect of imposing structured sparsity can be clearly seen for the HGL, PC and FAM models, where the high values of the coefficients tend to cluster around the root of the wavelet tree (i.e., top-left corner of the image) and their intensity decreases descending the tree. 
The family model shows the grouping among the siblings, where four leaves are either all zero or all non-zero.
For the $256 \times 256$ image, despite being coded in \texttt{C}, the discrete model is approximately $160$ times slower than the other methods, which are computationally equivalent: e.g., in our tests, the family model took around $60$ seconds, while the RC one required almost $2$ hours. 
We therefore did not use the RC model on the larger $2048 \times 2048$ mountain image, but we compared also against Total Variation (TV) pursuit, which obtains the best performance on this image.

\subsection{Neuronal spike detection from compressed data}

\begin{figure}[tb]
\captionsetup{width=0.8\textwidth}
  \begin{center}
    \includegraphics[width=0.4\textwidth]{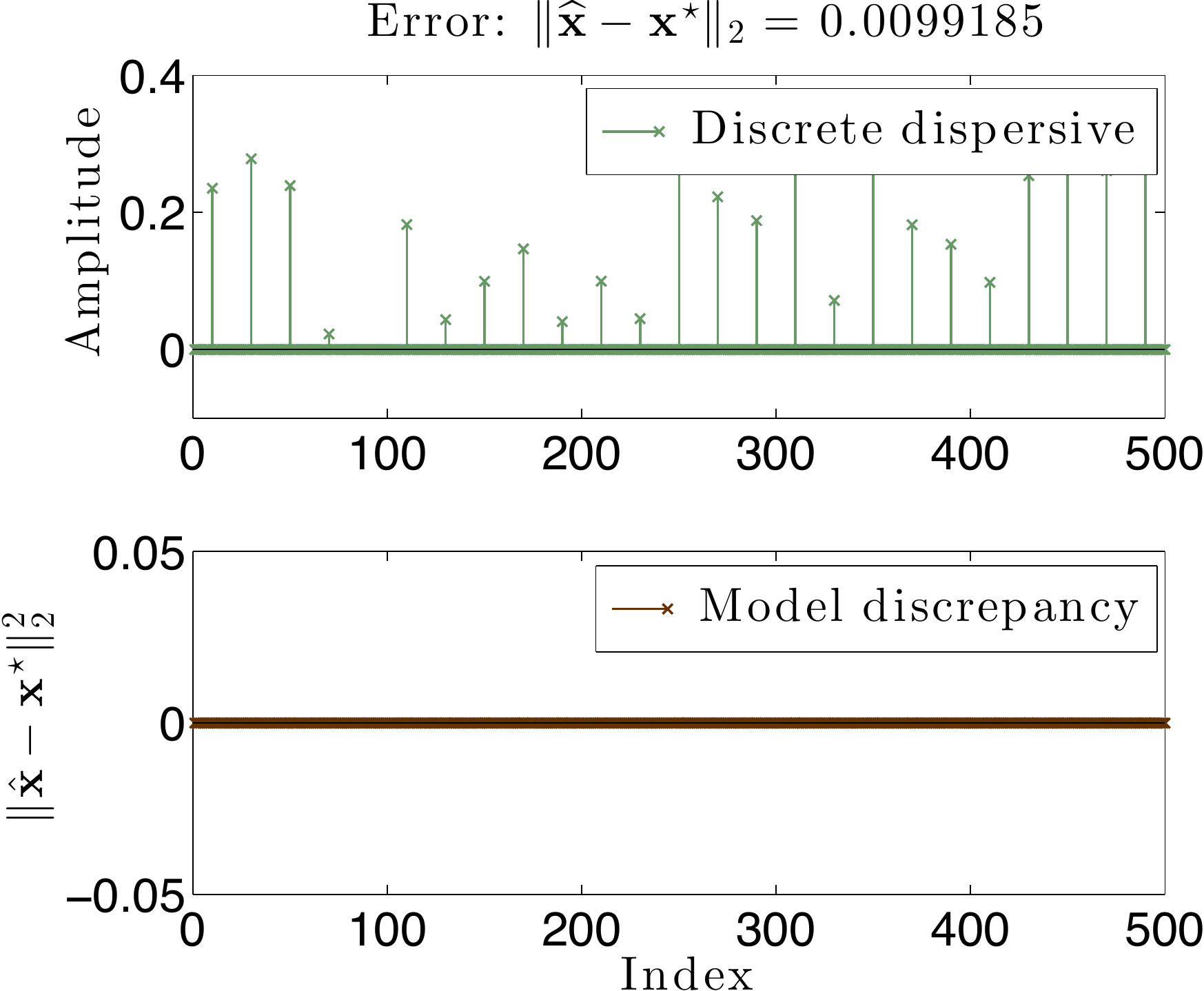}
  \end{center}
  \caption{Performance of the discrete dispersive approach for the problem spike train recovery from a limited set of linear measurements. }
  \label{fig:dispersive22}
\end{figure}

\begin{figure}[!ht]
\captionsetup{width=0.8\textwidth}
\centering
		\begin{subfigure}[b]{0.4\textwidth}
                \includegraphics[width=0.8\textwidth]{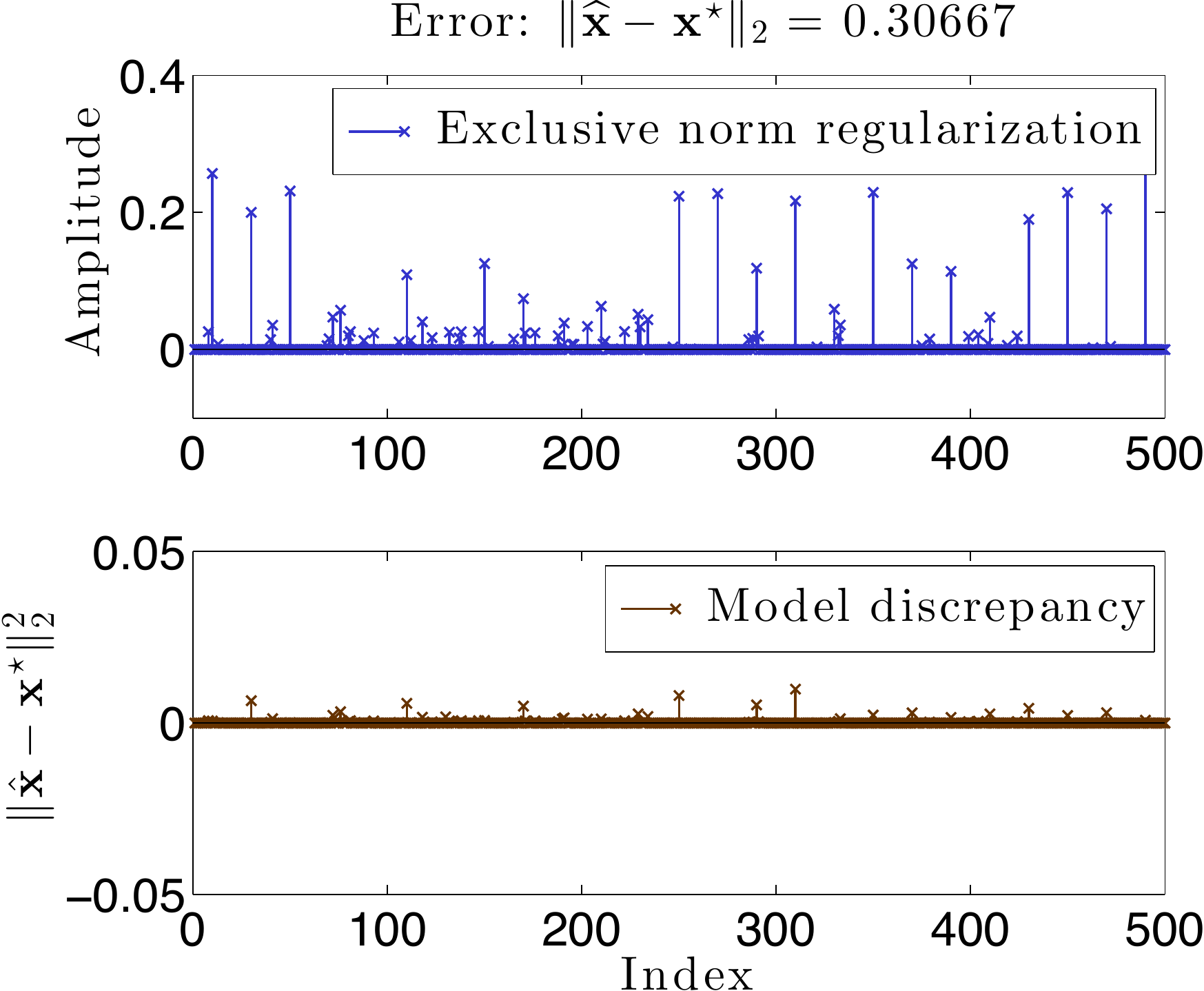} 
                \caption{Exclusive norm regularization}
                \label{fig:dispersive2b}
        \end{subfigure} 
		\begin{subfigure}[b]{0.4\textwidth}
                \includegraphics[width=0.8\textwidth]{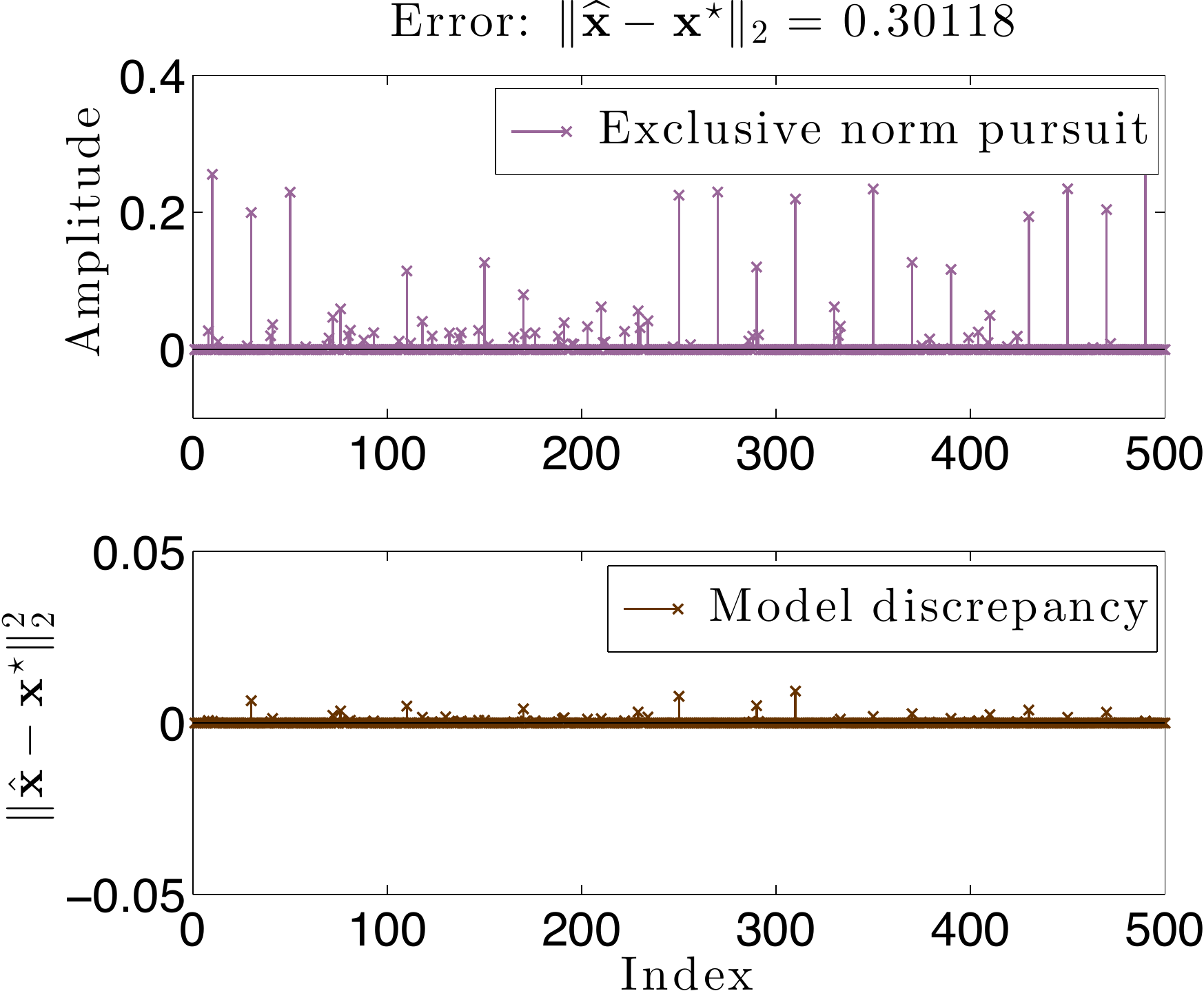} 
                \caption{Exclusive norm pursuit}
                \label{fig:dispersive2c}
        \end{subfigure}
\caption{\label{fig:dispersive2} Performance of dispersive convex relaxations for the problem of spike train recovery from a limited set of linear measurements. (\textbf{Left panel:}) Exclusive norm regularization approach. (\textbf{Right panel:}) Exclusive norm pursuit approach.}
\end{figure}

In the experiments that follow, we compare the performance of the following three optimization criteria, assuming the dispersive model $\dispersive$. 

\begin{equation}
	\begin{aligned}
	& \underset{\x \in \R^\dim}{\text{minimize}}
	& & f(\x)
	& \text{subject to} 
	&& \x \in \dispersive.   \quad \quad \quad \quad \quad \quad 
	\end{aligned} \label{eq:disp_discrete} \tag{Discrete dispersive}
\end{equation}
\begin{equation}
	\begin{aligned}
	& \underset{\x \in \R^\dim}{\text{minimize}}
	& & f(\x) + \lambda\cdot \Omega_{\text{exclusive}}(\x).   ~ \quad \quad
	\end{aligned} \label{eq:disp_exclusive} \tag{Exclusive norm regularization}
\end{equation}  
\begin{equation}
	\begin{aligned}
	& \underset{\x \in \R^\dim}{\text{minimize}}
	& & \Omega_{\text{exclusive}}(\x)
	& \text{subject to} 
	&& f(\x) \leq \sigma^2. 
	\end{aligned} \label{eq:disp_basis} \tag{Exclusive norm pursuit}
\end{equation}  
\noindent \textbf{Empirical performance on synthetic data:} 
Figures \ref{fig:dispersive22}-\ref{fig:dispersive2} illustrate the utility of each approach in the compressed sensing setting where $f(\x) := \frac{1}{2} \|\obs -\sensing\x\|_2^2$. That is, we observe $\x^{\star} \in \reals^\dim$ through a limited set of linear sketches $\obs = \sensing \x^{\star} + \noise \in \R^\numsam$ where $ \sensing \in \R^{\numsam \times \dim}$ is a known linear sketch matrix. 
Here, we assume $n = 500$ and $m = 70$ for $\|\x^{\star}\|_0 = 25$. Without loss of generality, we assume $\left(\x^{\star}\right)_i \geq 0, ~\forall i$ 
and $\Delta^{\star} = 20$. 

In the discrete case, we relax the refractory period $\Delta$ to model signal structure deviations; here, we assume $\Delta = 15$. The discrete exclusive model \cite{baraniuk2010model, needell2009cosamp} clearly outperforms the rest of the approaches under comparison; such behavior is also observed on average over the set of experiments conducted (Figure \ref{fig:dispersive22}). This also implies that the discrete model usually requires fewer measurements  
for accurate recovery compared to conventional sparse approximation, as long as the underlying signal approximately follows $\dispersive$. 

On the other hand, due to convex relaxations, convex approaches introduce unnecessary nonzero coefficients that do not comply with the underlying model. However, both approaches show good performance in recovering $\x^{\star}$ from limited measurements; see Figure \ref{fig:dispersive2}.

\medskip
\noindent \textbf{Real neuronal spike data:} In order to understand the functioning of the human brain, it is necessary to identify and study the behavior of neuronal
cell membranes under rapid change in the electric potential. However, to observe such phenomena, electrical activities on neurons need to be recorded using specialized equipment. In this experiment, we perform somatic spike detection of a tufted L5 pyramidal cell responding to in-vivo-like current injected in the apical dendrites and the soma simultaneously (see \cite{incf} for the experimental details). 

\begin{figure}[!htp]
\captionsetup{width=0.8\textwidth}
\centering
		\begin{subfigure}[b]{0.33\textwidth}
                \includegraphics[width=1\textwidth]{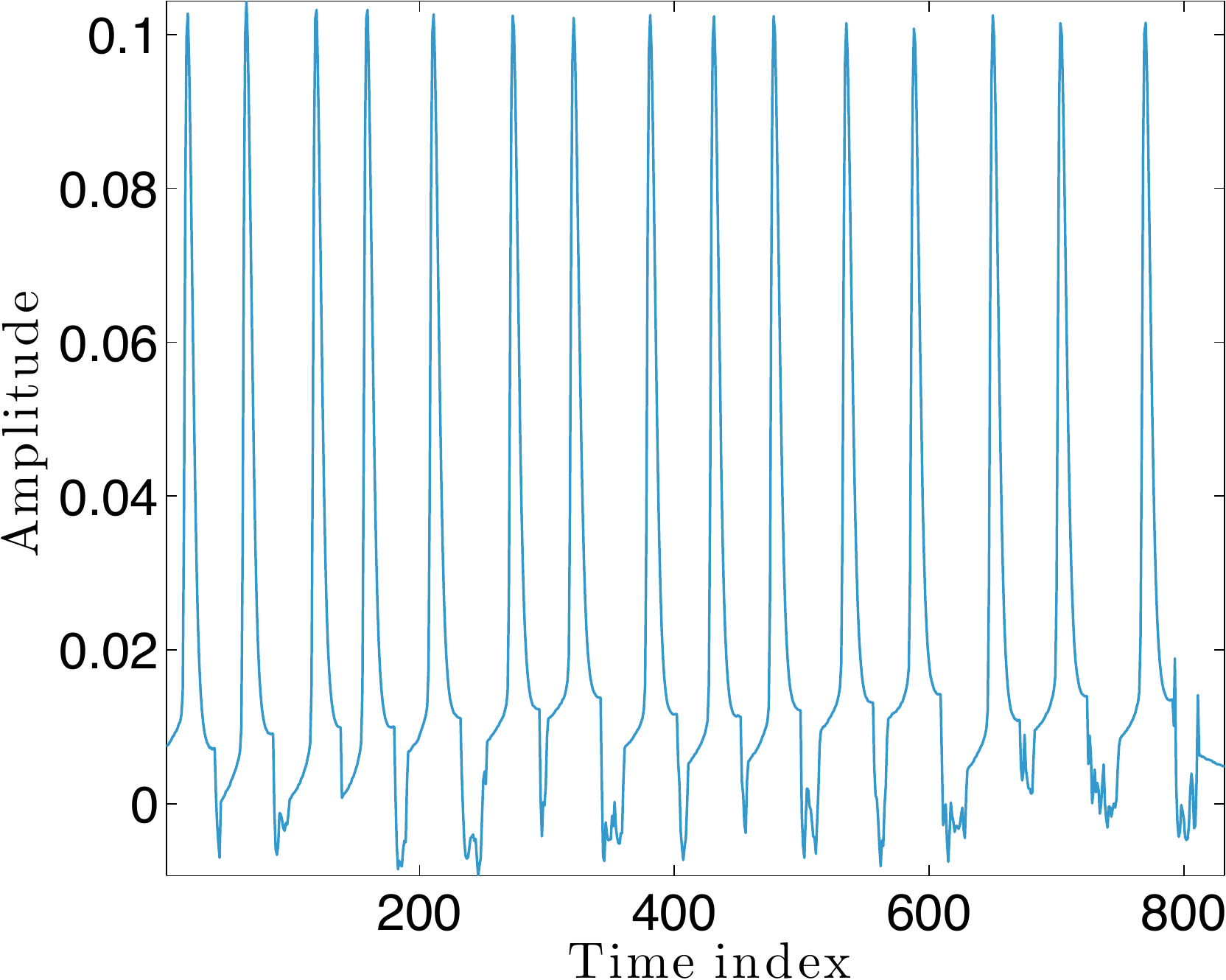} 
                \caption{Original neuronal spike train signal.}
                \label{fig:neuronal1a}
        \end{subfigure} \hspace{0.2cm}
		\begin{subfigure}[b]{0.33\textwidth}
                \includegraphics[width=1\textwidth]{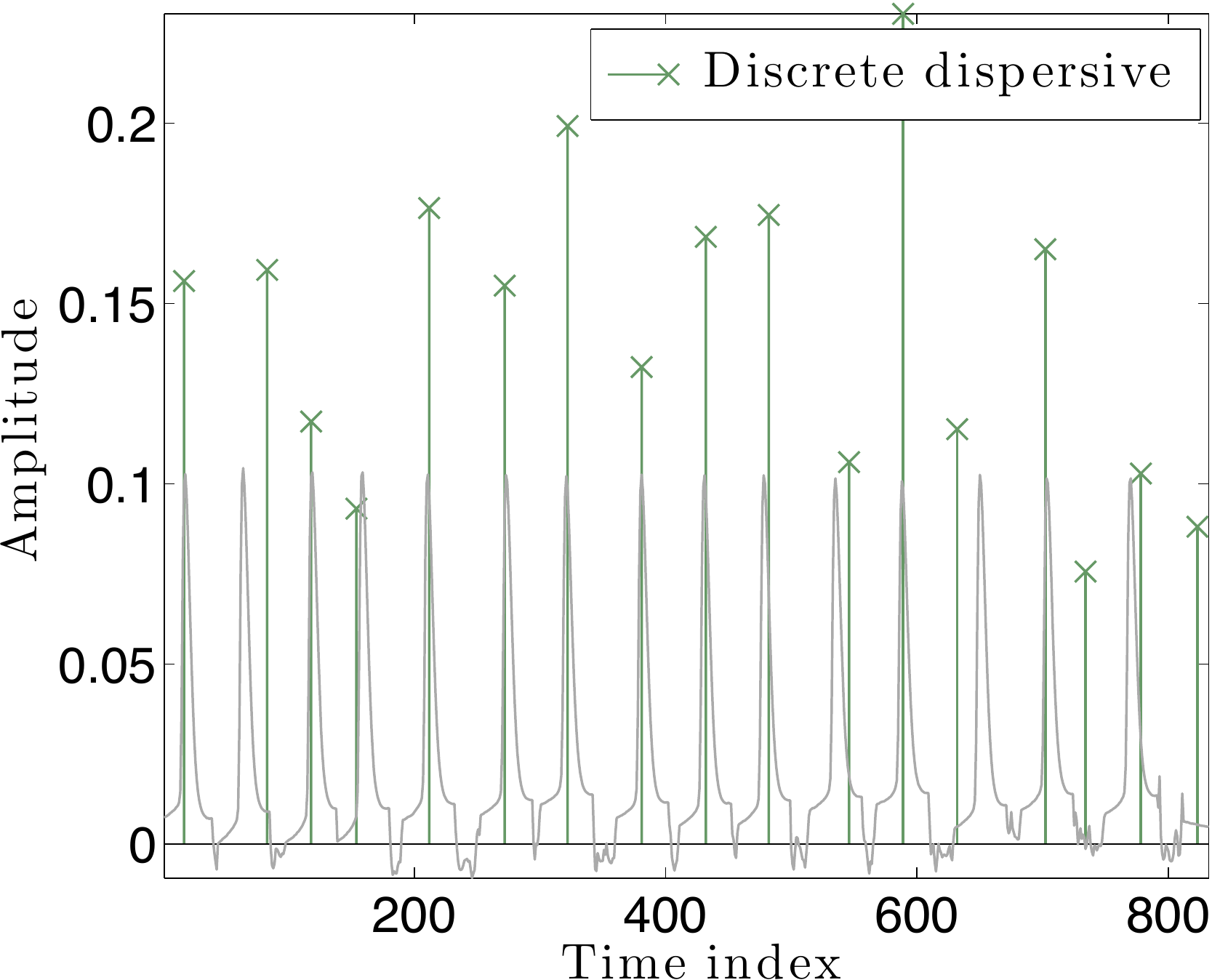}
                \caption{Spike detection using discrete model.}
                \label{fig:neuronal1b}
        \end{subfigure} \\ \medskip
		\begin{subfigure}[b]{0.33\textwidth}
                \includegraphics[width=1\textwidth]{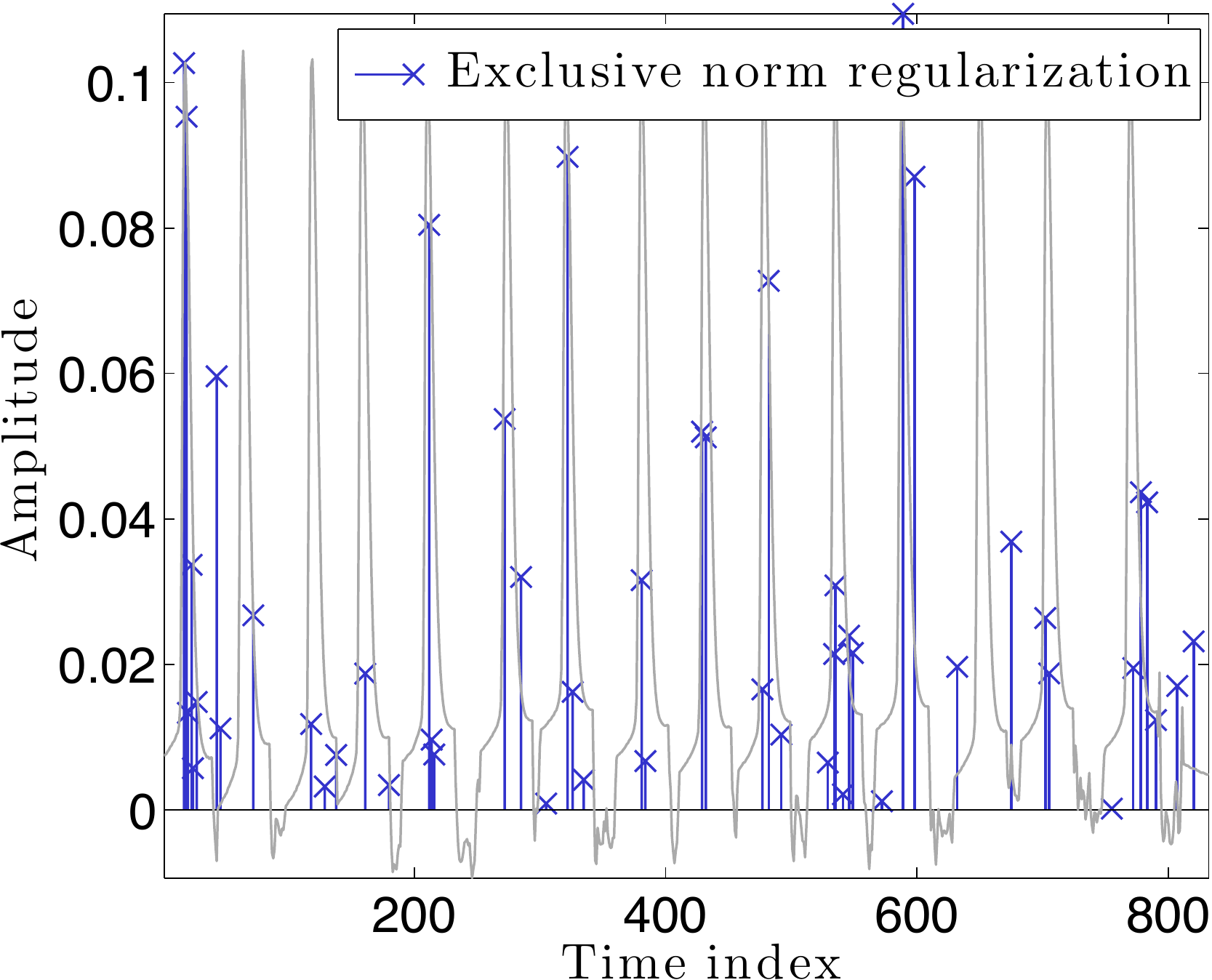} 
                \caption{Spike detection using the exclusive norm-regularized convex approach.}
                \label{fig:neuronal1c}
        \end{subfigure} \hspace{0.2cm}
		\begin{subfigure}[b]{0.33\textwidth}
                \includegraphics[width=1\textwidth]{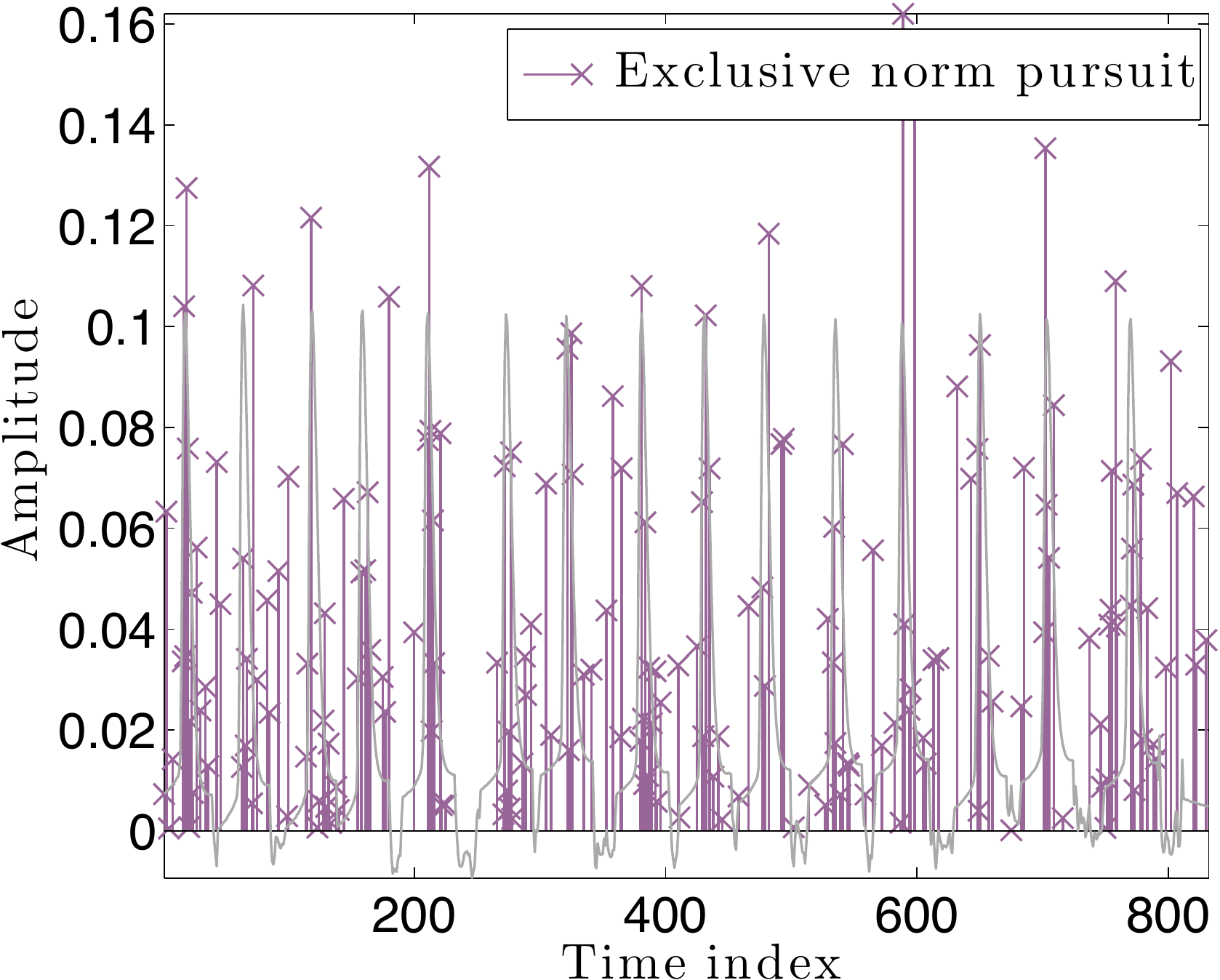} 
                \caption{Spike detection using the Model-based Basis pursuit.}
                \label{fig:neuronal1d}
        \end{subfigure} \\ \medskip
		\begin{subfigure}[b]{0.33\textwidth}
                \includegraphics[width=1\textwidth]{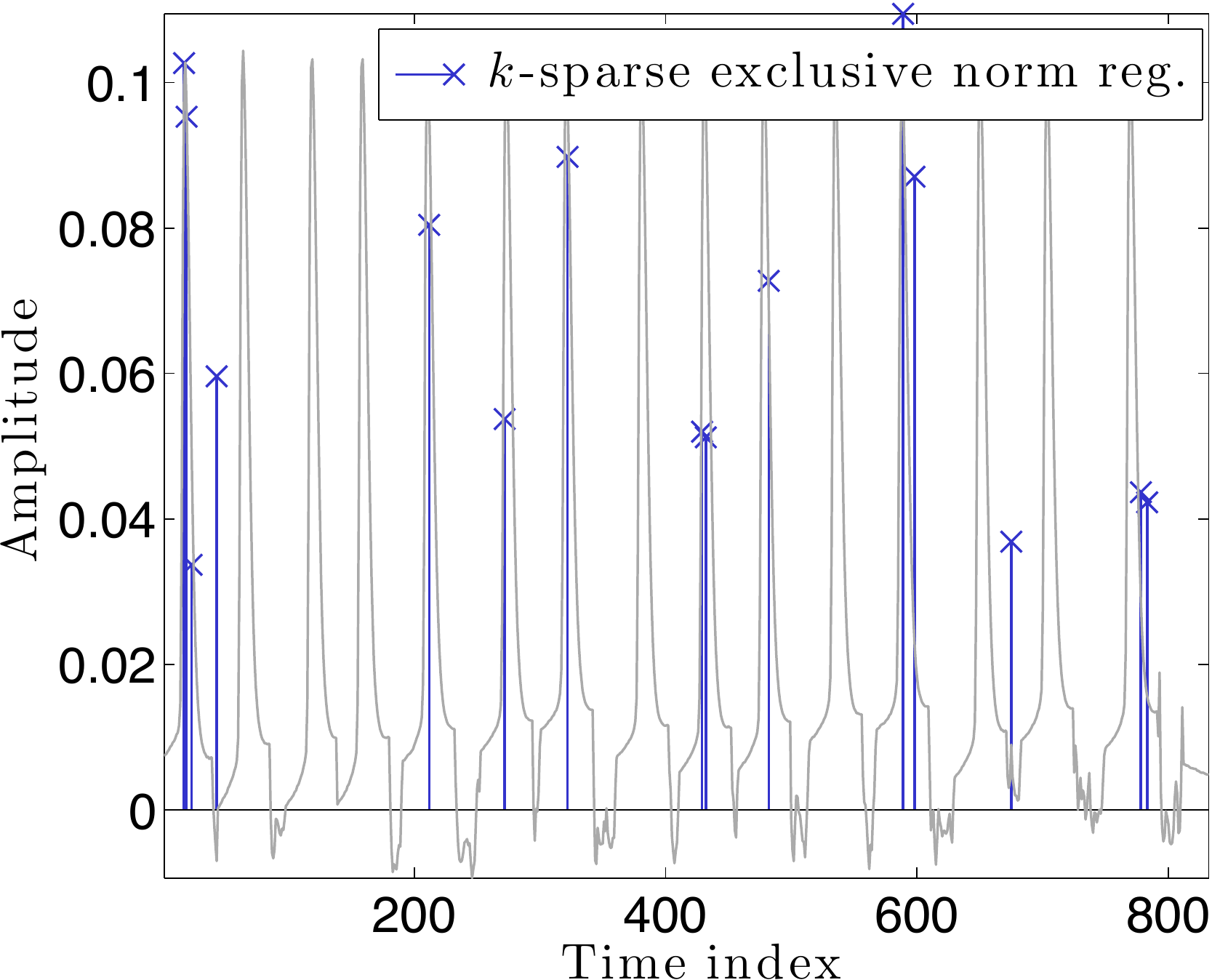} 
                \caption{$\sparsity$-sparse approximation of the exclusive norm-regularized convex solution.}
                \label{fig:neuronal1e}
        \end{subfigure} \hspace{0.2cm}s
		\begin{subfigure}[b]{0.33\textwidth}
                \includegraphics[width=1\textwidth]{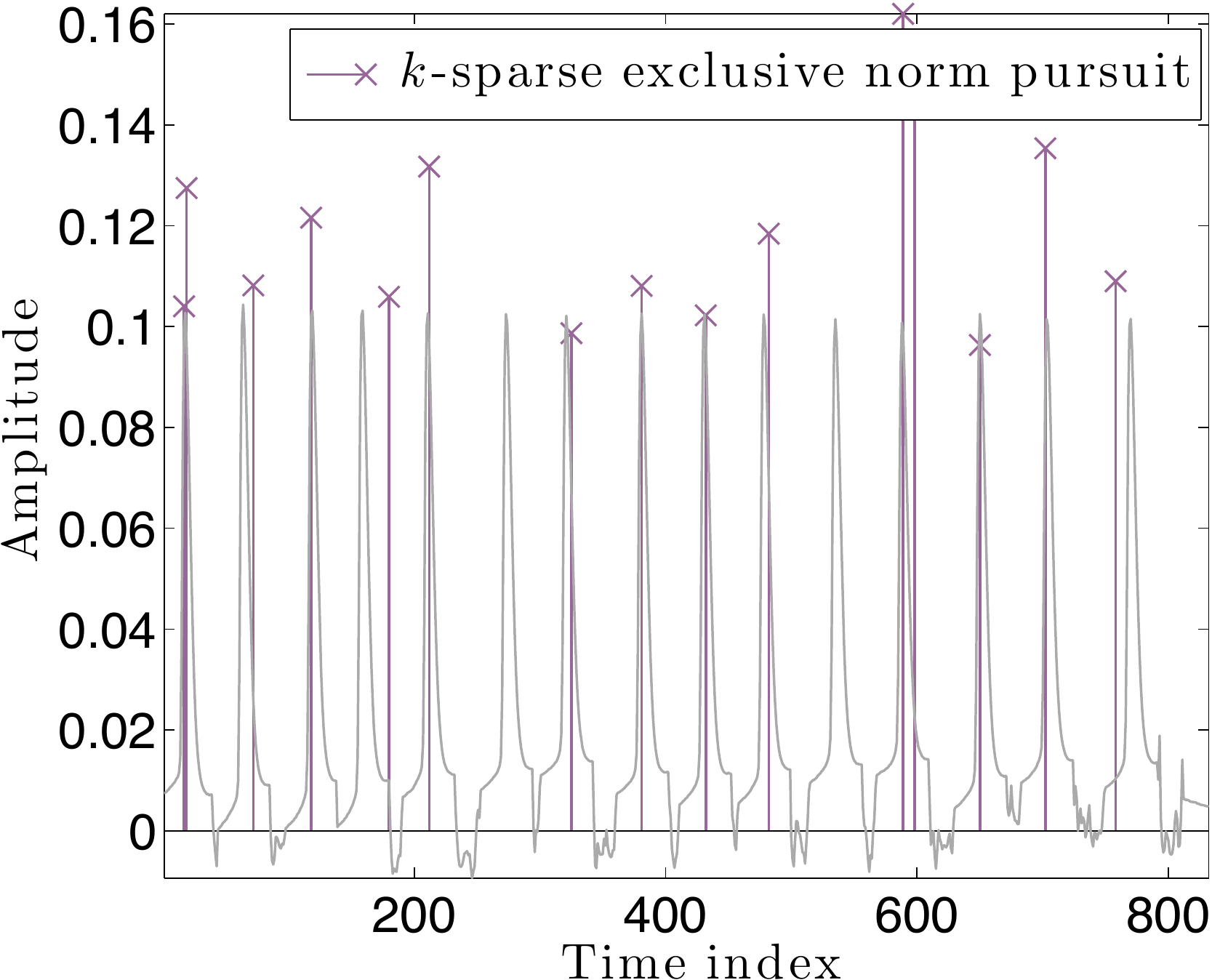} 
                \caption{$\sparsity$-sparse approximation of the model-based Basis pursuit solution.}
                \label{fig:neuronal1f}
        \end{subfigure}         
\caption{\label{fig:neuronal1} Spike detection in real neuronal data using the dispersive model. Figure \ref{fig:neuronal1a} depicts the original signal $\x^{\star}$. We observe $\obs = \sensing \x^{\star}$ using only $25\%$ measurements through a linear sketch $\sensing$. Figures \ref{fig:neuronal1b}-\ref{fig:neuronal1d} illustrate the performance of the three approaches under comparison. Figures \ref{fig:neuronal1e}-\ref{fig:neuronal1f} show the convex solutions, sparsified to be $\sparsity$-sparse.}
\end{figure}

A snapshot of the neuronal spikes waveforms is shown in Figure \ref{fig:neuronal1}. In order to accurately detect the neuronal spikes, a high-frequency sample acquisition equipment is required. Within this context, we apply CS ideas to decrease the number of samples needed to approximately detect the \emph{positions} of the spike train. Let $\x^{\star} \in \R^{\dim}$ with $\dim = 832$ represent the signal in Figure \ref{fig:neuronal1a}; furthermore, let $\sensing \in \R^{\numsam \times \dim}$ be the sensing matrix where \emph{$\numsam = 0.25 \cdot \dim$, i.e., we perform a $75\%$ compression. }

We use the proposed schemes to recover the locations of the neuronal spikes under the assumption of the dispersive model with refractory period $\Delta$. Here, $\Delta$ is set equal to the \emph{average} period between two consecutive spikes. 

Figure \ref{fig:neuronal1b} represents the recovered $\sparsity$-sparse approximation using the discrete dispersive model $\dispersive$. Here, $\sparsity$ is set to the number of spikes expected to appear for a given time period---such number can be easily deduced by observing the behavior of a specific neuron type. From Figure \ref{fig:neuronal1b}, we observe that the discrete model approximates the locations of the spikes quite accurately: most of the spike locations are exactly recovered. However, due to the ``strictness'' of the discrete model, we observe that small deviations from $\dispersive$ lead to imprecise estimations; e.g., between the $12^{\text{th}}$ and $13^{\text{th}}$ spike of the sequence, a larger (than usual) refractory period is observed that leads to mis-location of the next spike estimation.

Figures \ref{fig:neuronal1c}-\ref{fig:neuronal1d} depict the performance of convex solvers using the exclusive norm as $(i)$ regularizer and $(ii)$ objective function. Tweaking the $\lambda$ parameter in the $(i)$ case, one can achieve \emph{sparse} solutions that approximate the underlying model (Figure \ref{fig:neuronal1c}); however, one can observe multiple detected spikes with separation less than $\Delta$, violating the assumed model. In the model-based Basis pursuit case, the solver tries to \emph{fit} the solution to the data, which usually leads to less sparse solutions (Figure \ref{fig:neuronal1d}). One can further sparsity the convex solutions to obtain a $\sparsity$-sparse answer as in Figures \ref{fig:neuronal1e}-\ref{fig:neuronal1f}: however, in most cases, further processing of the returned signal is required to maintain a $\dispersive$-modeled solution. E.g., in this case, due to the fact that convex norms force the solution to fully \emph{explain} the observations, the sparsified solution includes more than one spike per true spike location.


\subsection{Digital Confocal Imaging}

Confocal imaging \cite{minsky1961microscopy,sheppard1997confocal} has become one of the best techniques for 3D imaging, due to its ability to reduce the blur caused by out-of-focus scattering by means of focus light and a physical pinhole.
The recent work of Goy et al. \cite{goy2012digital} combines this technique with digital holography which allows access to the complex field at the recording CCD sensor and enables a novel and flexible way of eliminating the out-of-focus contribution via a so-called ``virtual pinhole" in the digital domain. 
Under the assumption that the scattered light is not distorted by the matter in the light cones around the focus, this makes it possible to obtain very sharp images of the sample at all depths.
 
We are interested in modeling the scattering process so that we can instead leverage the information carried by the out-of-focus scattering for reducing the number of focusing locations that are needed for a faithful imaging of the sample.
We create a linear model by adopting the classical first Born approximation for the scattering \cite{born1999principles}. 
Our model takes into account the specific form of the Gaussian beam used to focus the light inside the sample, the numerical aperture of the transmission lens that determines which fraction of the scattered light is captured and the resolution and size of the CCD that records the hologram. For simplicity, we model the scattered field as it would arrive on the CCD and not its interaction with the reference field normally used to obtain the hologram.
We use a 3D discrete model for representing the imaged sample, where one 3D pixel, or voxel, contains the average value of the scattering potential in the volume contained therein. 

For a specific focus location $i$, the measurement model ${\bf A}_i$ maps a 3D object represented by its scattering potential $\bold{F}$ to a 2D complex field $\bold{S}$.
The complete measurement model ${\bf A}$ is then given by the concatenation of ${\bf A}_i$ for all focus points
$$
{\bf A}({\bf F}) = \left [ \begin{array}{c} \mathbf{A}_1({\bf F}) \\ \mathbf{A}_2({\bf F}) \\ \vdots \\ \mathbf{A}_P({\bf F}) \end{array} \right ] \; ,
$$
where $P$ is the total number of focus points.

For the experiments in this section, we consider a $64 \times 8 \times 64$ volume filled with oil with refractive index $1.52$, containing two concentric cylinders  with axis aligned with the $y$ axis. Their refractive indexes are $1.54$ for the outer one and $1.56$ for the inner one. A $x$--$z$ cross section of their scattering potential is shown in Figure \ref{fig:cylinder} (Left). Note that $\bf F$ is equal to zero outside the cylinder and has constant value inside each. These properties could be captured via sparsity and clusteredness, for example via $\ell_1$ norm or Total Variation minimization or via the Ising discrete model; see Section \ref{sec:submodular}.
We choose $16$ focus locations taken from a uniform grid over the central $x$--$z$ plane. The focusing lens and the transmission lens have numerical aperture $1$ and $0.65$ respectively, while the coherent light of the Gaussian beam has wavelength $\lambda=405 \cdot 10^{-9}$.
We consider a $91 \times 91$ CCD that covers the entire field of view of the transmission lens. 
We simulate the measurements by applying the measurement model to the synthetic scattering potential $\bf F$, that is $\obs = {\bf A}({\bf F})$, where we have ${\bf F} \in \reals^{64 \times 8 \times 64}$ and $\obs \in \mathbb{C}^{16 \times 91 \times 91}$.
Despite the apparent oversampling ratio, the problem is still ill-posed, because each recorded field carries information only about a small neighborhood around each focus point.

We compare four methods for estimating ${\bf F}$ from the measurements $\obs$. Let the data-fit term be the square loss $L({\bf F}) = \frac{1}{2} \|\obs - {\bf A}({\bf F})\|_2^2$.
\begin{equation}
\begin{aligned}
& \underset{\bf F}{\text{minimize}} &&L({\bf F}) \; .
\end{aligned} ~~\quad \quad \quad \quad \quad \quad \quad \quad  \quad \quad\tag{Conjugate Gradient (CG)}
\end{equation}
\begin{equation}
\begin{aligned} 
& \underset{\bf F}{\text{minimize}} && \|{\bf F}\|_1 &  \text{subject to} && \obs = {\bf A}({\bf F}) \; .
\end{aligned} ~~\quad \quad \quad \quad \tag{Basis Pursuit (BP)}
\end{equation}
\begin{equation}
\begin{aligned}
& \underset{\bf F}{\text{minimize}} && \|{\bf F}\|_{TV} &  \text{subject to} && \obs = {\bf A}({\bf F}) \; .
\end{aligned} \tag{Total Variation (TV) pursuit}
\end{equation}
\begin{small}
\begin{equation}
\begin{aligned}
& \underset{\bf F}{\text{minimize}} &&  L({\bf F}) + \lambda R_{\text{\textsc{Ising}}}(\supp({\bf F})) + \tau |\supp({\bf F})| \; ,
\end{aligned} \tag{Ising plus Cardinality (IC)}
\end{equation}
\end{small}
In the last approach, one has two regularization parameters $\lambda, \tau \geq 0$ to be set.

\subsubsection{Results}

We evaluate the recovery performance of the methods using the relative recovery error $\|{\bf \widehat{F}} - {\bf F}^*\|_2/\|{\bf F}^*\|_2$, where ${\bf F}^*$ is the synthetic scattering potential used to simulate the output ${\bf y}$.
We explored several pairs of regularization parameters $\lambda$ and $\tau$ for the discrete IC model, selecting the pairs that gave the best performance. Of course, in practice one cannot resort to this type of parameter exploration and needs to chose them according to knowledge regarding the noise level and the desired amount of clusteredness. Furthermore, this method is very sensitive to changes in the regularization parameters, making it difficult to find the best value for them.

\begin{figure}
\captionsetup{width=0.8\textwidth}
    \centering
        \includegraphics[width=0.7\columnwidth]{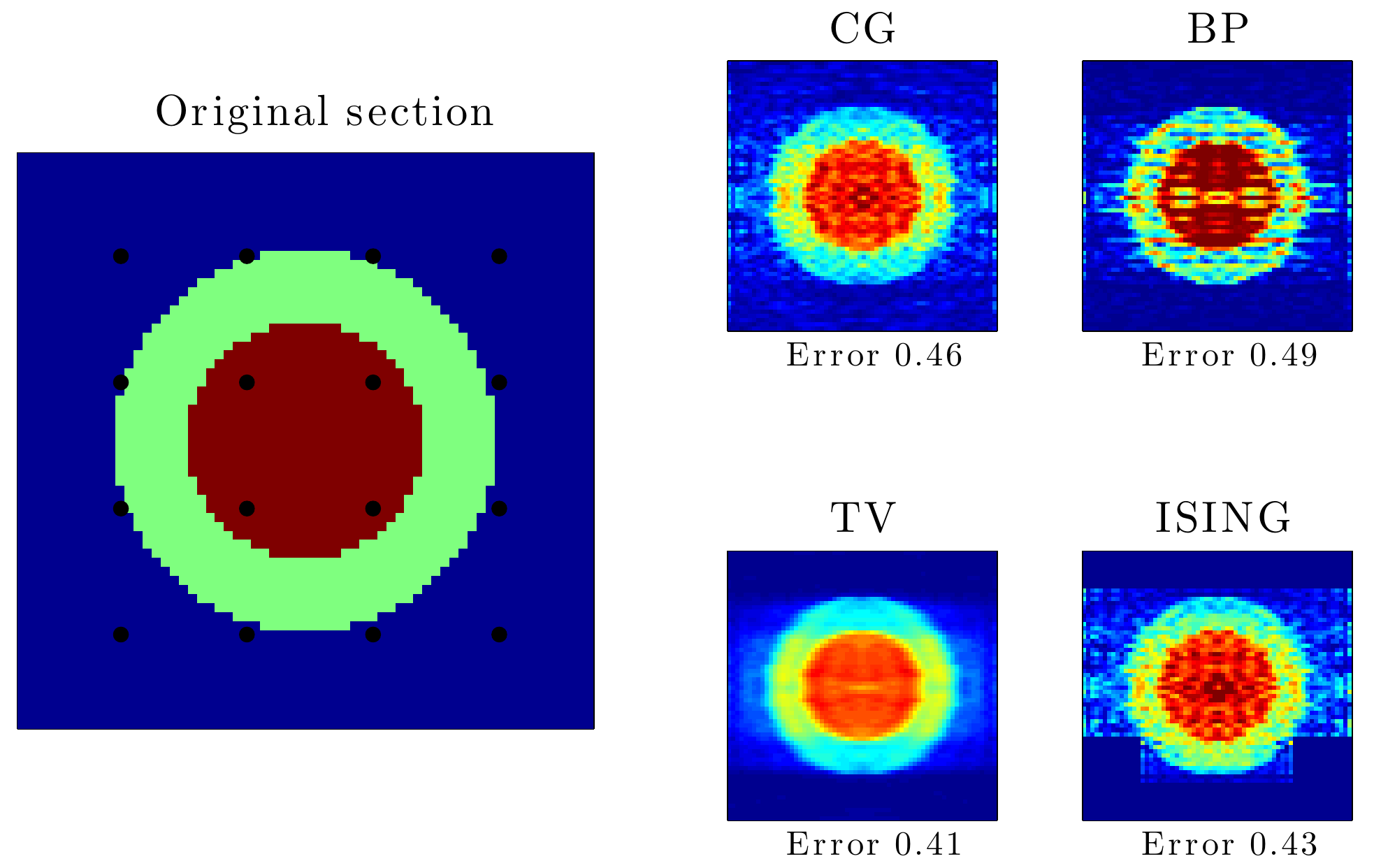}
    \caption{(Left) Central section of the sample to be reconstructed. The black dots represent the $16$ focus locations. (Right) Reconstruction results for the four methods.}
     \label{fig:cylinder}
\end{figure}

Figure \ref{fig:cylinder} (right) reports the results on this experimental setup. Total Variation pursuit achieves the best performance without having to tune any parameter, closely followed by the discrete model.
Despite the scattering potential is also sparse, Basis Pursuit enforces too much sparsity, especially in the regions of the sample that lie farthest away from the focus points and hence do not contribute to the the output $\obs$. It is interesting to note, that in this case, enforcing the incorrect structure leads to poorer performance than unstructured recovery.

%

%
%
